\documentclass[prx,aps,eqsecnum
,nofootinbib
,superscriptaddress
,letterpaper
,floatfix
,print
 ]{revtex4-2}
\usepackage{myStyle}
\usepackage{slashed}
\usepackage{framed}
\usepackage{hhline}
\usepackage{footnotebackref}
\usepackage{subcaption}
\usepackage{makecell}
\usepackage[normalem]{ulem}
\usepackage{wasysym}
\usepackage{mathbbol}

\graphicspath{{Images/},{Images/SpinOrbitOrders/}}

\def\arraystretch{1.0}
\usepackage{setspace} 
\newcolumntype{L}[1]{>{\raggedright\arraybackslash}p{#1}}
\newcolumntype{C}[1]{>{\centering\arraybackslash}p{#1}}
\newcolumntype{R}[1]{>{\raggedleft\arraybackslash}p{#1}}

\makeatletter
\def\hlinewd#1{%
	\noalign{\ifnum0=`}\fi\hrule \@height #1 %
	\futurelet\reserved@a\@xhline}
\makeatother

\newcommand{\matTone}{
\left( \begin{smallmatrix} 
	1 & 0 & 0\\
	0 & -1 & 0\\
	0 & 0 & -1\\
\end{smallmatrix} \right) }

\newcommand{\matTtwo}{\left( \begin{smallmatrix} 
	-1 & 0 & 0\\
	0 & -1 & 0\\
	0 & 0 & 1\\
\end{smallmatrix} \right)}

\newcommand{\matRy}{\left( \begin{smallmatrix} 
		1 & 0 & 0\\
		0 & 0 & 1\\
		0 & 1 & 0\\
	\end{smallmatrix} \right)}

\newcommand{\matCsix}{\left( \begin{smallmatrix} 
		0 & 0 & 1\\
		1 & 0 & 0\\
		0 & 1 & 0\\
	\end{smallmatrix} \right) }

\newcommand{\JJ}{\mathds{J}}
\newcommand{\Sq}{\mathrm{Sq}}

\newcommand{\psus}{\PSU(N)_{\mmrm{s}}}
\newcommand{\psuv}{\PSU(M)_{\mmrm{v}}}
\newcommand{\sus}{\SU(N)_{\mmrm{s}}}
\newcommand{\suv}{\SU(M)_{\mmrm{v}}}
\renewcommand{\sun}{ \SU(\Nf)_{}}
\newcommand{\gmicro}{G_{\mmrm{micro}}}
\newcommand{\glat}{G_{\mmrm{lat}}}
\newcommand{\gint}{G_{\mmrm{int}}}
\newcommand{\ug}{\U(1)_{\mmrm{g}}}

\newcommand{\be}{\begin{equation}}
\newcommand{\ee}{\end{equation}}
\newcommand{\bea}{\begin{eqnarray}}
\newcommand{\eea}{\end{eqnarray}}

\begin{document}

\title{\texorpdfstring{Theory of Dirac Spin-Orbital Liquids: \\		monopoles, anomalies, and applications to $\SU(4)$ honeycomb models}{DSL}}

\author{Vladimir Calvera}
\email{fvcalvera@stanford.edu}
\affiliation{Department of Physics, Stanford University, Stanford, CA 94305}
\affiliation{%
	Perimeter Institute for Theoretical Physics, Waterloo, Ontario N2L 2Y5, Canada
}%

\author{Chong Wang}%
\email{cwang4@pitp.ca}
\affiliation{%
	Perimeter Institute for Theoretical Physics, Waterloo, Ontario N2L 2Y5, Canada
}%

\date{\today}

\begin{abstract}
	Dirac spin liquids represent a class of highly-entangled quantum phases in two dimensional Mott insulators, featuring exotic properties such as critical correlation functions and absence of well-defined low energy quasi-particles. Existing numerical works suggest that the spin-orbital $SU(4)$ symmetric Kugel-Khomskii model of Mott insulators on the honeycomb lattice realizes a Dirac spin-orbital liquid, described at low energy by $(2+1)d$ quantum electrodynamics (QED$_3$) with $N_f=8$ Dirac fermions. In this work we generalize methods previously developed for $SU(2)$ spin systems to analyze the symmetry properties and stability of the Dirac spin-orbital liquid. We conclude that the standard Dirac state in the $SU(4)$ honeycomb system, based on a simple parton mean-field ansatz, is intrinsically unstable at low energy due to the existence of a monopole perturbation that is allowed by physical symmetries and relevant under renormalization group flow. We propose two plausible alternative scenarios compatible with existing numerics. In the first scenario, we propose an alternative $U(1)$ Dirac spin-orbital liquid, which is similar to the standard one except for its monopole symmetry quantum numbers. This alternative $U(1)$ state represents a stable gapless phase. In the second scenario, we start from the standard $U(1)$ Dirac liquid and Higgs the $U(1)$ gauge symmetry down to $\mathbb{Z}_4$. The resulting $\mathbb{Z}_4$ Dirac spin-orbital liquid is stable. We also discuss the continuous quantum phase transitions from the $\mathbb{Z}_4$ Dirac liquids to conventional symmetry-breaking orders, described by the QED$_3$ theory with $N_f=8$ supplemented with a critical charge-$4$ Higgs field. We discuss possible ways to distinguish these scenarios in numerics. We also extend previous calculations of the quantum anomalies of QED$_3$ and match with generalized lattice Lieb-Schultz-Mattis constraints.
\end{abstract}



\maketitle

\newpage
\tableofcontents
\newpage

\section{Introduction}

 Quantum spin liquids represent a class of exotic quantum phases that can emerge out of interacting spin systems. They host appealing features such as long-range quantum entanglement, absence of Landau symmetry-breaking order, and the emergence of gauge theories\cite{wen2004quantumBook}. Spin liquid phases have been detected in more and more systems in recent years, both experimentally in realistic materials, and numerically in simple lattice models\cite{savary_2017,Zhou_QSL_review}.

 Spin liquids can be roughly divided into two categories according to their behavior at low energy (temperature). Some spin liquids are well described by nearly free quasi-particles at low energy, even though the microscopic system may be strongly interacting. Examples include all gapped spin liquids, typically pocessing intrinsic topological orders, and spin liquids with gapless free low energy excitations, such as three dimensional $U(1)$ quantum spin liquids with free photons and the Kitaev spin liquid with free Majorana fermions.

 A more intriguing type of quantum spin liquids consist of those that do not have quasi-particle description. The simplest such spin liquids are described by some interacting critical (more precisely conformal) field theories (CFT) at low energy, and are some times called critical (or algebraic) spin liquids. The most well known critical spin liquid is perhaps the spin-$1/2$ antiferromagnetic Heisenberg model in one dimension, described by the Bethe ansatz solution and at low energy by the $SU(2)_1$ CFT. In two space dimensions, the simplest known critical spin liquids are $U(1)$ Dirac spin liquids (DSLs), described at low energy by gapless Dirac fermions $\psi$ coupled with an emergent $U(1)$ gauge field, also known as QED$_3$:
 \begin{equation}
 \label{QEDL}
 		\Lmc = \sum_{\a=1}^{N_{f}} \bpsi_{\a}(i\slashed{\partial}-\slashed{a})\psi_{\a} + \frac{1}{4g^{2}}f_{\m\n}^{2},
 \end{equation}
 where $f_{\m\n}$ is the field strength of the $U(1)$ gauge field $a$, $\psi_{\a}$ is a Dirac fermion with flavor indexed by $\a$, and $N_f$ is the total flavor number determined by further details of the states. It is known that at large enough $N_f$ the above theory flows to an interacting CFT at low energy\cite{Hermele04Stability}. The current numerical estimation of the lower critical $N_f$ is $N_c\sim 2$\cite{KarthikQED,KarthikQEDa}, which implies that DSLs, typically with $N_f\geq 4$, can indeed exist as critical spin liquid states. An important feature of Dirac spin liquids is that they are automatically gapless (critical) as long as certain symmetries are unbroken. This is in sharp contrast with more familiar critical field theories such as $O(N)$ Wilson-Fisher theories, and is the reason why Dirac spin liquids can be stable critical phases rather than just critical points.

Recent numerical computations suggest that DSLs may describe the ground states for several simple lattice spin systems: the spin-$1/2$ Heisenberg model on Kagome lattice\cite{he_2017}, the $J_1$-$J_2$ spin-$1/2$ model on triangular lattice\cite{Hu_2019}, and the spin-orbital $SU(4)$ symmetric Kugel-Khomskii model (the natural generalization of Heisenberg model to $SU(4)$ spins) on honeycomb lattice with one fundamental representation per site\cite{PhysRevX.2.041013}. The last one is different from the others in that the on-site continuous symmetry is $SU(4)$ rather than $SU(2)$, and is perhaps the simplest realizations of DSL in spin-orbital systems. The potential relevance of the $SU(4)$ model to realistic materials has also been discussed in Ref.~\cite{PhysRevLett.121.097201}. At low energy, all these systems are believed to be effectively described by the QED$_3$ theory, with $N_f=8$ for the $SU(4)$ system and $N_f=4$ for the $SU(2)$ spin systems. Related experimental systems appear to be promising although detailed comparisons with theories remain nontrivial and controversial\cite{savary_2017}. 

The QED$_3$ theory contains an important class of critical fluctuations known as the monopoles. Monopoles are defined as operators that insert $U(1)$ gauge flux into the system. By Dirac quantization the minimum flux that can be inserted is $\int_{S}da=2\pi$ where $S$ is a two-dimensional surface enclosing the point of flux-insertion in space-time. As the QED$_3$ theory flows to the CFT regime, the scaling dimension of the most relevant monopole is given by\cite{kapustin_2002,DyerMonopoleTaxonomy}
\be
\Delta_{\mathcal{M}}=0.265N_f-0.0383+O(1/N_f).
\ee
This result is calculated using the $1/N_f$ expansion. It turns out to be quite accurate even for moderate values of $N_f$\cite{KarthikMonopole}: for $N_f=4$ the error is only about $\sim 10\%$. An important consequence is that  for $N_f<12$ the monopole is relevant under RG  and leads to instabilities. Since the DSLs of our interest have either $N_f=4$ ($SU(2)$ spins) or $N_f=8$ ($SU(4)$ spins), the monopole is always a relevant operator that requires special attention. If the microscopic symmetries allow the monopole as a perturbation, the DSL will not be a stable gapless phase -- this turns out to be true for $SU(2)$ spins on the square lattice (also known as the staggered flux state)\cite{Alicea08,song2018spinon,song2018spinonNumeric}. On triangular and Kagome lattices, on the other hand, the relevant monopoles carry nontrivial symmetry quantum numbers and are forbidden as perturbations\cite{song2018spinon,song2018spinonNumeric}. This makes the DSLs stable on triangular and Kagome lattices, consistent with existing numerics. 

The monopole symmetry properties for the $SU(4)$ Dirac spin-orbital liquid have not been analyzed in the literature so far. In this work we will generalize the analytical methods previously developed mainly for $SU(2)$ spin systems and apply to the $SU(4)$ honeycomb Dirac spin-orbital liquid. We find that for the standard Dirac liquid based on a simple parton mean field ansatz, there is a relevant monopole operator that transforms trivially under all microscopic symmetries. This monopole will then leads to an instability, possibly towards symmetry breaking orders. This means that the numerics in Ref.~\cite{PhysRevX.2.041013}, which found a gapless spin liquid well described by the Gutzwiller projected wavefunction of the Dirac spin liquid, cannot be explained in the most straightforward manner. A relatively boring explanation is that the apparent DSL is merely a finite-size effect, and the system will eventually form a conventional symmetry-breaking order in the thermodynamic limit.

There are (at least) two alternative scenarios that are more interesting than just symmetry breaking, which we discuss in this paper:
\begin{enumerate}
\item In the first scenario, we propose an alternative $U(1)$ Dirac spin-orbital liquid. This alternative state is very similar to the standard one, also described at low energy by QED$_3$ theory with $N_f=8$. It however differs from the standard state in terms of monopole symmetry quantum numbers. In particular, all the relevant monopoles in the alternative DSL state carry some nontrivial symmetry quantum number, so there is no symmetry-allowed monopole perturbation to destabilize the gapless state. This alternative $U(1)$ DSL can be constructed using a parton construction, albeit slightly more complicated than the standard one. {One unsatisfactory aspect of this scenario is that the symmetry breaking orders in proximity to the alternative Dirac spin-orbital liquid do not seem to match with those observed in numerical simulations. For this reason this scenario may not be the most natural one.}

\item In the second scenario, we propose a $\mathbb{Z}_4$ Dirac spin-orbital liquid, which can be obtained from the standard $U(1)$ DSL state by Higgsing the $U(1)$ gauge symmetry down to $\mathbb{Z}_4$. Here $\mathbb{Z}_4$ is the minimum gauge symmetry that could preserve the $SU(4)$ global symmetry -- the analogous statement for $SU(2)$ spins with $\mathbb{Z}_2$ gauge symmetry is familiar. The $\mathbb{Z}_4$ DSL is free at low energy, so it represents a stable gapless phase. In terms of mean field ansatz, the $\mathbb{Z}_4$ state is indistinguishable from the $U(1)$ state -- this feature makes it compatible with the numerics based on Gutzwiller projected wavefunction\cite{PhysRevX.2.041013}. {We argue that the $\mathbb{Z}_4$ DSL can go through a continuous quantum phase transitions to certain conventional symmetry breaking orders, including some VBS states previously observed in numerical simulations. The transition is described by a QED$_3$ theory with $N_f=8$ supplemented with a critical charge-$4$ Higgs field. For this reason we consider this scenario more natural than the first one. }

\end{enumerate}

An interesting feauture of DSLs is the quantum anomalies associated with the QED$_3$ field theory. Such anomalies provide strong constraints on possible low energy fates of the theory. It is also appreciated recently\cite{Oshikawa2015,Cheng2016,ChoHsiehRyu,Jian2017,MetlitskiThorngren} as related to (generalized) Lieb-Schultz-Mattis theorems on the lattice scale\cite{lieb1961two,oshikawa2000commensurability,hastings2004lieb,Po2017}. The quantum anomalies for DSL have been partially calculated in the simplest cases in Ref.~\cite{song2018spinon}. In this work we extend the calculation to more general situations and obtain more complete results.

The rest of the paper is organized as follows. In Sec.~\ref{Generality} we discuss general aspects of $U(1)$ Dirac spin liquids with $N_f$ flavours, focusing on various symmetry properties including those of the monopoles. In Sec.~\ref{Nf=8Honeycomb} we focus on the honeycomb lattice system with $SU(4)$-fundamental spins. We carefully analyze the symmetry properties of the standard Dirac spin-orbital liquid state and conclude that it is unstable due to a symmetry-allowed monopole perturbation. In Sec.~\ref{sec:Alternatives} we discuss the two alternative scenarios mentioned above to realize gapless Dirac spin-orbital liquids in the $SU(4)$ honeycomb system. In Sec.~\ref{sec:GlobalSymmetriesLSM} we calculate the quantum anomalies of QED$_3$ for various symmetries and compare with the requirements from generalized Lieb-Schultz-Mattis theorems. We end with some discussions in Sec.~\ref{Discussions} and various details are included in the Appendices. 

\section{Dirac Spin Liquids with \texorpdfstring{$N_f$}{Nf} flavours}
\label{Generality}
\subsection{Emergent compact \texorpdfstring{$\QED_3$}{QED3}}
\label{DiracParton}

We are interested in describing $\sus$ spin systems on different lattices (the subscript is a label to keep track of the origin of the symmetry.). We consider systems where spins (with operators $ \Sbs^{}_{r} $) are in a  fully-antisymmetric representation of $\sus$, i.e. the Young tableau (YT) has only one column. These correspond to antisymmetric products of fundamental representations of $\sus$. 
We further restrict to systems that additionally have an anti-unitary time reversal symmetry (TRS) and lattice symmetries that include translations, rotations and reflections. We shall denote the former by $ \Tmc $ and the latter by $ \glat $. Then the microscopic symmetry group is 
\begin{equation}
	\begin{split}
		\gmicro= \glat \times ( \sus \rtimes \Tmc),
	\end{split}
\end{equation}
Note the semi-direct product between the spin symmetry and TRS as they may not commute.
 
We use the standard parton decomposition of $\SU(N)$ spins by introducing auxiliary degrees of freedom, called partons, by writing (for a general overview see Ref.~\cite{wen2004quantumBook})
\begin{align}\label{eq:partonConstruction}
\Sbf^{A}_{r}= \sum_{a,b=1,\dots,N} f_{\,r,}^{\dag \,a} \, (T^{A})_{a }^{\,\,b} \, f^{}_{r,b}\,,
\end{align}
where $f_{r,}^{\dag\,a}(f_{r,a}^{})$ are fermionic creation (annihilation) operators at site $r$ with $\sus$ flavour $a \in \{1,\dots,N\}$. The $f_{r,a}$ operators transform as $SU(N)_s$ fundamentals. $(T^{A})_{a}^{\,b}$ are the matrix representation of the (hermitian) $\sus$ generators acting on the fundamental representation. $A$ is an index in the adjoint representation and we normalize $T^{A}$ such that $\Tr[T^{A}T^{B}] = \half \d^{AB}$. 

In order to reproduce the original Hilbert space, we need to impose the constraint $Q_{r}= 0 $ on all physical states, where
\begin{align}
\label{eq:GaugeCharge}
Q_r \equiv \sum_{\alpha} f_{\,r,}^{\dag \,a}f^{}_{r,a} -n_{box} .
\end{align}
and $ n_{box} $ is the number of boxes in the representation's YT. The constraint makes the physical states to have exactly $n_{box}$ occupied states. 
Due to the anticommuting nature of fermions, the states will be a representation of the $n_{box}$-fold antisymmetric product of the fundamental representation as we want.

The next step is to write the spin Hamiltonian in terms of the partons and perform a mean-field (MF) decomposition that preserves the constraint only as an expectation value $\mel{GS_{MF}}{Q_{r}}{GS_{MF}} =n_{box}$. The constraint will later be enforced. We specialize to the case where the low-energy description of the MF Hamiltonian is of the form 
\begin{align}
H_{MF} = - \sum_{r,s}t_{rs}f^{\dag\,a}_{r, }f^{}_{s,a} 
\end{align}
such that the fermion bands only have Dirac cones at the filling imposed by the constraint $\expval{Q_{r}}_{MF}=0$ (i.e. there is no Fermi surface at this filling). In the rest of the paper the number of Dirac cones (also referred as valleys) on the MF spectrum at the appropriate filling shall be denoted by $M$ (recall that $M$ is even due to the parity anomaly).  We further assume that $\gmicro$ is projectively realized in $ H_{MF} $ so that no physical symmetry is  explicitly broken.

The decomposition in Eq.~\ref{eq:partonConstruction} has a gauge redundancy generated by $Q_{r}$ that maps $G_{g}:f_{r} \longrightarrow e^{i\phi_{r}}f_{r}$. In order to recover gauge invariance to $H_{MF}$, we introduce a $\U(1)$ gauge field $a_{\m}$ whose temporal and spatial components come from Lagrangian multiplier that imposes the constraint and $t_{rs} \longrightarrow t_{rs}e^{i a_{r,s}}$, respectively. The purpose of the gauge field $a_{\m}$ is to impose the constraint $Q_{r}=0$. 

The low energy physics of our model is described by the effective field theory with Lagrangian $\Lmc_{\mathrm{DSL}} = \Lmc_{\QED_{2+1}} + \d \Lmc_{}$, with
\begin{equation}
\label{eq:LagrangianQED}
		\Lmc_{\QED_{2+1}} = \sum_{\a=1}^{N_{f}} \bpsi^{\a}\Ds_{a}\psi_{\a} - \frac{1}{4g^{2}}f_{\m\n}^{2},
\end{equation}
where $\psi_{\a}$ are two-component Dirac fermions with a flavour index $\a = 1,\dots, \Nf =N\times M$. We work in the mostly plus signature $\h_{\m\n} =\diag(-1,+1,+1)$ and $(\g^{0},\g^{1},\g^{2}) = (-i\s^{2},\s^{3},\s^{1})$ are 3-dimensional gamma matrices. The adjoint Dirac fermion is defined as $\bpsi^{\a} = \lrRb{\psi^{\dag}}^{\a}C$ with $C =-i\g^{0}$. $f_{\m\n} = \pd_{\m }a_{\n}-\pd_{\n}a_{\m}$ and $g$ is the gauge coupling. $\Lmc_{\QED_{2+1}}$ corresponds to Quantum-Electrodynamics in 2+1 dimensions ($\QED_{2+1}$) and $\d\Lmc$ will correspond to operators of $\QED_{2+1}$ allowed by the microscopic symmetries of the original Hamiltonian. 

In addition to fermionic excitations, $\Lmc_{\QED_{2+1}}$ also allow for non-trivial topological configurations of the gauge field. These configurations carry charge under an extra $\U(1)$ symmetry of Eq.~\ref{eq:LagrangianQED} with conserved current $j^{\m} =\frac{1}{2\p}\ve^{\m\n\l} \pd_{\n}a_{\l}$. This symmetry is usually denoted by $\Utop$. We define \textit{local} operators $\Mmc^{\dag}_{q}$ that carry charge $q$ under $\Utop$. The insertion of this operators in the path-integral can be interpreted as the insertion of a $2\p q$-flux around a space-time point. $\Mmc^{\dag}_{q}$ will be referred as the bare monopole. In order to have gauge invariant opeartors, the bare monopoles need to be dressed by fermion zero modes which transform non-trivially under the flavour symmetry that mixes the Dirac fermions. 

We next focus on the symmetry properties of the building blocks of the simplest operators of $\QED_{2+1}$, namely fermion bilinears and $q=1$ monopoles. Large $N$ \cite{RantnerWen,Hermele05Mother,kapustin_2002} calculations of $\QED_{2+1}$ find that these are the lightest operators and therefore potentially the most important to describe the stability or near-by phases of the DSL. 

In order to understand how the microscopic symmetries act on the effective degrees of freedom, it is helpful to take detour and recognize the symmetries of $\Lmc_{\QED_{2+1}}$. The theory has the continuous Lorentz group ($\Spin(2,1)_{L}$) and discrete Lorentz symmetries (charge conjugation $\Cmc_{0}$, time reversal $ \Tmc_{0} $ and space reflection $\Rmc_{0y}$ \footnote{The subscripts denote that these are the 'bare' discrete symmetries that do not necessary match the corresponding physical symmetry.}). In addition to this, the theory naively has an internal symmetry group
\begin{align}
G_{\mathrm{int}}^{(\text{naive})}  = {\sun}_{\mmrm{f}}\times \Utop 
\end{align}
where ${\sun}_{\mmrm{f}}$ mixes the Dirac fermions and $\Utop$ acts only on the monopoles. In a later section, will see that the actual faithful symmetry acting on gauge invariant operators is a quotient of $G_{\mathrm{int}}^{(\text{naive})} $ by a discrete subgroup.

\subsection{Symmetries of the effective theory I: partons}\label{sec:SymmPartons}

As partons are not gauge-invariant operators, the microscopic symmetry do not necessarily act linearly on them. In order to understand this, Wen \cite{SpinLiquids} introduced the notion of the projective symmetry group (PSG): an element in the PSG corresponds to a unitary or anti-unitary transformation that maps partons into partons and leave the MF Hamiltonian invariant. The microscopic symmetries are then realized as elements of the PSG that make the original spin operators transform as they should when written in terms of partons. In general, the PSG has a subgroup that also leave the spin operators unmodified, commonly called the invariant gauge group (IGG). This subgroup ends up being the gauge group of the effective field theory. Because of the IGG, for every element $ g $ in the microscopic group there is a whole orbit of elements in the PSG that correspond $ g $.  For actual computations we fix an element in the PSG for each $ g\in \gmicro$ but we pay the price as the commutation relations are only satisfied up to IGG elements.

Once we know the action of $\gmicro$ on the partons, we can project the action of this symmetries down to the low-energy degrees of freedom read how $\gmicro$ is embedded in symmetry group of $\QED_3$. In general, we expect that $\sus$ will be embedded as a sub-group of the $\sun$ of $\QED_3$. On the other hand, elements in $ \glat $ and $  \Tmc $ will include discrete symmetries of $\QED_{3}$ in addition to $\sun$ elements. 

We now review how the symmetries of $ \QED_{3} $ act on Dirac fermions. The ${\sun}_{\mmrm{f}}$ and $\Spin(2,1)_{L}$ symmetries act on $\psi$ as usual ($\Umc \in \SU(\Nf)_{\mmrm{f}}$ and $\L \in \Spin(2,1)_{L}$):
\begin{subequations}
		\begin{align}
			\Umc: \psi_{\a}(x^{\m}) &\longrightarrow \Umc_{\a}^{\,\b}\psi_{\b}(x^{\m})\\
			\L: \psi_{\a}(x^{\m}) &\longrightarrow L[\L]\psi_{\a}(\L^{\m}_{\,\,\n} x^{\n})
		\end{align}
\end{subequations}
where $L[\L]$ is the matrix element of the spin-1/2 representation of $\Spin(2,1)_{L}$. The discrete symmetries can be chosen to act as 
\begin{equation}\label{eq:DiscreteLorentz}
	\begin{split}
		\Tmc_{0}: \psi_{\a}^{}(t,\rbs) & \longrightarrow  + \g^{0} \psi_{\a}^{}(-t,\rbs), \hspace{1cm} i \rightarrow -i , \\
		\Rmc_{0y}: \psi_{\a}^{}(t,\rbs) & \longrightarrow  i\g^{2}\psi_{\a}^{}(t,R_{y}\rbs), \\
		\Cmc_{0}: \psi_{\a}^{}(t,\rbs)  & \longrightarrow  +(\bar{\psi}^{\alpha}_{}(t,\rbf) C)^{\top}.\\
	\end{split}
\end{equation}
where $R_{y}= \diag(+1,-1)$ and $C = -i\g^{0}$ such that $C(\g^{\m})^{\top} C = -\g^{\m}$.

\subsection{Symmetries of the effective theory II: Monopoles}\label{sec:SymII:Monopoles}

\subsubsection{Introduction and continuous symmetries}
In the presence of a charge one bare monopole $\Mmc^{\dag}$, each Dirac mode contributes one fermion zero mode. The zero modes transform as Lorentz scalars and as the parent Dirac fermion under $\sun$. Gauge invariance requires half of the zero modes to be filled \cite{borokhov2003topological}. Therefore, there are $\binom{\Nf}{n}$, $n = \Nf/2$, different (complex) monopole operators that can be schematically written as 
\begin{equation}
\begin{split}
(\Phi_{}^{\dag})^{A} &= F^{\dag A }_{} \mathcal{M}_{}^{\dag} \\
F^{\dag A}_{} &= f_{}^{\dag [\alpha_1} \dots  f_{}^{\dag \alpha_n]}
\end{split}
\end{equation}
where $f_{}^{\dag \,\a}$ is the creation operator of the zero mode corresponding to the Dirac mode $\psi_{}^{\dag\,\a}$ and $A=[\alpha_1 ,\dots, \alpha_{\Nf/2}]$ is an antisymmetric multi-index\footnote{ Here as usual square brackets $[\dots]$ means antisymmetrization of the indices. For example, $T^{[\a\b]} \equiv \frac{1}{2}\lrRb{T^{\a\b}-T^{\b\a}}$. }. As $f$ transform in the fundamental representation of $\SU(\Nf)_{\mmrm{f}}$, $\Phi^{\dag}$ transform in $\Nf/2$-fold antisymmetric product of the fundamental representation. This turns out to be the irreducible representation of $\sun$ whose YT has one column and $\Nf/2$ rows. This representation is always self-conjugate with an invariant bilinear given by 
\begin{align}
E_{AB}=E_{[\a_{1}\dots\a_{n}][\b_{1}\dots\b_{n}]} \equiv \frac{1}{(\Nf/2)!}\ve_{\a_{1}\dots\a_{n}\b_{1}\dots\b_{n}}.
\end{align}
It is convenient to define $E^{AB} = E_{AB}$ so that we can use $E^{AB}$ and $E_{AB}$ to raise and lower the antisymmetric indices $A,B,\dots$. The overall factor is chosen such that $\sum_{\b_{1}\dots\b_{n}}E_{A[\b_{1}\dots\b_{n}]}E^{C[\b_{1}\dots\b_{n}]} = \hat{\d}^{C}_{A}$, with $\hat{\d}^{A}_{B}$ an identity tensor for the antisymmetric indices\footnote{i.e. \[\sum_{\b_{1}\dots\b_{n}}  \hat{\d}^{[\b_{1}\dots \b_{n}]}_{[\a_{1}\dots\a_{n}]}X_{\b_{1}\dots\b_{n}}=X_{[\a_{1}\dots\a_{n}]}. \] }. Notice that $E_{AB} = (-)^{n}E_{BA}$.  This means that for $\Nf \equiv 0 \mod 4$, the monopoles are in a orthogonal representation while for $ \Nf \equiv 2 \mod 4 $ the representation is simplectic.

As $\Mmc^{\dag}$ and the zero modes are Lorentz scalars, $\Phi^{\dag}$ are Lorentz scalars as well. From now on, it will be useful to think of $\Phi^{\dag}$ as abstract operators that are Lorentz scalars, have charge one under $\Utop$ and transform in the self-conjugate fully-antisymmetric representation of $\sun$. 

Before proceeding further, we discuss the faithful continuous symmetries acting on gauge invariant operators. Recall that the center of $\sun$ is made of matrices of the form $\w^{k} \id$ for $\w = e^{\frac{2\p i}{\Nf}}$ and integers $k= 0,\dots, \Nf-1$. Then there is no way to distinguish between the action of $\w\id$ and $-1 \in \Utop$ on the $\Phi^{\dag}$ monopoles. On the other hand, the whole center of $\sun$ is equivalent to the $\ug$ when acting on fermions. Therefore, the faithful internal symmetry group is 
\begin{align}
\gint = \frac{\sun \times \Utop}{\ZZ_{\Nf}}, 
\end{align}
where $\ZZ_{\Nf}$ is generated by  $(\w\id, -1) \in \sun \times \Utop$.

\subsubsection{Discrete Lorentz symmetries}

To understand the implementation of the discrete Lorentz symmetries ($\Cmc_{0},\,\Rmc_{0y}$ and $ \Tmc_{0} $), we look at how this symmetry operations affect gauge flux and their commutation properties with $\SU(\Nf)_{\mmrm{f}}$. To simplify our expressions, we only restrict to cases where $\Nf$ is divisible by $4$, which is the case in most of the previously studied DSL candidates. For more details and the $\Nf \equiv 2 \mod 4$ case see App.~\ref{app:MonopoleOperators}.

Notice that for all three discrete Lorentz symmetries, a gauge flux $\phi$ is mapped to $-\phi$ or in other words they flip the $\Utop$ charge. Therefore, we must have $\Phi \longleftrightarrow \Phi^{\dag}$. From the action of $\Cmc_{0},\,\Rmc_{0y}$ and $ \Tmc_{0} $ in Eq.~\ref{eq:DiscreteLorentz}, we see that $\Tmc_{0}$ and $ \Rmc_{0y} $ does not change the $\SU(\Nf)_{\mmrm{f}}$ representation while $\Cmc_{0}$ does exchange the fundamental and anti-fundamental representations. By appropriately choosing the phases in the definition of the monopole operators we can assume (see Appendix~\ref{app:MonopoleOperators} for details)
\begin{equation}
	\begin{split}
			\Tmc_{0}: \Phi^{\dag\,A} &\longrightarrow E^{AB}\Phi_{B} \\
		\Cmc_{0}: \Phi^{\dag\,A} &\longrightarrow \,\,\Phi_{A} \\
		\Rmc_{0y}: \Phi^{\dag\,A} &\longrightarrow E^{AB}\Phi_{B}
	\end{split}\quad,
\end{equation}
where the space-time arguments have been omitted. The phases were chosen such that $\Tmc_{0}^{2}=1$ and $\lrRb{\Cmc_{0}\Rmc_{0y}}^{2}=1$ when acting on monopole operators. 

\subsubsection{Microscopic symmetries, Berry phases and all that }

We now want to find how $ \gmicro $ are embedded in $\gint$. It is convenient to break $\sun \rightarrow \sus\times \suv$ because $\gint $ becomes 
\begin{align}
\gint'  = \PSU(N)_{\mmrm{s}}\times \PSU(M)_{\mmrm{v}}\times \Utop
\end{align}
where $\PSU(K)\equiv \SU(K)/Z(\SU(K))$, for $Z(K)$ the center of $ K $. We do this because we expect the action of spin symmetry and lattice symmetries to decouple inside of $\gint$. An advantage of this restriction is that there is no ambiguity in what we call an element of $\Utop$ and of $ \SU(\Nf) $. 

The spin rotations are determined solely by the action of $\psus\subset \gint'$ on $\psi$. Similarly, we also know how lattice symmetries are embedded in  $ \psuv\times \SO(2,1)_{L}\rtimes <\Tmc_{0},\Cmc_{0},\Rmc_{0y} > $ from the action on $\psi$. The only missing information is the $\Utop$ factor. These can be thought of as Berry phases the monopole acquires by moving around the lattice with the partons fixed. It can be found either analytically, by using techniques from band topology \cite{song2018spinon}, or numerically, by calculating expectation values of free fermions hopping in a magnetic field \cite{Alicea08,song2018spinonNumeric}. 

Once we have the symmetry properties of the monopole operators under the microscopic symmetries, we know which operators will generally be present in the Lagrangian. For small enough $\Nf$ (less than about $12$), the single monopole operator is expected to be a relevant perturbation to $\Lmc_{\QED_{2+1}}$ and thus if there is a 'trivial' monopole, i.e. a monopole that is invariant under microscopic symmetries, the theory will flow to strong coupling were a Dirac mass is expected to be generated in addition to monopole proliferation that will gap $a_{\m}$. 

Another prediction we can make is which symmetry breaking phases are proximate to the DSL. We assume that transitions happen by condensation of a Dirac mass with a subsequent monopole proliferation for small enough $\Nf$. This can be captured by introducing additional bosonic degrees of freedom that couple linearly to the Dirac mass or equivalently we can add a Gross-Neveu term to the Lagrangian. If the generated mass is $\bpsi^{\a}\psi_{\a}$, a Chern-Simons term is generated, monopole proliferation is suppressed and we are left with $(\Spin_\CC)_{\Nf/2}$ topological order -- {in the standard $U(1)$ Chern-Simons language the $K$ matrix is a $(\Nf/2-1)\times (\Nf/2-1)$ matrix with $-2$ on the diagonal and $-1$ everywhere else \cite{Geraedts2017}}. On the other hand, any of the other adjoint masses will not generate a Chern-Simons term.
Instead, it breaks the degeneracy between the Dirac fermions and the zero-modes. If we restrict to adjoint masses of the form $\bpsi^{\a}T_{\a}^{\,\,\b}\psi_{\b}$, with $T^{2}=\id$\footnote{This ensures that $T$ has no zero eigenvalue.}, the degeneracy is completely broken and there is a unique preferred monopole that proliferates. This could break more symmetries than the adjoint mass. In particular, the lightest monopole $\Phi^{\dag}$ will have the modes with negative $T$ eigenvalues filled.

\subsection{Monopole Wavefunctions}\label{sec:monopoleWF}

When discussing actual calculations it is useful to have extra notation for monopoles. We introduce orthogonal tensors\footnote{$ \chi_{\r}^{A}\lrRb{\chi_{\s}^{A}}^{*} = \d_{\s}^{\r} $} $\{\chi_{\r}^{A}\}_{\r = 1,\dots, \mathrm{dim}R}$ with $\mathrm{dim}R = \binom{\Nf}{\Nf/2}$ and define $\Phi^{\dag}_{\r} = \chi_{\r}^{A}E_{AB}\Phi^{\dag\,B}$. We shall refer to $\chi_{\r}^{A}$ as monopole wave-functions (MWF). 

For $\Nf= 0\!\mod 4$, the discrete bare symmetries act as \footnote{For $\Nf= 2\!\mod 4$ we just need to introduce the appropiate extra factors of $i$ or $-1$ as detailed in App.~\ref{app:MonopoleOperators}. }
\begin{equation}
\begin{split}
\Tmc_{0}: \Phi^{\dag}_{\r} \longrightarrow &  \lrRb{\chi_{\r}^{A*}E_{AB}\chi_{\s}^{B}}\Phi_{\s} \\
\Rmc_{0y}:		\Phi^{\dag}_{\r} \longrightarrow  & \lrRb{\chi_{\r}^{A}E_{AB}\chi_{\s}^{B}}\Phi_{\s}^{}\\
\Cmc_{0}:		\Phi^{\dag}_{\r} \longrightarrow  	&  \lrRb{\chi_{\r}^{A}\chi_{\s}^{A}}\Phi_{\s} 
\end{split}\,\,
\end{equation}
If an element  $U \in \SU(\Nf)_{\mmrm{f}}$ acts on Dirac fermions as $\psi_{\a} \longrightarrow U_{a}^{\,\,\b}\psi_{b}$, then $\Phi^{\dag}_{\r} \longrightarrow  U_{\r\s}\Phi^{\dag}_{\s}$ with
\begin{align}
U_{\r\s} = \chi_{\r}^{[\a_{1}\dots \a_{n}]}			U_{\a_{1}}^{\,\,\b_{1}}\cdots U_{\a_{n}}^{\,\,\b_{n}} \lrRb{\chi_{\s}^{[\b_{1}\dots \b_{n}]}}^{*}.
\end{align}

As in the discussion of microscopic symmetries we break $\sun \rightarrow \sus\times\suv$, it is convenient to also decompose the  MWF.  In order to achieve this, first notice that under the branching $\SU(\Nf)_{\mmrm{f}}  \rightarrow \SU(N)_{\mmrm{s}}\otimes \SU(M)_{\mmrm{v}}$, the antisymmetric representations with $k$ boxes, $\varpi_{k}$, decomposes as  \footnote{This is a consequence of the dual Cauchy identity and the the expression of the characters of the irreps of the $\SU$ groups in terms of Schur polynomials \cite{bump2013lie}.}
\begin{align}\label{eq:Decomp}
(\varpi_{k})\longrightarrow \bigoplus_{\l: \abs{\l}=k} (\l,\l^{\top}),
\end{align}
where the sum is over all YTs $ \l $ with $k$ boxes\footnote{{As usual we discard YTs with more than $N$ or $M$ columns.}}. $\l^{\top}$ denote the transposed of $\l$ along the main diagonal. The first (second) factor on RHS correspond to the representation of $\sus (\suv)$. 
Then we introduce tensors with $\sus$ and $\suv$ indices, $\chi_{\rho^{(s)}}^{[s] A_{s}}$ and $\chi_{\rho^{(v)}}^{[v] A_{v}}$ that transform in the various $(\l)$ representations and define
\[ \tilde{\chi}_{\rho}^{\,\, \alpha_1 \dots \alpha_n} = \chi_{\rho^{(s)}}^{[s]\,\, \alpha_1^{(s)} \dots \alpha_n^{(s)}}  \chi_{\rho^{(v)}}^{[v] \,\, \alpha_1^{(v)} \dots \alpha_n^{(v)}}. \]
The $\SU(N_f)$ indexes have been split into $\SU(N)_\mmrm{s} \times \SU(M)_\mmrm{v}$ indexes as $\alpha=(\alpha^{(s)},\alpha^{(v)})$. We can then antisymmetrize in $\a_{1},\dots,\a_{n}$ and introduce extra normalization factors to make them orthogonal. A nice feature of this prescription is that it is easier to find the transformations of the monopoles under $\SU(N)_\mmrm{s} \times \SU(M)_\mmrm{v}$ and the discrete symmetries.

We remark that the introduction of MWF is just a calculation tool that allow us to better see how the symmetry acts on the monopoles and confirm analytical arguments. When presenting the results it is not necessary to specify the MWF as we can still think of the monopoles as abstract operators with the corresponding symmetry transformations.

\section{Candidate quantum spin-orbit liquid on the Honeycomb lattice and compact \texorpdfstring{$\QED_3 \,\, N_f=8$}{QED3 Nf=8}}
\label{Nf=8Honeycomb}
Ref.~\cite{PhysRevX.2.041013} found some signatures of a stable algebraic-spin liquid on a $\SU(4)$ symmetric Heisenberg model on the Honeycomb lattice where the spins are in the fundamental repreentation. The authors performed a Gutzwiller projection on fermionic partons on a $\pi$-flux quadratic Hamiltonian that displays Dirac cones at quarter filling. Inspired by this results, we analytically study the low-energy properties of this parton construction by introducing gauge field fluctuations in lieu of the Gutzwiller projection.

\subsection{Model and symmetries}
The model considered in Ref. \cite{PhysRevX.2.041013} had a Hamiltonian 
\begin{equation}
H=\sum_{<i,j>} \left(2 \Sbf_i\cdot \Sbf_j +\frac{1}{2} \right) \left(2 \Tbf_i\cdot \Tbf_j +\frac{1}{2} \right) 
\end{equation}
where the sum is over nearest neighbours. $\Sbf$ are spin $1/2$ operators ($\uparrow$ and $\downarrow$ states) and $\Tbf$ are pseudo-spin $1/2$ operators (states $a$ and $b$). We can group this flavours together to a $\SU(4)$ basis: $\ket{\tikzcircle[red,fill]{3pt}} = \ket{\uparrow a}$, $\ket{\tikzcircle[green,fill]{3pt}} = \ket{\downarrow a}$, $\ket{\tikzcircle[blue,fill]{3pt}} = \ket{\uparrow b}$ and $\ket{\tikzcircle[yellow,fill]{3pt}} = \ket{\downarrow b}$. In this new basis, the Hamiltonian can be rewritten as an exchange operation $H=\sum_{<i,j>}P_{i,j}$ where the $\SU(4)_s$ symmetry is evident. 

The microscopic symmetries of the model are the $\SU(4)_s$ spin symmetry, time reversal $\Tmc$ and lattice symmetries. We will use two sets of Pauli matrices $\s^{a}$ and $\t^{a}$ to generate the $\mathfrak{su}(4)_s$ algebra by setting $\Sbf^a=\frac{1}{2}\s^a $ and $\Tbf^a=\frac{1}{2}\t^a $. Time reversal is an anti-linear operator that satisfies $\Tmc:\Sbf\rightarrow-\Sbf$ and $\Tmc:\Tbf\rightarrow-\Tbf$. This relation can be written as $ \Tmc[T^{A}] = (O_{\Tmc})^{A}_{B}T^{B} $, for $T^{A}\sim \frac{1}{4}\s^{a}\t^{b}$ the generators of $\mathfrak{su}(4)_{s}$. In this basis for $\mathfrak{su}(4)_{s}$, $O_{\Tmc}$ is a diagonal matrix.

\begin{figure}[h]
	\centering
	\begin{subfigure}[t]{0.45\textwidth}
		\centering
	\includegraphics[height=6cm]{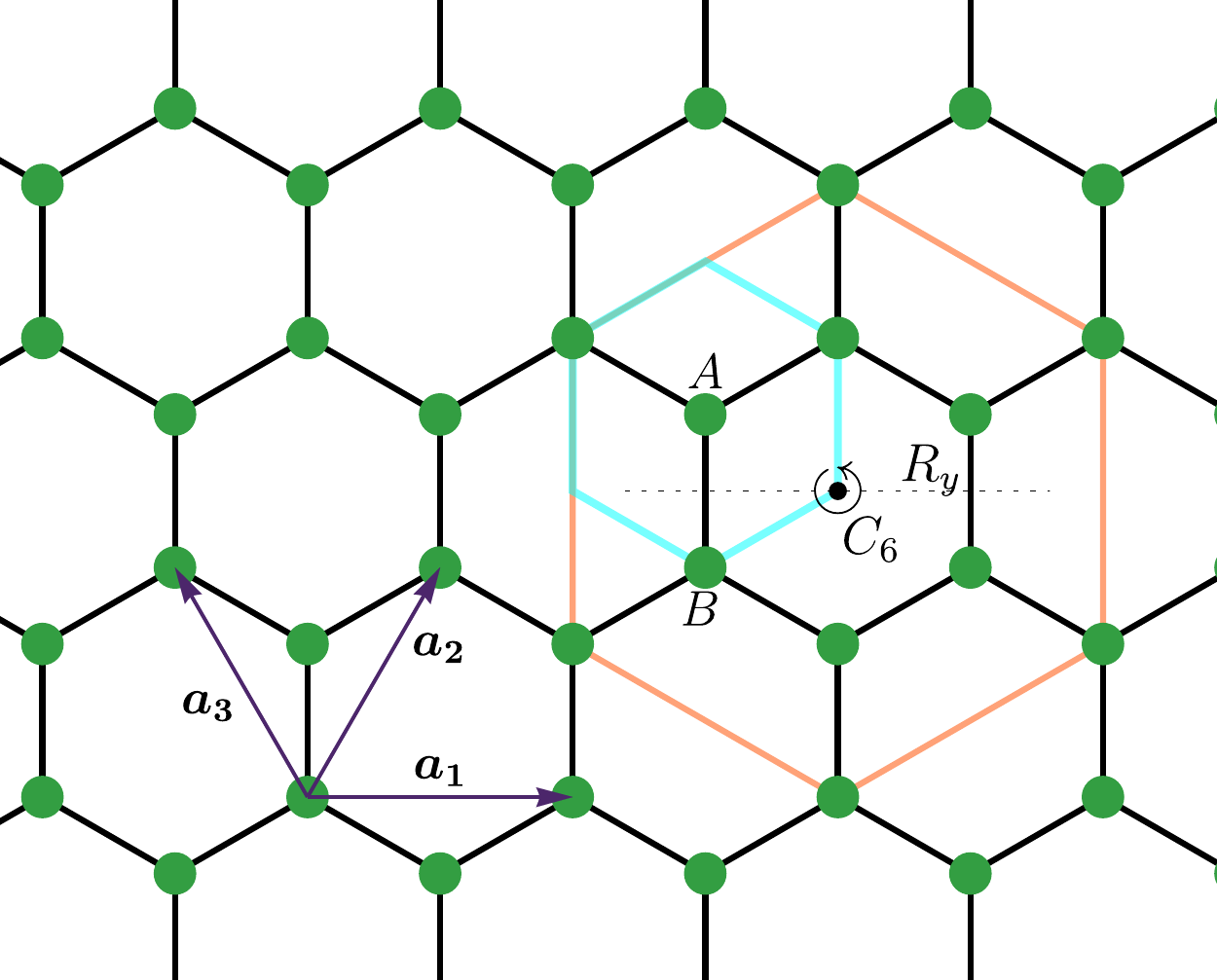}
	\caption{\label{fig:HClattice symmetries}  }
	\end{subfigure}
	\begin{subfigure}[t]{0.45\textwidth}
		\centering
		\includegraphics[height=6cm]{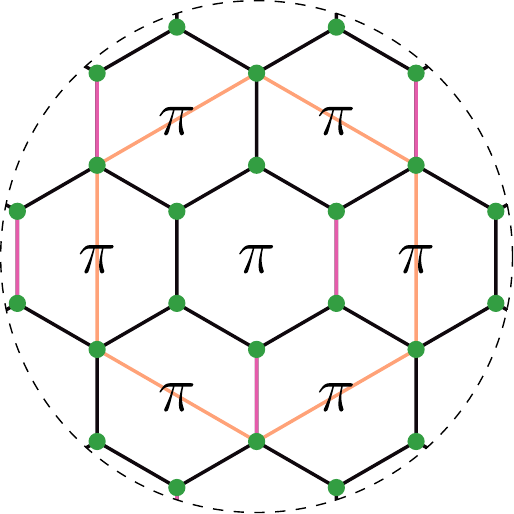}
		\caption{\label{fig:piFluxAnsatz}}
	\end{subfigure}
	\caption{\label{fig:HoneycombLattice} (a) Conventions for the honeycomb lattice. $\abss_{i}$ are the lattice unit vectors. The choice of the boundary of a (enlarged) unit cell is indicated in \textcolor{cyan}{cyan} (\textcolor{orange}{orange}). The symmetry generators are $ R_y $ which corresponds to a reflection perpendicular to the $ y $ axis and $C_6$ is a $60^{\circ}$ around plaquette center. (b) Depiction of the $\p$-flux ansatz where the $t_{ij}$ are positive (negative) on the black (pink) bonds. }
\end{figure}

The lattice symmetries are generated by translations of two unit lattice vectors ($T_{1,2}$), a reflection ($R_y$) and an hexagon centred $\pi/3$ rotation ($C_{6}$). These act as 
\begin{equation}
\begin{split}
T_1:&\, (r_1, r_2, u) \, \longrightarrow \, (r_1+1,\,\,r_2,u),\\
R_{y}:&\, (r_1, r_2, u) \, \longrightarrow \, (r_1+r_2\,,-r_2,1-u),\\
\end{split}
\hspace{2cm}
\begin{split}
T_2:&\, (r_1, r_2, u) \, \longrightarrow \, (r_1,r_2+1,u),\\
C_{6}:&\, (r_1, r_2, u) \, \longrightarrow \,  (u-r_2,r_1+r_2-u,1-u).\\
\end{split}
\end{equation}
where $\rbf= r_1 \abf_1 + r_2 \abf_2$ ($\abf_{i} $ are lattice unit vectors) and $u=0,1$ is the sublattice index with the identification $(A,B) \leftrightarrow (0,1)$. The lattice conventions and the action of the symmetries are summarized in Fig. \ref{fig:HoneycombLattice}. 

\subsection{\texorpdfstring{$\pi$}{pi}-flux state and PSG}
We consider $\SU(4)_s$ fermionic partons at quarter filling with a $\p$-flux mean-field Hamiltonian on the honeycomb lattice given by 
\begin{equation} \label{eq:HpiFluxMF}
\begin{split}
H_{\pi}&=\sum_{\alpha=1}^{4}\sum_{\rbs,\sbs } t_{\rbs \sbs}^{AB} f^{\dag\, a}_{\rbs A}f^{}_{\sbs B a} + h.c., \\
t_{\rbs \sbs}^{AB}&= (-)^{ r_1 } \delta_{\rbs ,\sbs} +\delta_{\rbs+\abss_2,\sbs}+\delta_{\rbs+\abss_3,\sbs},
\end{split}
\end{equation}
where $\rbs$ and $\sbs$ are the position of the unit cells. After a Fourier transforming ($f_{\rbs} = \frac{1}{\sqrt{N_{cells}}} \sum_{\kbs} e^{i \kbs \cdot \rbs} f_{\kbs}$ and $\kbs$ is in the original Brillouin zone), we obtain a doubly degenerated Dirac cone at quarter filling (more details in App. \ref{app:piFluxHC}). In an appropiate basis, the low-energy Hamiltonian becomes the Dirac Hamiltonian
\begin{equation}\label{eq:HpiDirac}
H_{\pi D}= \sum_{\qbs}\sum_{i=1}^{2}\sum_{i=1}^{4}  \bar{\psi}(\qbs)^{a i} i (q_x\gamma^1+q_y\gamma^2) \psi(\qbs)_{a i}
\end{equation}
with the conventions of previous sections for gamma matrices. Here $a=1,2,3,4$ is an $\SU(4)_s$ spin index and $i=1,2$ is an $\SU(2)_v$ valley index. We combine these indices into $\SU(8)$ indices (denoted by greek leters). Once we introduce the gauge fluctuations, the Lagrangian of this theory becomes Eq. \ref{eq:LagrangianQED} with $N_f=8$ ($N=4$ and $M=2$). We span $\mathfrak{su}(8)$ by three sets of Pauli matrices: the old $\s^a$ and $\t^a$ that span $\mathfrak{su}(4)_s $ and a new set of Pauli matrices $\mu^a $ that span  $\mathfrak{su}(2)_v $. 

We find that the lattice symmetries are realized as:
\begin{equation}\label{eq:piFluxPSG}
\begin{split}
T_1: \,\, \psi \,\, &\longrightarrow \,\, i\m^1 \, \psi, \\
T_2: \,\, \psi \,\, &\longrightarrow \,\, i\m^3 \, \psi, \\
R_y: \,\, \psi \,\, &\longrightarrow \,\, W_{ry}  \mathcal{R}_{0y}[\psi] = i\g^2 W_{ry}  \psi, \\
C_{6h}: \,\, \psi \,\, &\longrightarrow \,\, L_{c6} W_{c6} \, \psi, \\
\mathcal{T}: \,\, \psi \,\, & \longrightarrow \,\, \h_{T} \Tmc_0[\psi]= \gamma^0 \eta_T  \psi,
\end{split}
\end{equation}
where the space-time dependence of the operators have been omitted and 
\begin{equation}
\begin{split}
L_{c6} &= \exp(-\frac{\pi}{6} \gamma^0), \\
W_{ry} &=i \frac{\mu^3-\mu^2}{\sqrt{2}}=i\m^{3}\exp(\frac{i\pi}{4}\m^{1} ), \\
W_{c6} &= \exp(i\frac{2\pi}{3} \vec{n} \cdot \vec{\mu}) \, \, , \, \vec{n}=\frac{1}{\sqrt{3}}(1,1,1),\\
\eta_{T} &=i^3\s^2 \t^2 \mu^2. 
\end{split}
\end{equation}
A nice way to interpret the PSG results (and useful in later sections) is to identify the $\SU(2)_v$ elements associated to the lattice symmetries and then study how the vector $\vec{\mu}$ transform under conjugation by these elements. It is convenient to talk about the extended point group $C_{6h}^{'''}$ that corresponds to the original $C_{6h}$ lattice point group augmented by translations that square to the identity. This only allows the $\Mbf$ points (middle points of the edges of the hexagonal BZ). For a more complete review of the representations of this group see Refs.~ \cite{serbyn2013spinon,venderbos2016symmetry,  basko2008theory, hermele2008properties, fernandes2019intertwined} or App.~\ref{app:C6vppp}.

We can choose the restriction of the generators to $\SU(2)_{\mmrm{v}}$ to be 
\begin{equation}
\begin{split}
T_1 \,\, & : \,\, U_{t1}=i\mu^1 \\
T_2 \,\, & : \,\, U_{t1}=i\mu^3 \\
T_3 \,\, & :\,\, U_{t1}=i\mu^2 
\end{split}
\hspace{2cm}
\begin{split}
C_{6h} \,\, & : \,\, U_{c6}=W_{6h} \\
R_y \,\, & : \,\, U_{ry}=W_{ry} \\
\mathcal{T} \,\, & : \,\, U_{T}=i \mu^2 \mathcal{K} 
\end{split}
\end{equation}
where $\mathcal{K} $ is complex conjugation and for convinience of the reader we have included the translation by $\abf_3$: $T_3= T_1^{-1}\cdot T_2$. It is easy to see from here that $\vec{\mu}$ will transform as the $F_{2}$ representations, i.e. they transform as objects at momenta $ \Mbf_{1,2,3}$ that  pick a minus sign under reflections. Additionally $\vec{\mu}$ are odd under time reversal. Note that $C=-i\g^{0}$  transforms as $A_{2}$ (gets a minus sign from reflections) and is odd under time reversal. Combining this observations it is straightforward to find the symmetry properties of the Dirac masses. We shall denote them as
\begin{equation}
M^{00}=\bar{\psi}\psi, \hspace{1cm} M^{\imf 0}=\bar{\psi}\mu^{\imf}\psi, \hspace{1cm} M^{0A}=\bar{\psi}T^{A}\psi, \hspace{1cm} M^{\imf A}=\bar{\psi}\mu^{\imf}T^{A}\psi.
\end{equation}
with $\imf=1,2,3$  and $A=1,\dots,15$. $\imf$ correspond to vector indices of $\SO(3)_{v}$ and $ A $ to adjoint indices of $\SU(4)_{s}$. The symmetry properties under $\gmicro$ of the Dirac masses are summarized in Tab.~\ref{tab:piDSLDiracMasses}.

\begin{widetext}
	\begin{table*}
		\captionsetup{justification=raggedright}
		\begin{center}
		\centering
			\begin{tabular}{ | c | c || c | c| c |c |}
				\hline
				\fbox{Mass} & 	\fbox{Bilinear} &	\fbox{$ \mathrm{R}_{\SU(4)_s}$}	& 	\fbox{$ \mathrm{R}_{C_{6v}^{'''}}$} & \fbox{$ \Tmc$} & \fbox{Order}
				\\\hlinewd{.8pt}\hline
				
				\fbox{$M^{00}$} & $\bpsi_{}\psi_{}$ &$ \mmrm{Trivial} $ & \fbox{$A_2$}  & \fbox{$-1$}& \fbox{Quantum Hall}  \\ \hline
				
				$M^{\imf 0}$ & $\bpsi_{}\m^{\imf}\psi_{}$  & $\mmrm{Trivial}$	& $F_1$&	
				\fbox{$ +1 $}	&	\fbox{Valence Bond}
				\\
				\hline	
				
				$M^{0 A}$ & $\bpsi_{}T^{A}\psi_{}$ & $\mmrm{Adjoint}$	&$A_2$&	
				\fbox{$ -O_{\Tmc} $}	&	\fbox{Quantum Spin-orbit Hall}
				\\
				\hline	
				
				$M^{\imf A}$ & $\bpsi_{}\m^{\imf}T^{A}\psi_{}$ & $\mmrm{Adjoint}$	&$ F_1$&	
				\fbox{$ +O_{\Tmc} $}	&	\fbox{Spin-Orbit N\'eel}
				\\
				\hline	
				
			\end{tabular}
			
			\caption{\label{tab:piDSLDiracMasses} Symmetry properties of the Dirac masses. $\mathrm{R}_{G}$ denote the representation of the mass under the symmetry group $G$. $O_{\Tmc}$ comes from the action of $\Tmc$ on the original spin-orbit operators.} 
		\end{center}
	\end{table*}
\end{widetext}

\subsection{Monopole operators}\label{sec:MonopoleOperators1}
Following the previous discussion, we identify the monopole transformation properties in two steps. First, we focus on the transformation coming from the zero modes. Second, we extract the contributions from the Dirac sea that is encapsulated by the Berry phase obtained numerically and/or analytically. 

\subsubsection{Symmetry properties from zero modes}\label{sec:Notation}

Using the results in Sec \ref{sec:monopoleWF}, the $\SU(8)$ representation of the monopoles $\mmbf{70}_8$ breaks into $\SU(4)_s \times \SU(2)_v$ representations as
\begin{equation}\label{eq:Rep708}
	\begin{alignedat}{6}
		\yng(1,1,1,1) &= &&\left(\rule{0cm}{1.2cm} \yng(2,2) \, ,\, \yng(2,2) \right)  &&\oplus &&\left(\rule{0cm}{1.2cm}\yng(2,1,1)\,,\,\yng(3,1) \right) &&\oplus  &&\left(\rule{0cm}{1.2cm}\yng(1,1,1,1)\,,\,\yng(4) \right)\\
		\mmbf{70}_8 &= && \hspace{0.5cm}\left(\mmbf{20}_4   , \mmbf{1}_2 \right) &&\oplus &&\hspace{0.5cm}\left(\mmbf{15}_4,\mmbf{3}_2 \right) &&\oplus &&\hspace{0.7cm}\left(\mmbf{1}_4 ,\mmbf{5}_2 \right)
	\end{alignedat}
\end{equation}
The $\SU(4)_s$ representations can be understood by the identification $\SU(4)/\mathds{Z}_2 = \SO(6)$. $\mmbf{1}_4$, $ \mmbf{15}_4$ and $\mmbf{20}_4$ are the trace (scalar), antisymmetric part (adjoint) and traceless symmetric part of an $\SO(6)$ matrix obtained by taking the tensor product of two vectors. For the $\SU(2)$ case, we can similarly use the identification $\SU(2)/\mathds{Z}_2=\SO(3)$. 

We can then think of the monopoles as three sets of operators 
\begin{equation}
	\begin{split}
	\Vmc_{\imf \jmf}^{\dag} \sim  {(\bf{1}_4,\bf{5}_2) }\quad , \quad  \Nmc_{{A} \imf }^{\dag} \sim  (\bf{15}_4,\bf{3}_2) \quad , \quad \Qmc^{\dag}_{\amf\bmfk}\sim (\bf{20}_4,\mmbf{1}_2)
	\end{split}
\end{equation}
where $\imf,\jmf=1,2,3$ are $\SO(3)_{s}$ vector indices. $ \amf,\bmfk = 1,\dots,6 $ and $A=1,\dots,15$ are $\SO(6)_{s}$ vector and adjoint indices, respectively. $\Vmc^{\dag}_{\imf\jmf}$ and $ \Qmc_{\amf\bmfk}^{\dag} $ are traceless symmetric in their indices. Alternatively, we outline an explicit construction for the monopoles in App.~\ref{app:HCSU4_explicit}.

We find that $\Tmc$ acts as 
\begin{equation}
	\begin{split}
					 \Vmc^{\dag}_{\imf\jmf} &\longrightarrow \h^{\Tmc}_{\mmrm{top}} \Vmc^{}_{\imf\jmf}\\
		\Tmc:  \quad \Nmc^{\dag}_{\imf A} &\longrightarrow \h^{\Tmc}_{\mmrm{top}} (O_{\Tmc})^{B}_{A}\Nmc^{}_{\imf B}\\
		\quad \Qmc^{\dag}_{\amf  \bmfk} &\longrightarrow \h^{\Tmc}_{\mmrm{top}} \h_{\abf}^{\Tmc}\h^{\Tmc}_{\bmfk} \Qmc^{}_{\amf \bmfk}
	\end{split}
\end{equation}
where $\O_{\Tmc}$ is the matrix that encodes how $ \Tmc $ acts on $\mathfrak{su}(4)_{s}$ (the original spin-orbit operators), $ \h^{\Tmc}_{\amf=1,2,3} = +1 $ and $ \h^{\Tmc}_{\amf=4,5,6} = -1 $. The $\SO(6)_{s}$ indices have been ordered so that the first three components transform under the $\SU(2)$ spanned by $\s^{a}/2$ while the last three transform under the $ \SU(2) $ generated by $ \t^{a}/2 $. $ \h^{\Tmc}_{\mmrm{top}} $ is the common $\Utop$ phases that we will shortly determine. 
In this basis, the bare reflection $\Rmc_{0y}$  acts as $(\Vmc^{\dag},\Nmc^{\dag},\Qmc^{\dag}) \longrightarrow (\Vmc^{},\Nmc^{},\Qmc^{})$. 

\subsubsection{Time reversal and Berry phases}\label{sec:TRSandBP}
We use techniques from band topology to identify the gauge charge distribution and from this calculate the Berry phases. We follow the recipe described in Ref.~\cite{song2018spinon}. First, we introduce the Dirac mass $ -\bar{\psi} \sigma^3 \psi$ that corresponds to a quantum spin Hall (QSH) insulator. This mass favors a monopole $\Phi_{\ua}^{\dag}$ that  transforms as $ \Phi_{\ua}^{\dag} \longrightarrow e^{2i\q}\Phi_{\ua}^{\dag} $ under a spin rotation by angle $\q$ around the $\s^{3} = 1$ axis.
This mass is invariant under $\Tmc_0$, the $\SU(2)$ orbital rotations  and lattice symmetries that do not involve reflections. We can consider the following Kramers time reversal $\mathcal{T}_{K}=i\tau^2 \Tmc$, $\Tmc_{K}^2 =(-)^F$, to use the fact that a QSH insulator is also a $\ZZ_2$ topological insulator. Opposite to the $N_f=4$ case, this QSH phase is trivial because we now have two copies (from the degeneracy in the pseudo-spin quantum number) of a non-trivial QSH that become trivial. Then the monopole must be trivial:
\begin{equation}\label{eq:monopoleTrivial}
\mathcal{T}_K:\,\, \Phi^{\dag}_{\,\, \ua}\,\, \longrightarrow \,\,  (\Phi^{\dag}_{\ua})^{\dag}.
\end{equation}
This operator has charge 4 under the $\U(1)$ subgroup of $ \sus $ generated by $ \s^{3}/2 $ which means that $\Phi^{\dag}_{\ua}$ is a linear combination of the $\Qmc_{\amf\bmfk}^{\dag}$ with $\amf,\bmfk \in \{1,2,3\}$ \footnote{$ \Phi^{\dag}_{\ua} \propto(\Qmc^{\dag}_{11}-i\Qmc^{\dag}_{12}-i\Qmc^{\dag}_{21}- \Qmc^{\dag}_{22} $)} . As $\Phi^{\dag}_{\ua}$ is trivial under $i\t^2$, Eq. \ref{eq:monopoleTrivial} also holds for $\Tmc$. Thus $ \h_{\mmrm{top}}^{\Tmc}=1  $.

The next step is to decompose the band structure of the $\pi$-flux state (with the quantum spin Hall gap) in terms of "elemental bands", i.e. Wannier insulator bands, {using techniques from Refs~\cite{Po17,Po18,Cano2018}}. This is done by comparing the eigenvalues of some PSG rotations \footnote{i.e., the rotations with the corresponding gauge transformation that leave Eq. \ref{eq:HpiFluxMF} invariant.} at high-symmetry points of the elemental bands with the corresponding eigenvalues of the parton Fourier mode. The final result is  (see App. \ref{app:Wannier} for details)
\begin{equation}
\Gamma_{Occupied}=2 \Gamma^v-2\Gamma^{h}.
\end{equation}
where $\Gamma^{v(h)}$ are Wannier insulators localized at lattice sites (hexagon centers). {The minus sign in the above equation comes from the notion of ``fragile topology''\cite{Po18}, and for our purpose can be treated simply as having negative gauge charges.} This means that there is $ q_{C_{6h}}=-2 $ gauge charge at the $ C_{6h} $ rotation center which implies that the $2\pi$ monopole will get a $C_{6h}$ Berry phase of $ \exp(\frac{-2\pi}{3})=\omega^{-1}=\omega^2 $, i.e., $ C_{6h}: $ $ \Phi^\dag_{\,\, \ua} \longrightarrow \omega^{2}\Phi^\dag_{\,\, \ua} $. As the zero modes are trivial under $\SU(2)_{\mmrm{v}}$, this transformation must come from $\Utop$, $C_{6h}\eval_{\Utop}: {\Phi}^{\dag}_{} \,\, \longrightarrow \,\, \omega^{2} \, {\Phi}^{\dag}_{}$. Translations can be obtained by a $C_3$ rotation around sublattice $A$ with another $C_3$ rotation in the opposite direction around sublattice $B$. As the gauge charge in both sublattices is the same there is no Berry phase for translations. Then we can say that the bare monopole $\Mmc^{\dag}$ transform as $E_2$ under the point group. We performed a numerical calculation of the rotation and translation Berry phases \cite{song2018spinonNumeric} and find consistent results (see App. \ref{app:berryNumerics} for details). 

Now that we have all the symmetry transformations of monopoles, we can find the representation they belong to by recalling that the $\imf$ indices ($\mmbf{3}_2$ of $\SU(2)_{\mmrm{v}}$) transform as  $F_2$ when restricted to $\glat$. Then under the restriction $\SU(2)_{\mmrm{v}} \rightarrow \glat$, we have $\mmbf{1}_{2} \rightarrow A_{1}$ , $\mmbf{3}_{2} \rightarrow F_{2}$  and $\mmbf{5}_{2} \rightarrow E_{2}\oplus F_{1}$, where we have used properties of the symmetric and anti-symmetric squares of representations. Finally by tensoring with $E_{2}$ we find the representations of the monopoles under $\glat$. 

The symmetry properties of the monopoles under $\gmicro$ is summarized in Tab.~\ref{tab:repMonopoles}. We want to emphasize the existence of a trivial monopole due to the cancellation of $C_3$ angular momentum coming from the zero modes and from the filled bands. We outline a more explicit calculation of the monopole quantum numbers in App.~\ref{app:HCSU4_explicit}.

\begin{table}[t]
	\centering
	\begin{tabular}{ |c||c|| c| c |c | }
		\hline
		\fbox{Type} 	&	\fbox{$ \mmrm{R}_{\SU(4)_{\mmrm{s}}}$}	& \fbox{$ \mmrm{R}_{C_{6v}^{'''}}$} &	{\fbox{$\Tmc$}}	&	{\fbox{Comments}}	 
		\\\hlinewd{.8pt}\hline
		
		\fbox{$\Qmc^{\dag}_{\amf\bmfk}$} & \fbox{Quadrupolar} & \fbox{$E_2$}  & \fbox{$\h^{\Tmc}\otimes \h^{\Tmc}$}& \fbox{\makecell{Quadrupolar order with $ C_3 $ angular momentum at $ \GGbs $.}  }\\ \hline
		
		\fbox{$ \Nmc_{\imf{A}  }^{\dag} $} & Adjoint	&$F_1 \oplus i F_2$&	
		\fbox{$ \Omc_{\Tmc} $}	&\fbox{	\makecell{Real part same as $M^{\imf A}$ (Spin-orbit order at $\Mbs$).\\ Imaginary part additionally breaks reflections.	}}
		\\
		\hline
		&		&\fbox{$ A_1 \oplus i A_2$}&	
\fbox{$+1$}	&	\fbox{	\makecell{Real part is trivial.\\ Imaginary part only breaks reflections.	}}

		\\\cline{3-5}
		&		&$ E_2$&	
\fbox{$+1$}	&	\fbox{Bond order with $C_3$ angular momentum at $\GGbs$.}	

		\\\cline{3-5}
		
		\multirow{-3}{*}{$\Vmc_{\imf \jmf}^{\dag} $}	    & \multirow{-3}{*}{Trivial}
		&   $F_1 \oplus i F_2$ &   $+1$  &  	\fbox{	\makecell{Real part same as $M^{\ibf 0}$(Bond order at $\Mbs$).\\ Imaginary part additionally breaks reflections.	}}
		
		\\\hline
		
	\end{tabular}
	
	\caption{\label{tab:repMonopoles} Symmetry properties of the monopoles operators of the $\p$-flux state. $\mathrm{R}_{G}$ denote the representation of the mass under the symmetry group $G$. $\Omc_{\Tmc}$ is the matrix that encodes how TRS acts on the original spin-orbit operators. $\h^{\Tmc}=+1$ when acting on spin indices while $\h^{\Tmc}=-1$ on orbit indices.  } 
\end{table}

\subsection{Instability of the DSL and nearby phases}
\label{sec:Instability}
\subsubsection{Trivial monopole}

Dirac masses are not allowed because they transform non-trivially under $\gmicro$. On the other hand, the existence of a trivial monopole could mean that the Dirac spin liquid is ultimately non-stable at the lowest energies if the monopole is relevant. The large $N_f$ estimation of the scaling dimension is\cite{kapustin_2002,DyerMonopoleTaxonomy}  $\Delta_1 = 0.265N_f-0.0383 + O(1/N_f )$. For $N_f=8$, $\Delta_1=2.0817<3$. Recent Monte Carlo simulation\cite{KarthikMonopole} shows that the finite-$N_f$ correction to this value at $N_f=8$ is small and the monopole is indeed a relevant perturbation. The proliferation of the trivial monopole breaks the $\SU(8)$ symmetry down to the original microscopic symmetry group $\SU(4)_v \times G_{Lat} \times \mathcal{T}$.

Nevertheless, the Lieb-Schultz-Mattis theorem survives and predicts a non-trivial ground state. We then expect that a Dirac mass is spontaneously generated thus breaking some symmetry. The question now is which mass? The most likely masses will be the ones in the "smallest" representations , which usually have the smallest scaling dimensions, which in present case correspond to masses of the form $ {\bpsi}\vec{n}\cdot \vec{\mu}\psi $ for some vector $\vec{n}$. 

The masses $ {\bpsi}\,\lrRb{\vec{n}\cdot \vec{\mu}}\,\psi $ transform as valence bond solid (VBS) orders. In particular for $\vec{n}=\tfrac{1}{\sqrt{3}}(1,1,1)$, the mass has the same symmetry properties as the $\mmbf{R}$ tetramerization order parameter of Ref \cite{Penc_TetrametrizationSU4} under the enlarged point group $C_{6v}^{'''}$ ($O_h$ in the cited work.). This order is schematically represented in Fig. \ref{fig:HCVBSRefOrder} where the light blue lines correspond to spins that form an $\SU(4)_{\mmrm{s}}$ singlet. In the following subsection we explore this and other possible symmetry breaking phases accessible from the $ \p$-flux state.

The monopole proliferation seems to contradict the numerical results \cite{PhysRevX.2.041013} of an stable algebraic spin liquid phase and stability against tetramerization \cite{Penc_TetrametrizationSU4}.  One explanation could be that the numerics are not probing large enough system sizes. Another possible explantion is that we are looking at a different gapless spin liquid. We explore two different cases in the next two sections and find that a  $\ZZ_{4}$ Dirac spin liquid is the most plausible.

\subsubsection{Phase near the \texorpdfstring{$\p$}{pi}-flux state}\label{sec:SpinOrbitOrdersPi}

Before delving into the other SL candidates, we identify natural symmetry breaking phases near the $\p$-flux state by looking for Dirac masses and/or monopoles that break the same symmetries.

\begin{figure}[ht]\label{fig:SpinOrbitVBS}
	\begin{subfigure}[t]{0.30\textwidth}
		\centering
		\includegraphics[width=0.9\columnwidth]{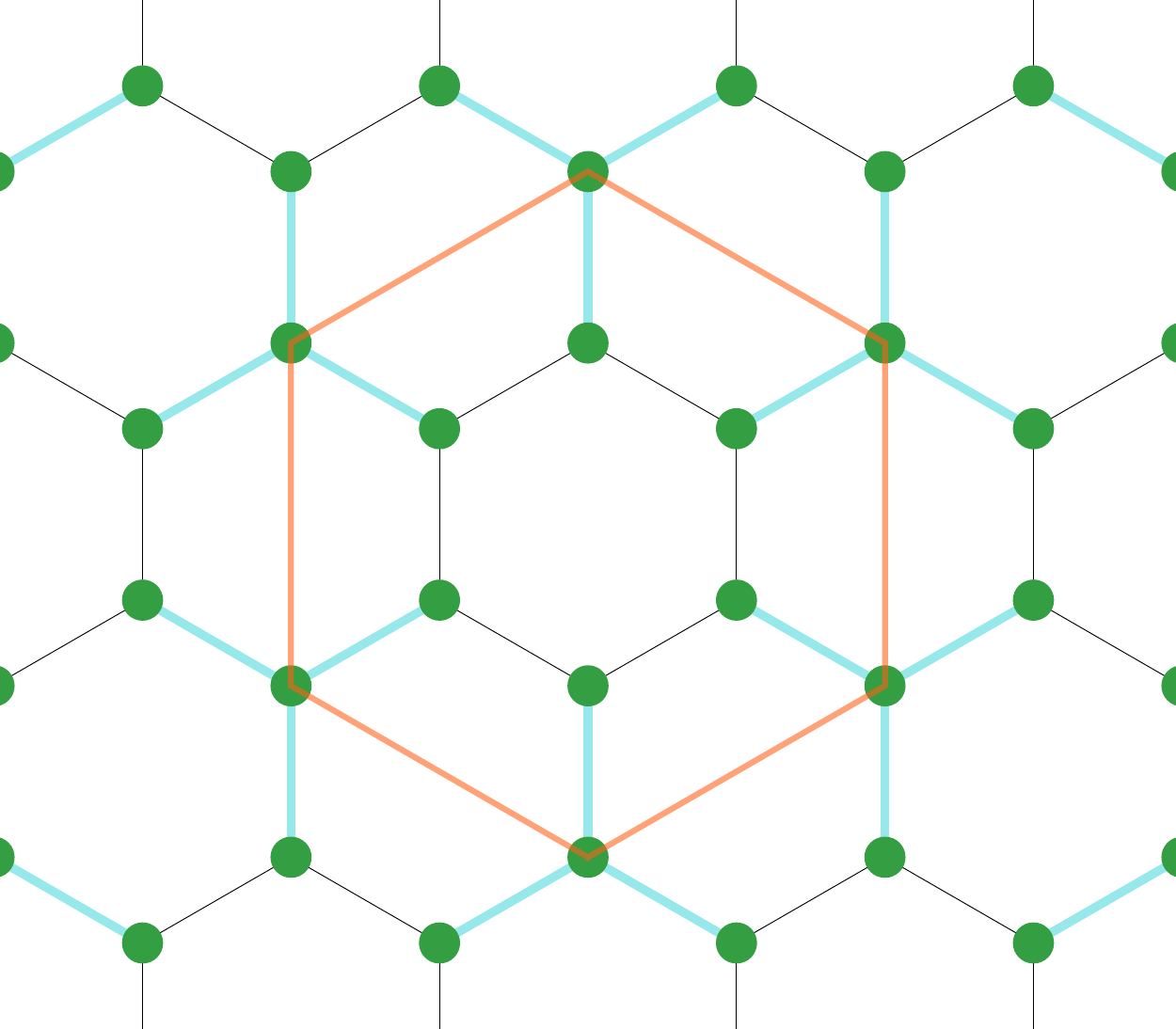}
		\caption{\label{fig:HCVBSRefOrder}  }
	\end{subfigure}
	\begin{subfigure}[t]{0.30\textwidth}
		\centering
		\includegraphics[width=0.9\columnwidth]{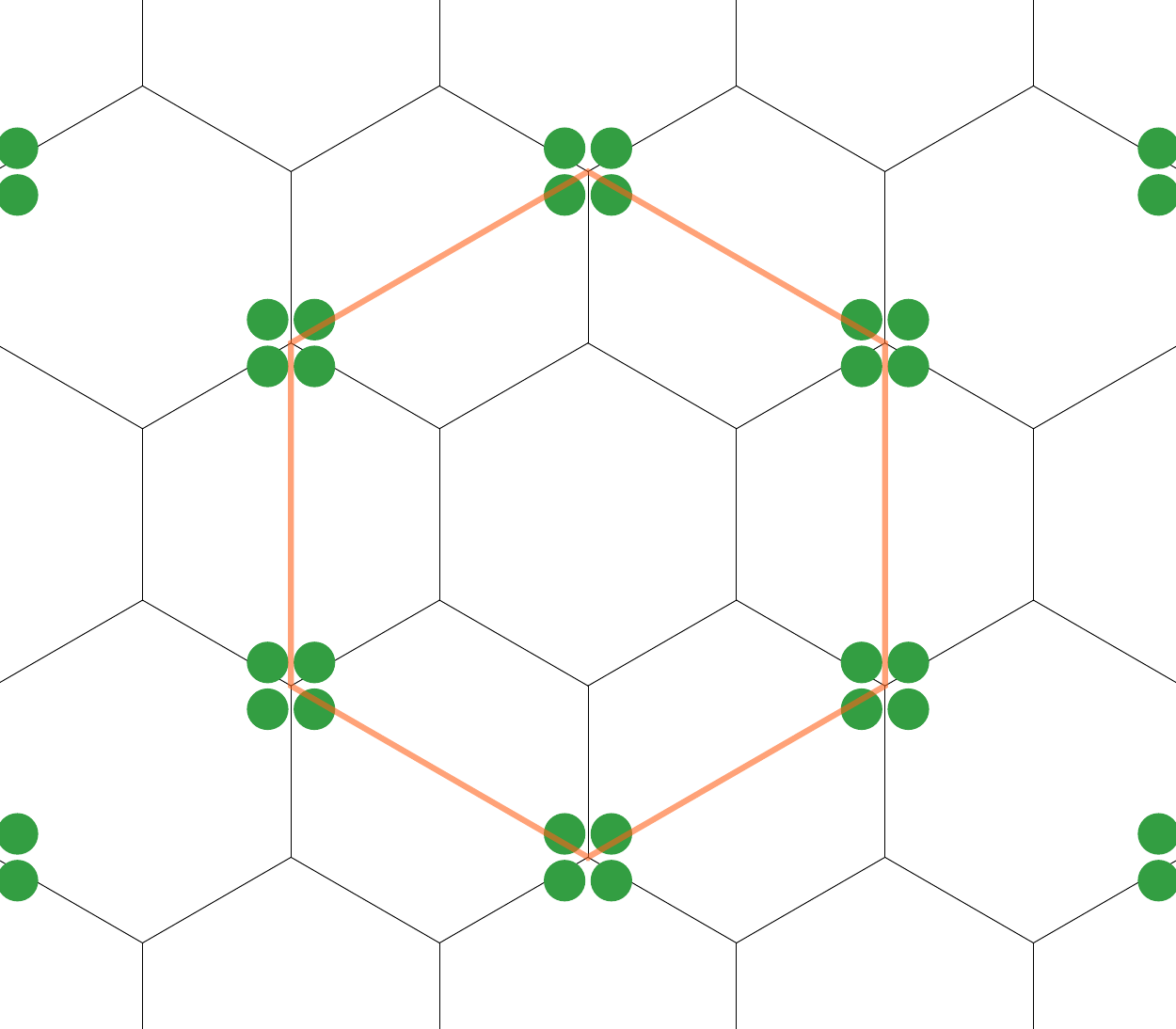}
		\caption{\label{fig:HCVBSRefOrderBand}}
	\end{subfigure}
	\begin{subfigure}[t]{0.30\textwidth}
		\centering
		\includegraphics[width=0.9\columnwidth]{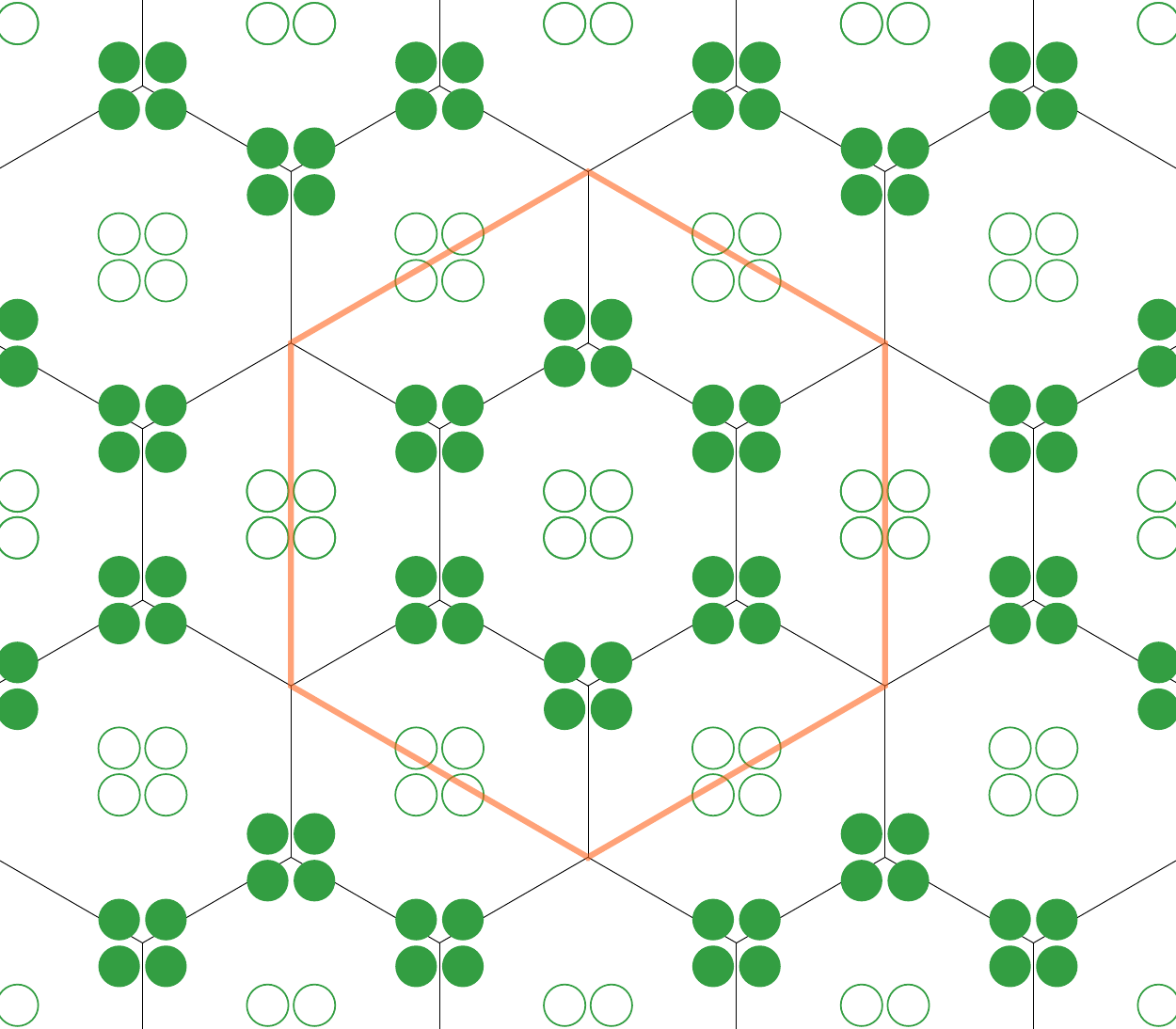}
		\caption{\label{fig:HCrVBSRefOrderBand}}
	\end{subfigure}
	\caption{  On \textcolor{blue}{(a)}: VBS-like state of Ref. \cite{Penc_TetrametrizationSU4} where the spins connected by light-blue lines form $\SU(4)_s$ singlets in groups of four. \textcolor{blue}{(b-c)} Candidate band decomposition for a Dirac mass $m \bar{\psi} \frac{\mu^1+\mu^2+\mu^3}{\sqrt{3}}\psi$ with large positive and negative, respectively value of $m$. Addition (substraction) of bands is represented by filled (empty) circles.}
	\label{fig:VBS}
\end{figure}

Consider the VBS phase mentioned previously (See Fig. \ref{fig:VBS}). This order is four-fold degenerated and breaks the $T_{1}$, $T_{2}$ and $T_{3}$ translations. A fermion bilinear with the same symmetries is $\frac{1}{\sqrt{3}} \bar{\psi}\left(\mu^1+\mu^2+\mu^3 \right) \psi$. In the presence of the previous mass, the lightest monopole 
can have a non-trivial $C_3$ angular momentum depending on the sign of the mass $m$ -- {note that opposite signs of $m$ are not related by any physical symmetry in this case, so they could correspond to very different symmetry breaking patterns}.  Let's take the sign convention such that $m>0$ corresponds to a trivial $C_3$ angular momentum. The mass can be written as the sum of minus the three translation related masses. This leads us to a interpretation of the $m<0$ phase as a short range resonating VBS composed of the superposition of the other three VBS states. To further backup this claim, we can look at the information about the parton band structure  gained from the monopoles quantum numbers. The $C_{6h}$ angular momentum of the monopole tells us that there should be $(1-\sign(m)) \mod 6$ gauge charge at the hexagon center. In Fig.~\ref{fig:HCVBSRefOrderBand} we show a possible band structure of the partons compatible with the simple VBS. For the $m<0$ case, we can put the parton in a band structure as in Fig.~\ref{fig:HCrVBSRefOrderBand}. We can understand this by first translating Fig.~\ref{fig:HCVBSRefOrderBand} by $T_{1}$, $T_{2}$ and $T_{3}$ and then putting negative charges on the hexagon centers in order to have net zero gauge charge. In this configuration we have $-4=2 \mod 6$ gauge charge at the hexagon center as we needed.

Let us now see what happens if the lightest monopole, $ \Wmc^{\dag}_{}$ , proliferates for both signs of $m$. For $m>0$, if $\expval{\Wmc^{\dag}_{}}$ has a non-zero imaginary part we break the remaining reflection symmetries otherwise there is no further symmetry breaking.

%
For $m<0$, the monopole breaks the remaining $C_3$ rotations and may break reflection symmetries. For example, $\expval{\Wmc^{\dag}_{}}=1$ preserves $R_y$ and $C_2R_y$ which could be understood as an enhanced correlation along horizontal chains on top of the superposition of the tetramer coverings. 
\begin{figure}[hb]
	\begin{subfigure}[t]{0.35\textwidth}
		\centering
		\includegraphics[width=0.95\columnwidth]{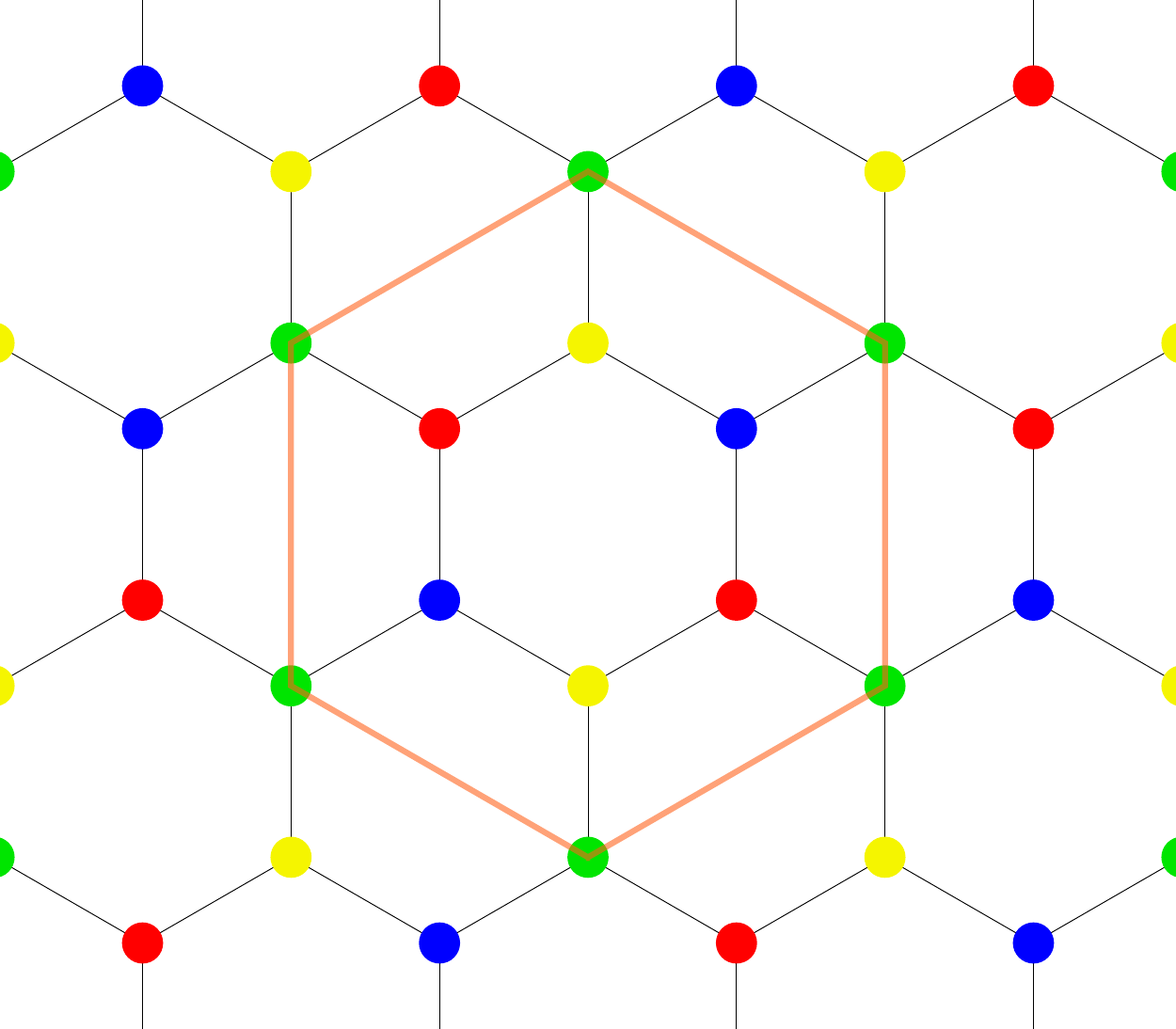}
		\caption{\label{fig:HCNeelOrder}}
	\end{subfigure}
	\hspace{1cm}
	\begin{subfigure}[t]{0.35\textwidth}
		\centering
		\includegraphics[width=0.95\columnwidth]{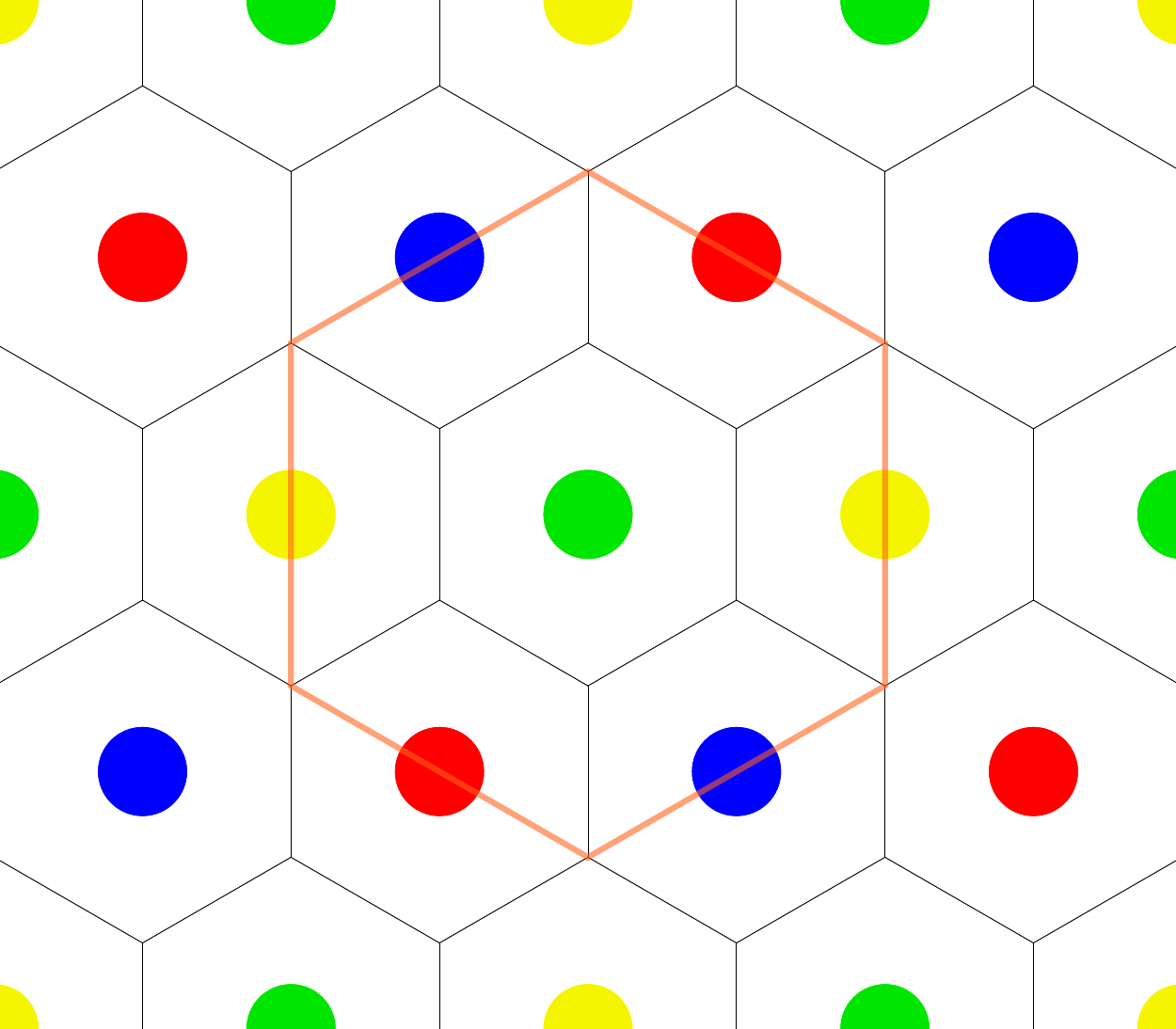}
		\caption{\label{fig:HCAntiNeelOrderHex}}
	\end{subfigure}
	\caption{  \label{fig:SpinOrbitNeel}  Neel-like orders near the $\p$-flux DSL. The colors represent the different $ \SU(4)_{s} $ flavors.}
\end{figure}

We now proceed to consider N\'{e}el like-orders. For $\SU(4)_{\mmrm{s}}$ the natural generalization would be to have different colors between neighbor site as in Fig. \ref{fig:HCNeelOrder}. This order breaks the $\SU(4)_s$ symmetry down to its maximal torus $\U(1)^3$ (Or equivalently $\SO(6)\rightarrow \SO(2)\oplus \SO(2)\oplus\SO(2)$). Lattice symmetries are broken alone but preserved if combined with broken elements of $\SU(4)_s$. The preserved translations are $\tilde{T}_1=i\tau^2 T_1$ and $\tilde{T}_2=i\sigma^2 T_2$. Unbroken reflections and rotations are  more complicated and given by
\begin{equation}
\tilde{R}_y=\left( \begin{smallmatrix} 
0 & 0 & -1 & 0\\
0 & 1 & 0 & 0\\
1 & 0 & 0 & 0\\
0 & 0 & 0 & 1\\
\end{smallmatrix} \right)  R_y = \exp(\frac{i\p\t^2}{2}\frac{1+\s^3}{2} ) R_y
\hspace{1cm} \text{and} \hspace{1cm}
\tilde{C}_{6h}=\left( \begin{smallmatrix} 
0 & 0 & 1 & 0\\
0 & 1 & 0 & 0\\
0 & 0 & 0 & 1\\
1 & 0 & 0 & 0\\
\end{smallmatrix} \right)  C_{6h},
\end{equation}
where the matrices act on the partons ($f_a \rightarrow M_a^{\,\, b} f_b$) in the basis ${\tikzcircle[red,fill]{3pt}} = ({\uparrow a})$, ${\tikzcircle[green,fill]{3pt}} = ({\downarrow a})$, ${\tikzcircle[blue,fill]{3pt}} = ({\uparrow b})$ and ${\tikzcircle[yellow,fill]{3pt}} = ({\downarrow b})$. $\tilde{\mathcal{T}}=\sigma^2\tau^2\mathcal{T}$ is a time reversal that leaves the pattern invariant. A mass that has the same broken/preserved symmetries is $\frac{1}{\sqrt{3}}\bar{\psi}\left(  \mu^1\sigma^3 +\mu^2\tau^3\sigma^3 -\mu^3\tau^3 \right) \psi$. As in the VBS case, the sign of the mass lead to different orders again as there is no microscopic symmetry that relates both signs. In order to identify the phase with opposite mass sign, we first note that in terms of partons the monopole quantum number tells us that there must be gauge charges at the center of the hexagons so we may interpret this new phase as polarizing the spin color on the hexagons instead of sites. We can schematically represent this order in Fig.~\ref{fig:HCAntiNeelOrderHex}. The monopole for the first case can additionally break the new reflections while the monopoles in the second case additionally (always) breaks the new rotations.

\section{Alternative scenarios for the \texorpdfstring{$\SU(4)$}{SU(4)} honeycomb model}
\label{sec:Alternatives}

\subsection{Scenario I: a different \texorpdfstring{$\U(1)$}{U(1)} Dirac spin-orbit liquid}

Instead of the standard parton condition, we can further fractionalize the $f_{}^{}$ fermions into three $g^{}_{s=1,2,3}$ fermions that transform in the anti-fundamental representation of $\SU(4)$ by writting 
\begin{align}
f_{a} = \sum_{bcd} \ve_{abcd} g^{b}_{1} g^{c}_{2}g^{c}_{3}.
\end{align}
This construction has a $\U(1)^{3} \rtimes S_{3}$ redundancy. The $\U(1)$ factors act as separate phase for the different flavours of $g_{s}$ and the $S_{3}$ is the permutation group that swaps the three flavors of the $ g $ fermions. In order to reproduce the original Hilbert space, we need to impose unit filling for each $g_{s}$ and further impose full symmetrization under the $s$ index.

Next we put the three $g$ fermions in the $\p$-flux state previously studied with slightly different hopping parameter to Higgs so the IGG is simply $\U(1)\times\U(1)\times\U(1)$. The low-energy theory then corresponds to three copies of $\Nf=8 $   $\QED_{3}$. At this point we have not gained anything.

 What we can do now is to introduce a $p$-wave hybridization between flavors 2 and 3:
 \begin{equation}\label{eq:pHopping}
 H_{2} = \x \sum_{\langle\langle i,j\rangle \rangle }  (-1)^{s_{ij}} g^{\dag}_{i,s=2}g^{\,}_{j,s=3} + \hc,
 \end{equation}
 where $\x$ is a real small parameter and $s_{ij}$ are $0$ or $1$ such that there is $\p$ flux around counterclockwise triangles made of three second neighbours or one second neighbour and two nearest neighbours. This term will gap the fermions and Higgs the last two $\U(1)$'s to a single $\U(1)$. Next, we introduce an onsite hybridization between flavors 1 and 2:
  \begin{equation}
 H_{3} = h \sum_{i}   g^{\dag}_{i,s=1}g^{\,}_{j,s=2}.
 \end{equation}
  This term leaves one gapless gauge flavor but Higgs the gauge group to a single $\U(1)$ such that all the initial $\U(1)$ gauge fields are aligned with no symmetry broken.

Then our effective field theory is simply $\Nf = 8$  $ \QED_{3}$ as before\footnote{Technically the partons now transform as the anti-fundamental representation of $\SU(4)_s$ but this not affect the overall result as the monopoles representations are self-conjugate.}. What we gained by this construction is that the fragile band structure will be three copies of the standard parton construction. This will render the $C_3$ angular momentum of the bare monopole to zero. Thus the bare monopole transforms as $A_1\oplus A_2$. The action of time-reversal is not modified as we still have an even number of fermions. The monopoles then transform as 
\begin{equation}
\begin{split}
\Vmc_{\imf \jmf}^{\dag} \sim & {(2E_{2} \oplus F_{1} \oplus F_{2} ,\bf{1})_{+} } \\
\Nmc_{{A} \imf }^{\dag} \sim &  (F_{1} \oplus F_{2}, \bf{15})_{+} \\
\Qmc^{\dag}_{\amf\bmfk}\sim & (A_{1}\oplus A_{2}, \mmbf{20})_{+}.
\end{split}
\end{equation}
In particular, there is no trivial monopole. 

As the Dirac masses quantum numbers are unchanged, we initially get the same symmetry breaking phases from mass proliferation. Nevertheless, once we include monopole proliferation, the symmetry breaking phases are different. 

Let's start with the VBS phase accessed by the proliferation of $m \bar{\psi} \frac{\mu^1+\mu^2+\mu^3}{\sqrt{3}}\psi$. For both signs of $ m $, the monopole operator will transform non-trivially under the rotation around the middle plaquette of the hexagonal enlarged unit cell. Therefore, we are not able to access the simple VBS phase observed in Ref.~\cite{Penc_TetrametrizationSU4} and all the $C_{3}$ rotation symmeries around hexagons will be broken.  

Similarly, if we start with the simple Neel phase using the mass $\frac{1}{\sqrt{3}}\bar{\psi}\left(  \mu^1\sigma^3 +\mu^2\tau^3\sigma^3 -\mu^3\tau^3 \right) \psi$. The symmetry breaking pattern is the same before monopole proliferation but regardless of the sign of $m$ even the magnetic $C_3$ rotations around the hexagons will be broken. Therefore we cannnot access the simple Neel order from this state. 

Therefore, even if the alternative $\U(1)$ DSL is stable it is not a good candidate to describe the numerics as it does not connect to other natural phases expected in the $\SU(4)$ Heisenberg model.

\subsection{Scenario II: \texorpdfstring{$ \ZZ_{4} $}{Z4} Dirac spin-orbit liquid}

The third option we consider is that of a $\ZZ_{4}$ DSL obtained by further condensing a (symmetric) charge-4 Higgs field, $\mathfrak{H}\sim \sum_{\rbs_{1,2,3,4}}\chi_{\rbs_1\rbs_2\rbs_3\rbs_4}f_{\rbs_1 \rCircle}f_{\rbs_2 \bCircle}f_{\rbs_3 \gCircle}f_{\rbs_4 \yCircle} + \hc$. Opposed to $\ZZ_2$ spin liquids, the condensation of charge four Higgs fields would not modify the mean field band structure by much, so at low energy we still have $N_f=8$ gapless Dirac fermions. The Dirac fermions are free in the IR since the gauge field is now gapped due to the Higgs condensate. The unmodified mean field theory means that in terms of the Gutzwiller projected wavefunction, the $\mathbb{Z}_4$ and $U(1)$ Dirac spin liquids are indistinguishable. This is consistent with the good performance of the Gutzwiller projection in the numerical work\cite{PhysRevX.2.041013}. {We note that a similar situation happens for the $J_1-J_2$ $SU(2)$ Heisenberg model on square lattice, where a gapless spin liquid was observed\cite{Capriotti2001,Hu2013,Ferrari2018,Ferrari2020,Nomura2020,Liu2020} in an intermediate range of $J_2/J_1$. This state cannot be the $U(1)$ Dirac spin liquid since it is unstable due to the existence of a symmetry-allowed monopole perturbation. Numerical evidences\cite{Capriotti2001,Hu2013,Ferrari2018,Ferrari2020} show that the state may be a $\mathbb{Z}_2$ Dirac spin liquid\cite{SpinLiquids}, which can be obtained from the $U(1)$ DSL through appropriate fermion pairing.}

Let us now consider possible phase transitions from the $\ZZ_4$ DSL to conventional symmetry breaking orders, for example the VBS or Neel orders discussed in Sec.~\ref{sec:SpinOrbitOrdersPi}. We can consider a two-step transition: first we drive the system through a spontaneous symmetry breaking transition, across which the Dirac fermions get a mass gap and the transition will be of the Gross-Neveu-Yukawa-Higgs type. This symmetry-breaking phase still has a deconfined $\ZZ_4$ gauge field and is topologically ordered. To obtain Landau symmetry breaking orders we need to go through another confinement transition, by condensing the $\ZZ_4$ gauge flux excitations. 

There is a more interesting transition directly from the $\ZZ_4$ DSL to a Landau symmetry breaking order, without passing through a gapped topological order. The continuum field theory is described by the following Lagrangian:
\begin{align}\label{eq:LQED3GNHiggs}
\Lmc = \Lmc_{\QED_{3}}^{\Nf=8} + \abs{D_{4a}\mathfrak{H}}^{2}  + r \abs{\mathfrak{H}}^{2} + u\abs{\mathfrak{H}}^{4} + \dots
\end{align}
where $\Lmc_{\QED_{3}}^{\Nf=8}$ is the QED$_3$ Lagangian for the standard $U(1)$ Dirac spin-orbital liquid, and $\mathfrak{H}$ is a continuum Higgs field with gauge charge $4$. The $...$ include the monopole perturbation that is allowed by physical symmetries as discussed in Sec.~\ref{sec:Instability}, as well as terms like the fermion quartics $\G_{\x\n}\lrRb{\bpsi  \TT^{\x} \psi }\lrRb{\bpsi  \TT^{\h} \psi }$ where $\TT^{\h=1,\dots,36}$ are the generators of $\mathfrak{u}(8)$. The transition is accessed by tuning $r$ across some critical value $r_c$. For $r<r_c$ we have the Higgs condensate $\langle\mathfrak{H} \rangle\neq0$ and we get the $\mathbb{Z}_4$ DSL. For $r>r_c$ the Higgs field is gapped and at low energy we are left with just the QED$_3$ theory, which as discussed in Sec.~\ref{sec:Instability}  is unstable due to the monopole perturbation. The resulting IR state is likely a confined symmetry breaking state. The exact pattern of the broken symmetry may depend on other ... terms in Eq.~\eqref{eq:LQED3GNHiggs}. As discussed in Sec.~\ref{sec:SpinOrbitOrdersPi} some of the possible orders are the VBS and Neel orders. At the critical point $r=r_c$, the monopole perturbation will become heavier than in the pure QED due to the extra critical Higgs field. In fact even without the Dirac fermions, the charge-$4$ Higgs field will render the monopole irrelevant: the pure Higgs model is dual to the $O(2)$ Wilson-Fisher transition and the monopole corresponds to the charge-$4$ operator $\phi^4$ in the $O(2)$ Wilson-Fisher theory, which is known to be irrelevant from numerics\cite{Carmona2000}. Therefore it seems reasonable to expect that with $N_f=8$ additional Dirac fermions the monopole will also be irrelevant. Other terms such as fermion quartics are also likely to be irrelevant. We therefore conclude that the QED-Higgs theory is likely stable at the critical point and describes a continuous transition. This mechanism for the direct nontrivial phase transition has been discussed, in a different context, in Ref.~\cite{Gazit2018}.

A question now is how we can differentiate between the $\ZZ_{4}$ DSL and the $\U(1)$ DSLs. The two types of DSL will have different operator scaling dimensions, since the $\ZZ_4$ DSL is IR free while the $U(1)$ DSL is a nontrivial CFT. Another way, possibly easier to implement in the short run, is to break time-reversal and reflection symmetries and see what phase is induced.  This can be done by adding to the original Heisenberg Hamiltonian  a spin-orbit chirality 
\begin{equation}
 H_{chiral} =  \x \sum_{\rbs \in A} f_{XYZ}T^{X}_{\rbs} T^{Y}_{\rbs +\abss_{1}} T^{Z}_{\rbs-\abss_{3}}  + \x  \sum_{\rbs \in B} f_{XYZ}T^{X}_{\rbs} T^{Y}_{\rbs +\abss_{2}} T^{Z}_{\rbs+\abss_{1}},
\end{equation}
where $f_{XYZ}$ are the structure constant of $\SU(4)$ defined by $[T^{X},T^{Y}]= if_{XYZ}T^{Z}$. For both $\U(1)$ and $\ZZ_{4}$ DSL, the chirality term will induce an $\SU(8)$ invariant Dirac mass $\pm\bpsi \psi$. This mass gaps out the fermions and induces a Chern-Simons term for the the gauge field. The resulting topological orders will have $ K $ matrices: 
\begin{equation}
K^{DSL}_{\U(1)} = 
\pm\begin{pmatrix}
2 & 1 & 1 \\
1 & 2 & 1 \\
1 & 1 & 2 \\
\end{pmatrix}
 \hspace{1cm} \text{and} \hspace{0.7cm} K_{\ZZ_4}^{DSL} = 
\pm\begin{pmatrix}
2 & 1 & 1 & 4 \\
1 & 2 & 1 & 4 \\
1 & 1 & 2 & 4 \\
4 & 4 & 4 & 16 \\
\end{pmatrix}.
\end{equation}
The differences can be easily distinguished by looking at the entanglement spectrum or edge modes using numerical techniques such as the density-matrix renormalization group (DMRG). In the $\U(1)$ case we have four superselection sectors spanned by an anyon $\psi$ such that $\psi^{4}=1$ and $\psi$ has a $\pm\frac{3\p}{4}$ self-statistic angle. In contrast, the $\ZZ_4$ case has $16$ superselection sector spanned by two anyons $\psi$ and $h$ such that $\psi^{4}=h^{4}=1$. The self-statistic angles are $\pm\frac{3\p}{4}$ and $\pm\frac{\p}{4}$, respectively, while the mutual phase is $0$.

\section{Anomalies of Dirac spin liquids and Lieb-Schultz-Mattis}
\label{sec:GlobalSymmetriesLSM}
 
 The Lieb-Schultz-Mattis-Oshikawa-Hastings theorems (LSMOH) \cite{lieb1961two, oshikawa2000commensurability,hastings2004lieb} forbid some systems to have a trivially gapped phase in the presence of symmetries. In Ref.~\cite{song2018spinon} it was shown that the anomaly of the $\Nf=4$ DSL together with LSMOH theorems restrict the possible values of the monopole phases under $\ZZ_2$-like symmetries. In this section we extent the calculation to DSLs with $N$ spin flavours and $M$ valleys (both $N$ and $M$ are even).

The anomaly of a symmetry $ G$ is calculated by trying to gauge $ G $ in addition to the original $\U(1)_{\mmrm{g}}$ of $\QED_3$. The resulting partition function will generally need to be interpreted as living on the boundary of a $(3+1)$d SPT with a topologically nontrivial bulk action. We will compute this anomaly later in this section. For now we shall only display the result of the bulk action:
\begin{equation}\label{eq:Anom}
\frac{S_{\mmrm{bulk}}}{2\pi i}= \int_{X_{4}}\,\lrSb{\frac{1}{2}w_2^s \cup w_2^v + \left( \frac{1}{N} w_2^s+\frac{1}{M} w_2^v + \frac{1}{2}\frac{dA^{top}}{2\pi} \right) \cup \frac{dA^{top}}{2\pi}}
\end{equation} 
where $w_2^s \in H^2(X,\mathds{Z}_N)$ and $w_2^v \in H^2(X,\mathds{Z}_M)$ are the Stiefel-Whitney (SW) classes of the $\PSU(N)$ and $\PSU(M)$ bundles respectively. They represent the obstruction to lift the respective projective group to its linear variant. $A_{\mmrm{top}}$ is the $ \Utop $ background gauge field. As in the $N=M=2$ case, the first term can be interpreted as a descendant of the parity anomaly of the fermions and the second term simply tells us that the magnetic particles of $\Utop$ (i.e. the Dirac fermions) anticommute between them and transform in the fundamental representations of $\sus$ and $ \suv$.

As that our system comes from a purely two dimensional system so the bulk action should be trivial when restricted to $\gmicro$. We can read how $w_v$ restricts to the gauge fields of $\glat$ from the transformation properties of the monopoles. We can then figure out what is the restriction of  $\frac{\dd{A}^{top}}{2\p}$ by requiring a cancellation of the bulk action. 

{
 For lattice $\SU(N)$ spins, the generalized LSM theorems (See Ref. \cite{PhysRevLett.121.097201} and references therein) state that there is an obstruction to featureless states by the number of fundamental spins (denote as $n$) in $\SU(N)$ per unit cell modulo $N$. This obstruction can be phrased as the anomaly
 \be
 \frac{S_{\rm{LSM}}}{2\pi i}=\int_{X_4}\frac{n}{N}w_2^s\cup x\cup y,
 \ee
 where $x,y\in H^1(X_4,\mathbb{Z})$ are the integer-valued forms associated with the translation gauge fields. In the honeycomb $\SU(4)$ system, we have $N=2n=4$. In general there can also be LSM anomalies associated with lattice rotation symmetries, but this does not happen in our honeycomb $\SU(4)$ case since there is no spin at the $C_6$ rotation center, and for $\SU(4)$ spins there cannot be any anomaly about on-site $C_{3}$ as it is impossible to have a non-trivial $\mathds{Z}_4$ cycle in $\mathds{Z}_3$.
 
 One can check that the two $U(1)$ Dirac spin-orbital liquids discussed in Sec.~\ref{Nf=8Honeycomb} and \ref{sec:Alternatives} indeed satisfy the anomaly matching condition. Notice that translations act projectively on the partons as $\ZZ_{2}$ groups that anticommute. Then if  $x$ and $y$ are the $\ZZ_2$ forms associated to translation gauge fields,  $w_{2}^{v}=x\cup y$. If we let $dA^{top}=0$, the anomaly becomes $S_{anom}=\pi w_{2}^{s}\cup x \cup y$ which is precisely what we expect from the LSM theorem as there are  $\tfrac{2}{4}=\tfrac{1}{2}$ fundamental spins per unit cells. Similarly if we gauge the $C_{2h}$ rotation symmetry which acts trivially on all fields and set $w_2^{v}=\dd A^{top}=0$, we get $S_{anom}=0$ which is consistent with the lack of anomaly as there is no spin at the inversion center.
 }

The aim of the rest of this section is to give some highlights of the derivation of the anomaly in three different cases. We start with the most straightforward generalization of the result of Ref.~\cite{song2018spinon} to $\PSUNs \times  \PSUMv \times \Utop$ { and show how to derive the anomaly Eq.~\eqref{eq:Anom}}. Next we consider gauging the complete $\frac{\PSU(\Nf)\times \Utop}{\ZZ_{\Nf}}$ continuous symmetry for the $\Nf  = 0 \mod 4$, {and derive the full anomaly of this continuous symmetry Eq.~\eqref{eq:AnomalyFullL}}. After this, we revisit the $\Nf= 4$ case and also include charge conjugation with caveat that we restrict to the case where we can lift the $\frac{\PSU(4)\times \Utop}{\ZZ_{4}}$ bundle to $\SO(6)\times \Utop${, and derive the anomaly Eq.~\eqref{eq:AnomalyChrg4}}. 

In Appendix \ref{app:math} we compile several mathematical constructions and identities used in the derivation of the various anomalies. 

\subsection{The \texorpdfstring{$\PSUNs \times  \PSUMv \times \Utop$}{PSU(N)s x PSU(M)v x U(1)top} anomaly} \label{app:anomaly}

We give a derivation of the anomaly of the $G=\PSUNs \times \PSUMv \times \Utop$ symmetry of $\QED_3$ following App.~{\blue{B}} of Ref.~\cite{song2018spinon} under the assumption that $N$ and $M$ are even. We introduce the background gauge fields $A_{s}$, $ A_{v} $ and $ A_{top} $ for each factor of $G$. The Dirac fermions are regularized using a Pauli-Villars regulator so that the regulated partition function in the presence of the gauge field and gravitational background $g$ is 
\begin{align}
Z[A,g]_{PV} =\abs{Z[A,g]}\exp(-\frac{i\p}{2} \h[A,g]),
\end{align}
where $\h[A,g]$ is the $\h$-invariant - a truly gauge invariant quantity that at classical level is equal to a Chern-Simons (CS) term\cite{Witten2015}. If we want to preserve the $\Tmc$ symmetry, our properly gauged $\QED_3$ should be written as 
\begin{align}\label{eq:S0_QED_Anom}
S = \int \lrSb{   \lrRb{\bpsi_{} \Ds_{a,A_{s},A_{v}} \psi}+\frac{iNM}{8\p}a\wedge\dd{a}+\frac{M}{2}i\CS[A_{s}]+i\frac{N}{2}\CS[A_{v}] + iNM\CS[g]   - \frac{1}{2\p} {a\wedge \dd{A^{top}} }}
\end{align}
where the $\CS$ terms are taken in the representations of $\psi$.  The $\CS$ terms are defined so that $\dd\CS[A^{\SU(L)}]=\frac{1}{4\p}\Tr_{\mathrm{fund}}[F^{\SU(L)}\wedge F^{\SU(L)}]$ with $F^{\SU(L)}$ the curvature of the $A^{\SU(L)}$ gauge field. Similarly, $\dd\CS[g] = \frac{1}{196\p}\Tr_{\mathrm{vec}}[R\wedge R]$. {Here $\Tr_{\mathrm{fund}}$ and $\Tr_{\mathrm{vec}}$ refer to trace in the fundamental and vector representations, respectively.} As the $\CS$ terms are not gauge invariant, they should really be though of as coming from a (3+1)D bulk whose boundary is our gauged $\QED_{3}$. The action of this bulk is given by
\begin{equation}\label{eq:Anom01}
	\begin{split}
		\frac{		S_{bulk}}{2\p i}= \lrRb{  \frac{NM}{4D^{2}} p_{1}[a] + \frac{M}{4N}p_{1}[A_{s}]+ \frac{N}{4M}p_{1}[A_{v}]   -\frac{NM}{16} \s[M_{4}]  - \frac{1}{D} \int_{M_{4}}\frac{\dd{a}}{2\p} \wedge \frac{\dd{A_{top}}}{2\p}}
	\end{split}
\end{equation}
where we have redefined $a\rightarrow a/D$ with $D =\mmrm{lcm}(M,N,2)$ for later convenience. $p_{1}[B]$ is the Pontryagin number of the $\PSU(L)$-gauge fields $ B $, given by 
\[p_{1}[B] = \frac{1}{8\p^{2}} \int_{M_{4}}\Tr_{\mathrm{Adj}}[F_{B} \wedge F_{B}] .\]
Notice that the trace is taken with respect to the adjoint representation of the linear group $\SU(L)$, that is why there is an extra factor of $2L$ when going from $\CS$ to $p_{1}$\footnote{ Notice that as the tensor product of the fundamental and its conjugate gives the adjoint and trivial representation, if $U\in \SU(L)$ \[  \Tr_{\mmrm{Adj}}[U] =  \Tr_{\mmrm{fund}}[U]\Tr_{\mmrm{fund}}[U^{-1}] - 1 .\] Take $U = e^{i\a_{a}T^{a}}e^{i\b_{b}T^{b}}$ for $T$ the Hermitian generators of $\mathfrak{su}(L)$. Then expanding to order $\a_{a}\b_{b}$ we find that \[ \Tr_{\mmrm{Adj}}[T^{a}T^{b}]=\Tr_{\mmrm{fund}}[T^{a}T^{b}]\Tr_{\mmrm{fund}}[1]+\Tr_{\mmrm{fund}}[1]\Tr_{\mmrm{fund}}[T^{b}T^{a}] =2N\Tr_{\mmrm{fund}}[T^{a}T^{b}]. \]}. For the $\U(1)_{\mmrm{g}}$, $p_{1}[a]= \frac{1}{4\p^{2}}\int_{M_{4}}\dd{a}\wedge\dd{a}$. The signature of the manifold 
\[  \s = -\frac{1}{24\p^{2}}\int_{M_{4}} \Tr_{\mathrm{vec}}[R\wedge R] \]
comes from extending the gravitational CS term to the bulk. $R$ and $F_{B}$ are the Riemannian tensor and field strength of $B$, respectively. 
%

The anomaly is present if the partition function depends on the choice of $M_{4}$ or the extension of the gauge fields to $M_{4}$. Note that it is sufficient to calculate $S_{bulk}$ over a closed manifold $X_4$ instead of $M_{4}$ because the difference of the partition function's phase between two choices of extension or bulk manifold can be calculated as $S_{bulk}$ over the closed manifold obtained by gluing one of the bulk manifolds to the inverted second manifold along the original $M_3$ manifold. Therefore, our task is to {eliminate the dynamical gauge field $a$ (which lives only on the boundary) in Eq.~\eqref{eq:Anom01} and } simplify the $S_{bulk}$ in terms of characteristic classes of the {probe gauge} fields integrated over a generic closed 4-manifold $X_{4}$.

In order to simplify the expression, we list conditions between the different characteristic classes of the gauge field. First, on any $2$-cycle we should satisfy
\begin{align}
\frac{1}{D}w_{2}^{a} + \frac{1}{N}w_{2}^{s} + \frac{1}{M}w_{2}^{v} + \frac{1}{2}w_{2}^{TM} = 0 \mod 1 .
\end{align}
Here $w_2^a \in H^2(M_{4},\mathds{Z}_D)$, $w_2^s \in H^2(M_{4},\mathds{Z}_N)$, $w_2^v \in H^2(M_{4},\mathds{Z}_M)$ and $w_2^{TM} \in H^2(M_{4},\mathds{Z}_2)$ are the second SW classes of $a$, $ \PSUNs $, $ \PSUMv  $ and tangent bundles respectively. As the only fields that have charge $1$ under $\U(1)_{\mmrm{g}}$, transform in the fundamental representation of the $  N(M) $-fold cover of $\PSUNs$ ($\PSUMv$) are fermions (spinors of the tangent bundle), this condition ensures that they are well-defined. We redefined $a$ to get a properly quantized gauge field ($\frac{\dd{a}}{2\p}\in H^2(M_{4},\mathds{Z})$) and $w_2^{a} =  \frac{\dd{a}}{2\p}\mod D$. {

Let $2d=\mmrm{gcd}(M,N)$, $M = 2dm$, $N=2dn$ and $2dD =MN$. The cocycle condition becomes 
\begin{align}\label{eq:CocyleCondition}
w_{2}^{a} +m w_{2}^{s} +nw_{2}^{v} +(dmn )w_{2}^{TM} =0 \mod D.
\end{align}

A useful relation for the $PSU(n)$ bundles ($n$ even) is
\begin{equation}
 p_{1}[\PSU(n)] = (n-1)\Pmc(w_{2}) \mod 2n,
\end{equation}
where $\Pmc: H^2(X_4,\mathbb{Z}_n)\to H^4(X_4,\mathbb{Z}_{2n})$ is the Pontryagin square operation (see Appendix~\ref{app:math} for a more general review).} Recall that if $w \in H^{2}(X_{4},\ZZ_{D}) $ and $\hat{w}\in H^{2}(X_{4},\ZZ)$ satisfy $w= \hat{w}\mod D$ we have $\Pmc(w)  =  w \cup w=  \hat{w} \cup  \hat{w} \mod 2D $\footnote{This can seen from the definition in Ref. \cite{kapustin2013topological} of $\Pmc(w) = w \cup w + w \cup_{1} d w$ as in our case $dw=0$.}. Then, as $w_2^{a}$ admits a lift to an integer class, namely $\frac{\dd{a}}{2\p}$, we have that $\Pmc(\w_{2}^{a}) = \w_{2}^{a} \cup \w_{2}^{a} \mod 2D$. We also have $\sigma = \Pmc(w^{TM}) \mod 4 $. 

{
Next, consider the first four terms in the integrand of Eq~\eqref{eq:Anom01}:
\begin{equation}\label{eq:anomalyCalulation}
	\begin{split}
		 I &= \frac{NM}{4D^{2}} p_{1}[a] + \frac{M}{4N}p_{1}[A_{s}]+ \frac{N}{4M}p_{1}[A_{v}]   -\frac{NM}{16} \s[M_{4}] \\
		 	&= \frac{NM}{4D^{2}} \Pmc(w_{2}^{a}) + M\frac{N-1}{4N}\Pmc(w_{2}^{s})+ N\frac{M-1}{4M}\Pmc(w_{2}^{s}) -\frac{NM}{16} \Pmc(w_{2}^{TM}) \\
		 	&= \frac{1}{4mn}\lrRb{\Pmc(w_{2}^{a}) - m^{2}\Pmc(w_{2}^{s})- n^{2}\Pmc(w_{2}^{v})- (dmn)^{2}\Pmc(w_{2}^{TM})}+\lrRb{\frac{M}{4} \Pmc(w_{2}^{s})+\frac{N}{4} \Pmc(w_{2}^{v})}\\
		 	&= \frac{1}{2}\lrRb{ \w_{2}^{s}\cup w_{2}^{v} + md\w_{2}^{s}\cup w_{2}^{TM}+ nd\w_{2}^{v}\cup w_{2}^{TM}}+\lrRb{\frac{md}{2} \Pmc(w_{2}^{s})+\frac{nd}{2} \Pmc(w_{2}^{v})}
	\end{split} \mod 1
\end{equation}
where we have used that $\Pmc(K w_{2}) = K^{2}\Pmc( w_{2})$ and in the last line we have used the cocycle condition Eq.~\eqref{eq:CocyleCondition} and the property of Pontryagin square $\Pmc(a + b) = \Pmc(a) +\Pmc(b) +2 a\cup b \mod 2L$ for $a,b\in H^{2}(X_{4},\ZZ_{L})$. The last terms in parenthesis is further simplified by the fact that $\Pmc(w_{2}^{s/v}) = w_{2}^{s/v}\cup w_{2}^{s/v} \mod 2$. For an oriented manifold, $w_{2}^{TM}$ is also the second Wu class. Thus if $w_{2}\in H^{2}(X_{4},\ZZ_2)$, $w_{2}^{}\cup w_{2}^{} =w_{2}^{}\cup w_{2}^{TM}$. We can use this relation to simplify $I$ to $I = \half\w_{2}^{s}\cup w_{2}^{v} \mod 1$.

Next, we use Eq.~\ref{eq:CocyleCondition} to replace $\frac{\dd{a}}{2\p}$ in the $\dd{a} \wedge \dd{A^{top}}$ term. The last step is to replace $\half w_{2}^{TM}\cup \frac{\dd{A_{top}}}{2\p} $ by $\half \frac{\dd{A_{top}}}{2\p}\cup \frac{\dd{A_{top}}}{2\p}$. Which is allowed because  $w_{2}^{top} = \frac{\dd{A_{top}}}{2\p} \mod 2$ then $ w_{2}^{TM}\cup \frac{\dd{A_{top}}}{2\p} = w_{2}^{top} \cup w_{2}^{TM} = w_{2}^{top} \cup w_{2}^{top} \mod 2$.
The anomaly Eq.~\ref{eq:Anom01} becomes
\begin{equation}
\frac{S_{\mmrm{bulk}}}{2\pi i}=\int_{X_{4}}\,\lrSb{\frac{1}{2}w_2^s \cup w_2^v + \left( \frac{1}{N} w_2^s+\frac{1}{M} w_2^v + \frac{1}{2}\frac{dA^{top}}{2\pi} \right) \cup \frac{dA^{top}}{2\pi}}.
\end{equation} 

}



\subsection{The \texorpdfstring{$ \frac{\SU(\Nf)\times \Utop}{\ZZ_{\Nf}} $}{(SU(Nf)xUtop)/ZNf} Anomaly}
In this section, we gauge the complete $\frac{\SU(\Nf)\times \Utop}{\ZZ_{\Nf}}$ symmetry group and see that the anomaly reduces to one obtained in the previous section under $\SU(\Nf)\longrightarrow \sus\times\SU(M)_v$. We first do the calculation for $\Nf=4$ and then consider the more general case $\Nf = L >4$ with $ L = 0 \mod 4$.

\subsubsection{\texorpdfstring{$ \Nf =4 $}{Nf=4}}
Let's start by considering $\frac{\SU(4)\times \U(1)_T}{\ZZ_4}$. {We first couple the theory to a $PSU(4)\times U(1)$ bundle.} There are classes $u_2 \in H^{2}(X_{4},\ZZ_{4})$ and $ \frac{\dd{A}}{2\p} \in  H^{2}(X_{4},\ZZ)$ which characterize the lifting from $\PSU(4)$ to $\SU(4)$ and the instanton number of $\U(1)_T$, respectively. We now impose two cocycle conditions 
\begin{subequations}
		\begin{align}
		\oint	\frac{1}{4}u_2 + \frac{1}{2}w_2^{TM} + \frac{da}{2\p} &\in\ZZ, \\
		\oint	\frac{2}{4}u_2 + \frac{dA}{2\p} & \in \ZZ.
		\end{align}
\end{subequations}
The first condition comes from the fermions {(as in Eq.~\eqref{eq:CocyleCondition})} and the second {condition ensures that the correct gauge group is $SU(4)\times U(1)_T/\mathbb{Z}_4$}. Before we start writing the anomaly polynomial, it is convenient to define $b = (2a - A)$ because the LHS of $  \frac{4a\wedge \dd{a}}{8\p} - \frac{a \wedge \dd{A}}{2\p}=\frac{b \wedge \dd{b}}{8\p} -\frac{A\wedge\dd{A}}{8\p}  $ appears in Eq.~\ref{eq:S0_QED_Anom}. The first cocycle condition now reads
\begin{equation}
	\begin{split}
		-\frac{db}{2\p} = \frac{1}{2}u_{2} + \frac{\dd{A}}{2\p} +w_{2}^{TM}  \mod 2 
	\end{split}
\end{equation}
Following similar derivations as in the previous case, the anomaly becomes
\begin{equation}
	\begin{split}
		\frac{S_{bulk}}{2\p i}&= \frac{1}{4}\lrRb{p_1[b] + \frac{1}{4}p_{1}[\PSU(4)] -\s-\frac{\dd{A}}{2\p} \cup \frac{\dd{A}}{2\p} } \\
		&= \frac{1}{4}\lrRb{\Pmc\lrRb{\frac{1}{2}u_{2} + \frac{\dd{A}}{2\p} +w_{2}^{TM}} + \frac{1}{4}p_{1}[\PSU(4)] -\Pmc(w_{2}^{TM}) -\frac{\dd{A}}{2\p} \cup \frac{\dd{A}}{2\p} } 
	\end{split}
\end{equation}
where we used that $p_1[b] = \frac{\dd{b}}{2\p} \cup \frac{\dd{b}}{2\p}= \Pmc(\r_2\lrSb{\frac{\dd{b}}{2\p}}) \mod 4$. Here $\r_{2}: H^{*}(X_{4},\ZZ) \rightarrow H^{*}(X_{4},\ZZ_{2})$ is mod $2$ reduction. 
%

Now we restrict to the case where we can lift the $\SO(6)\times \Utop /\ZZ_{2}\cong \SU(4)\times \Utop / \ZZ_4$ bundle to a $\SO(6)\times \Utop$ bundle, i.e. assume that $u_2 =2w_2^{\SO(6)}$ and $\frac{\dd{A_{top}}}{2\p}$ is an integer class. $w_{i}^{\SO(6)}$ are the i-th SW class of the $\SO(6)$ bundle. Then
\begin{equation}
\begin{split}
\frac{S_{bulk}}{2\p i}
&= \frac{1}{4}\lrRb{\Pmc\lrRb{w_{2} + \frac{\dd{A}}{2\p} +w_{2}^{TM}} + p_{1}[\SO(6)] -\Pmc(w_{2}^{TM}) -\frac{\dd{A}}{2\p} \cup \frac{\dd{A}}{2\p} } \\
& = \frac{1}{2}\lrRb{ w_{4}^{\SO(6)} + \lrRb{w_{2}^{\SO(6)}+\frac{\dd{A}}{2\p}} \cup \frac{\dd{A}}{2\p}}\\
& = \frac{1}{2}\lrRb{ w_{4}^{\SO(6)} + \lrRb{w_{2}^{\SO(6)}+w_{2}^{top}} \cup w_{2}^{top}}
\end{split}
\end{equation}
where we have introduced $w_{2}^{top} = \frac{\dd{A}}{2\p} \mod 2$.

We can now simply take $\SO(6) \longrightarrow \SO(3)_{s} \times \SO(3)_{v} \times \ZZ_{2}$  which can be interpreted as two $\Om(3)_{s}$ and $ \Om(3)_{v} $ bundles with $ w_{1}^{s} = w_{1}^{v} \equiv w_{1}$. Then we use the relations in App.~\ref{app:OmBundles} to obtain
\begin{equation}\label{eq:AnomChrg0}
\frac{S_{bulk}}{2\p i} = \frac{1}{2}\lrRb{ w_{2}^{s}\cup w_{2}^{v} + \lrRb{w_{2}^{s}+w_{2}^{v}+ w_{1}^{2}+w_{2}^{top}} \cup w_{2}^{top}}
\end{equation}

\subsubsection{\texorpdfstring{$ \Nf= L \, , \,$}{Nf=L}\texorpdfstring{$ 4|L $}{4|L}}
In the more general case of $\PSU(L)$ with $L = 0\mod 4$, the cocycle conditions are
\begin{equation}
	\begin{split}
			\oint	\frac{1}{L}u_2 + \frac{1}{2}w_2^{TM} + \frac{da}{2\p} &\in\ZZ \\
		\oint	\frac{1}{2}u_2 + \frac{dA}{2\p} & \in \ZZ
	\end{split}
\end{equation}
If we define $b =\frac{L}{2}a -A$, the first cocycle condition becomes
\begin{align}
	\oint	\frac{1}{L}u_2 + \frac{1}{2}w_2^{TM} + \frac{2}{L}\frac{db}{2\p} + \frac{2}{L} \frac{\dd{A}}{2\p}&\in\ZZ 
\end{align}
The bulk action now reads
\begin{equation}\label{eq:AnomalyFullL}
\begin{split}
\frac{S_{bulk}}{2\p i}&= \frac{1}{4}\lrRb{\frac{4}{L}p_1[b] + \frac{1}{L}p_{1}[\PSU(L)] -\frac{L}{4}\s-\frac{4}{L}\frac{\dd{A}}{2\p} \cup \frac{\dd{A}}{2\p} } \\
&= \frac{1}{4}\lrRb{\frac{4}{L}\Pmc\lrRb{{\frac{1}{2}u_{2} + \frac{\dd{A}}{2\p} + \frac{L}{4}w_{2}^{TM}}} + \frac{1}{L}p_{1}[\PSU(L)] -\frac{L}{4}\Pmc(w_{2}^{TM}) -\frac{4}{L}\frac{\dd{A}}{2\p} \cup \frac{\dd{A}}{2\p} } .
\end{split}
\end{equation}
{

This is our most general result on the anomaly of QED$_3$ associated with the continuous symmetries.}

As in the $\Nf =4$ case, we now restrict to a $\SU(L)/\ZZ_{L/2} \times \Utop$ bundle, i.e. we assume that $u_{2} = 2 w_2$ for some $w_{2} \in H^{2}(X;\ZZ_{L/2})$ and that $\frac{\dd{A}}{2\p}$ is an integer class. Then
\begin{equation}
\begin{split}
\frac{S_{bulk}}{2\p i}&= \frac{1}{4}\lrRb{\frac{4}{L}\Pmc\lrRb{{w_{2} + \frac{\dd{A}}{2\p} + \frac{L}{4}w_{2}^{TM}}} + \frac{1}{L}p_{1}[\PSU(L)] -\frac{L}{4}\Pmc(w_{2}^{TM}) -\frac{4}{L}\frac{\dd{A}}{2\p} \cup \frac{\dd{A}}{2\p} }  \\
& = \frac{1}{4}\lrRb{\frac{4}{L}\Pmc\lrRb{{w_{2} }} + \frac{1}{L}p_{1}[\PSU(L)] + 2( {\frac{4}{L}}w_{2}^{} + w_{2}^{top})\cup w_{2}^{TM} +2w_{2} \cup w_{2}^{top} }  \\
& = \frac{1}{2}\lrRb{\lrRb{1+\frac{2}{L}}\Pmc\lrRb{{w_{2} }} + \frac{2}{L}\frac{p_{1}[\PSU(L)]}{4} + w_{2}^{top}\cup w_{2}^{top} +  {\frac{4}{L}}w_{2} \cup w_{2}^{top}} 
\end{split}
\end{equation}

Notice now that 
\begin{alignat*}{2}
			p_{1}[\PSU(L)] & =  (L-1)\Pmc(u_{2}^{}) && \mod 2L \notag\\
			& =  (L-1)\Pmc(2w_{2}^{}) && \mod 2L \\
			& =  -4\Pmc(w_{2}^{}) && \mod 2L \notag
\end{alignat*}
which means that $p_{1}[\PSU(L)]$ under the restriction to $\PSU(L,2)\equiv\SU(L)/\ZZ_{L/2}$ bundles is always a multiple of 4. Define a $ q_{1}[\PSU(L,2)]\in H^{4}(X;\ZZ) $ by setting $p_{1}[\PSU(L)] = 4q_{1}[\PSU(L,2)]$. The previous calculation shows that $q_{1}[\PSU(L,2)] = - \Pmc(w_{2}) = - w_{2}\cup w_{2} \mod L/2$. We then define a class $w_{4}^{\PSU(L,2)} \in H^{4}(X;\ZZ_{2})$ by the condition $ q_{1}[\PSU(L,2)] = -\lrRb{\frac{L}{2}+1}\Pmc(w_{2}^{\PSU(L,2)}) + \frac{L}{2}w_{4}^{\PSU(L,2)} \mod L$. The anomaly then becomes
\begin{equation}
\begin{split}
\frac{S_{bulk}}{2\p i}
& = \frac{1}{2}\lrRb{w_{4}^{\PSU(L,2)} + \lrRb{ {\frac{4}{L}}w_{2}^{\PSU(L,2)}+w_{2}^{top}}\cup w_{2}^{top} } 
\end{split}
\end{equation}
where we have renamed $w_2$ as $w_2^{\PSU(L,2)}$.

We next want to compare this with out previous calculation of the anomaly polynomial under the restriction $ \psus \times \psuv \subset \PSU(L) $. For this it is easiest to return to Eq.~\ref{eq:AnomalyFullL} and recall that under the restrictions $ u_2 = M w_{2}^{s}+N w_{2}^{v} $ and $ p_{1}[\PSU(N)\otimes\PSU(M)] = M^{2}p_{1}[\PSU(N)] +N^{2}p_{1}[\PSU(M)]$. The first relation comes from the phases the fermions see under the subgroups and the second come from the expressions of $ p_1 $ in terms of the curvatures of the gauge fields. The anomaly then reads
\begin{equation}
\begin{split}
\frac{S_{bulk}}{2\p i}
&= \frac{1}{4}\lrSb{\frac{4}{L}\Pmc\lrRb{{\frac{M w_{2}^{s}+N w_{2}^{v}}{2} + \frac{\dd{A}}{2\p} + \frac{L}{4}w_{2}^{TM}}} + \frac{M^{2}p_{1}[\PSU(N)] +N^{2}p_{1}[\PSU(M)]}{MN} -\frac{L}{4}\Pmc(w_{2}^{TM}) -\frac{4}{L}\frac{\dd{A}}{2\p} \cup \frac{\dd{A}}{2\p} } \notag
\end{split}
\end{equation}
which simplifies to our previous result Eq.~\eqref{eq:Anom} after some algebra. For example, modulo 4 we have 
\begin{align*}
\frac{4}{L}\Pmc\lrRb{{\frac{M w_{2}^{s}+N w_{2}^{v}}{2}+ \frac{\dd{A}}{2\p} + \frac{L}{4}w_{2}^{TM}}}  =\frac{M(N+1)\Pmc(w_{2}^{s})}{N}& +2w_{2}^{s}\cup w_{2}^{v}+\frac{N(M+1)\Pmc(w_{2}^{v})}{M}+ \lrRb{\frac{4}{L}+2}\Pmc(w_{2}^{A})  \\
& +\frac{L}{4}\Pmc(w_{2}^{TM})+\frac{4}{L}(M w_{2}^{s}+N w_{2}^{v})\cup \frac{\dd{A}}{2\p}.
\end{align*}
The terms involving $w_{2}^{TM}$ cancel directly. Using the expressions for $p_{1}[\PSU(K)]$ in terms of $\Pmc(w_{2}^{\PSU(K)})$ cancel those terms. Finally, as $\frac{\dd{A}}{2\p}$ are integer classes we can replace $\Pmc(w_{2}^A) = \frac{\dd{A}}{2\p} \cup \frac{\dd{A}}{2\p} \mod L$ so only $\frac{1}{2} \frac{\dd{A}}{2\p} \cup \frac{\dd{A}}{2\p}$ remains in the anomaly. 
%
%
%

\subsection{The \texorpdfstring{$\lrRb{\SO(6)\times \Utop}\rtimes \ZZ_{2}^{\Cmc}$}{(SO(6)xUtop)xZ2C} Anomaly}
We want to gauge a charge conjugation $\Cmc$ that flips the charges of  $\Utop$, $ \U(1)_g $ and $ \SU(4) $. {
This is partly motivated by the fact that certain lattice symmetries act as charge conjugation in some cases -- for example the $N_f=4$ Dirac spin liquid on triangular lattice.}
The $\Cmc$ gauge field combines with the $ \PSU(4)$  gauge field to a $\PSU(4)\rtimes \ZZ_{2}^{\Cmc}$ gauge field. This gauge field has a Stiefel-Whitney class $u_2$ that corresponds to the obstruction to lift the bundle over to $\SU(4)\rtimes\ZZ_{2}^{\Cmc}$. In addition, we denote the $\ZZ_2^{\Cmc}$ piece of the gauge field by $ u_1 $. As $a$ and $A$ (the gauge fields for $\U(1)_{g}$ and $ \Utop $) are charged under $\Cmc$, all the differentials should be replaced by a covariant differential $D_{u_{1}}$. The cocycle condition now becomes
\begin{subequations}
		\begin{align}
			\oint	\frac{1}{4}u_2 + \frac{1}{2}w_2^{TM} + \half\frac{D_{u_{1}}b}{2\p} &\in\ZZ \\
			\oint	\frac{2}{4}u_2 +\frac{D_{u_{1}}A}{2\p}& \in \ZZ
		\end{align}
\end{subequations}
where $b=2a$.

To find the bulk action, we start by gauging the maximal $\SO(8)$ symmetry of the fermions and then break it down to $\lrRb{\SO(2)_{g}\times \SO(6)_{f}}\rtimes \ZZ_{2}^{\Cmc}$. The resulting bundle can be formulated as a $\Om(2)_{g}\times \Om(6)_{f} $ bundle where the first SW class of each factor are the same and corresponds to the $\Cmc$ gauge field $u_{1}$:{
\be
w_1^{O(6)}=w_1^{O(2)}=u_1.
\ee}
By using the formula for $ p_{1} $ under Whitney sum (see App.~\ref{app:OmBundles} for more information), we obtain 
\begin{align}
p_{1}[\SO(8)] = p_{1}[\Om(2)_{g}] + \b_{2}( u_{1}) \cup \b_{2}(u_{1})+ p_{1}[\Om(6)_{f}] ,
\end{align} 
where $\b_{2}: H^{*}(X;\ZZ_2) \longrightarrow H^{*+1}(X;\ZZ) $ is the Bockstein homomorphism. As $\b_{2}(u_{1})$ is an integer class that satisfies $\b_{2}(u_{1})  = \Sq^{1}(u_{1}) = u_{1} \cup u_{1} \mod 2$, we can replace $\b(u_{1}) \cup \b(u_{1}) =\Pmc(u_{1}^{2}) \mod 4$. Then 
\begin{equation}
\frac{S_{bulk}}{2\pi i}=\int_{X_{4}}\frac{1}{4}p_1[b]+\frac{1}{4}\Pmc(u_{1}^{2})+\frac{1}{4}p_1[O(6)]-\mathcal{P}(w_2^{TM})-\frac{1}{2} w_{2}^{b} \cup w_{2}^{\Om(2)}.
\end{equation}
After some algebra and noting that $u_{2} = 2 \cdot w_{2}$ (see App.~\ref{app:PSU4vsSO6} for more information), where $w_{i}^{\Om(M)}$ are the i-th SW class of the $\Om(M)$ bundle, we arrive at
\be\label{eq:AnomalyChrg4}
\frac{S_{bulk}}{2\pi i}=\frac{1}{2}\int_{X_{4}}\left[w_4^{O(6)}+w_{2}^{O(6)}\cup u_{1}^{2}+w_2^{O(2)}\cup(w_2^{O(6)}+w_2^{O(2)}+u_1^2) \right],
\ee
{which agrees with the result from a very different calculation\cite{Zou2021}.}

Now, we restrict to $	\Om(3)_{s}\times \Om(3)_{v} \subset \Om(6)$. After more algebra, we obtain
\be\label{eq:AnomalyChrg22}
\frac{S_{bulk}}{2\pi i}=\frac{1}{2}\int\left[w_2^{s}w_2^{v}+(w_{1}^{s})^{2}w_{2}^{s}+(w_{1}^{v})^{2}w_{2}^{v}+w_{1}^{s}w_{1}^{v}(\n_{2}^{s}+\n_{2}^{v})+w_2^{O(2)}(\n_2^{s}+\n_{2}^{v}+w_2^{O(2)}+w_{1}^s w_{1}^{v}) \right],
\ee
where we defined $\n_{2}=w_{2}+w_{1}\cup w_{1}$.  As a sanity check, if we set $w_{1}^{s} =w_{1}^{v}$ (which is equivalent to requiring $u_{1}=0$ ) in Eq.~\ref{eq:AnomalyChrg22}, it reduces to $S_{bulk}$ without gauging $\ZZ_{2}^{\Cmc}$ (Eq.~\ref{eq:AnomChrg0}).

{We can apply this result to the $N_f=4$ DSL on triangular lattice, where the lattice site-centered inversion symmetry acts as a charge conjugation symmetry in the DSL. Specifically\cite{song2018spinon,song2018spinonNumeric}, the inversion symmetry involves a charge conjugation in the $U(1)_{T}$ symmetry as well as a $-I_{3\times3}\in O(3)_v$. We denote the $\mathbb{Z}_2$ form associated with the inversion as $r$. Neglecting all other lattice symmetries, this means that in the anomaly Eq.~\eqref{eq:AnomalyChrg22} we set $w_1^v=r$, $w_2^v=r^2$, $w_1^s=0$, $w_2^{O(2)}=0$. Notice that $\int r^4=0$ (mod $2$) on orientable manifolds. Eq.~\eqref{eq:AnomalyChrg22} then reduces to
\be
\frac{1}{2}\int r^2w_2^s,
\ee
which is exactly the right result since there is one spin-$1/2$ moment on each rotation center.}

\section{Discussions}
\label{Discussions}

In this paper we performed a careful analysis of Dirac spin-orbital liquids (DSL), especially those that can emerge from a honeycomb lattice system with $SU(4)$ spins. The motivation comes from previous numerical study\cite{PhysRevX.2.041013} that suggests a realization of such DSL in the Kugel-Khomskii model on the honeycomb lattice. We found that the standard $U(1)$ DSL, constructed out of a simple $\pi$-flux parton mean field ansatz, is unstable due to a monopole perturbation that is symmetry-allowed and RG-relevant. We then proposed two alternative scenarios (other than spontaneous symmetry-breaking). In the first scenario, an alternative $U(1)$ DSL is proposed based on a different parton construction. This alternative DSL represents a stable phase since all relevant monopoles are disallowed by the physical symmetries. However certain aspects of this alternative DSL appear to be unnatural, for example the symmetry-breaking orders in proximity to this phase appear to be quite different from what has been observed numerically\cite{Penc_TetrametrizationSU4}. In the second scenario, we Higgs the $U(1)$ gauge symmetry in the standard DSL theory to a $\mathbb{Z}_4$ subgroup, through an $SU(4)$ singlet four-fermion condensation. The four-fermion condensate does not alter the mean field properties of the partons, but the gauge field will become massive at low energy and the theory becomes essentially non-interacting and therefore stable. This scenario appear to be more natural since some conventional symmetry-breaking orders can be accessed through a (likely) continuous quantum phase transition. We described the phase transition using an effective QED$_3$-Higgs field theory.

Along the way we have also obtained some general results for DSL with arbitrary (even) number of Dirac fermions $N_f$, extending some previous results for $N_f=4$. These include a more detailed analysis of symmetry properties of monopoles, as well as a more systematic analysis of the quantum (t'Hooft) anomalies of the QED$_3$ effective field theories. We expect these results to be relevant in future studies if a different DSL is found in another system. 

We can also contemplate on various ways to generalize our results, for example to different spin systems with spins forming higher representations of $\SU(N)$, or with on-site symmetries other than $\SU(N)$. We briefly discuss several such generalizations in Appendix~\ref{sec:DSLrect}. We note that for $\SU(2)$ spins, DSL states for higher representations (higher spins) have been theoretically discussed recently in Ref.~\cite{Calvera2020}, motivated by a recent experiment\cite{Liuetal20} on the $S=3/2$ system $\alpha$-CrOOH(D).

\section*{Acknowledgments}

We thank Yin-Chen He, Andreas L{\"a}uchli, Mark Mezei, Subir Sachdev, Ashvin Vishwanath and Liujun Zou for illuminating discussions. 
During the main stages of this work VC was supported by a Visiting Graduate Fellowship program at Perimeter Institute. 
Research at Perimeter Institute is supported in part by the Government of Canada through the Department of Innovation, Science and Industry Canada and by the Province of Ontario through the Ministry of Colleges and Universities.

\bibliography{References}
\newpage
\appendix                                                                                                                   
\begin{widetext}                                   
\section{Monopole operators and discrete symmetries}\label{app:MonopoleOperators} 

In this appendix we build the monopole operators $\Phi^{\dag}$ in the main text and study how the bare discrete symmetries of $\QED_{2+1}$ act on them. The symmetries in consideration are charge conjugation $\Cmc$, spatial reflection $\Rmc$ and time-reversal $\Tmc$, where these symmetries act as in Sec.~\ref{sec:SymmPartons} on fermions and we have dropped the subscripts for the rest of the Appendix. The best way to understand the definition of monopoles is in the large $\Nf$ limit, where gauge fluctuations are weak and the states can be described by free fermions and a Gauss law \cite{borokhov2003topological}.

Consider $\QED_{2+1}$ with $\Nf = 2n$ Dirac fermions , tlet $\ket{q}$ be states corresponding to the $2\p q$-flux inserted with all negative modes filled. Following Ref.~\cite{borokhov2003topological}, using the operator-state correspondence these states can be thought of as Dirac fermions living a sphere with $2\p q$ magnetic field. There will be  
$2N$ Dirac zero modes $a_{\a,q}^{}$ that are Lorentz scalars but transform as $\psi_{\a}^{}$ under $\SU(\Nf)_{\mmrm{f}}$. Then the ground states of the free fermion problem correspond to states 
\begin{align}
\ket{q}, \quad f^{\dag\,\a_{1}}_{\,\,q}\ket{q}, \quad f^{\dag\,\a_{1}}_{\,\,q}f^{\dag\,\a_{2}}_{\,\,q}\ket{q},\quad \dots, \qquad f^{\dag\,\a_{1}}_{\,\,q}f^{\dag\,\a_{2}}_{\,\,q}\dots f^{\dag\,\a_{2n}}_{\,\,q}\ket{q},
\end{align}
where $f_{\b q}^{}\ket{q}=0$ for all $\b$. It turns out that if require to quantize the zero modes in a consistent $\CRmc$ way, the gauge invariant state will then correspond to states with $N$ of the zero modes filled. There are $\binom{2n}{n}$ such states and they transform in the self-conjugate fundamental representation of $\SU(2n)_{\mmrm{f}}$, i.e., in the representation whose Young tableaux has one column and $n$ boxes\footnote{Or in terms of weights, its heighest weight correspond to $\varpi_{n}$}.

Define $\Mmf^{\dag\, [\a_{1}\dots \a_{n}]}_{q }$ as the operator associated with the state\footnote{$\Mmf_{+1}^{\dag\,[\cdot]}$ is $\Mmc$ in the main text.}
\begin{align}
\ket{\Mmf^{\dag\, [\a_{1}\dots \a_{n}]}_{q }  }= f^{\dag \a_{1}}_{q}\dots f^{\dag \a_{n}}_{q}\ket{q }.
\end{align}

We can use the result of Ref.~\cite{10.21468/SciPostPhys.5.1.006} to define a $\CTmc$  transformation\footnote{This is $\Tmc$ in the notation of Ref.~\cite{10.21468/SciPostPhys.5.1.006} .}
\begin{align}
\CTmc\cdot\Mmf^{\dag\, A}_{q} \cdot \CTmc^{-1} =(-1)^{\frac{n(n-1)}{2}} E_{AB}\Mmf^{\dag\, B}_{q}.
\end{align}
where we are using the conventions for the antisymmtric multiindices and $E_{AB}$ in the main text. This operation satisfy $\lrRb{\CTmc}^{2} = (-1)^{n M}$ when acting on monopoles. The monopole operators $\Phi^{\dag\,A} $ in the main text are the Lorentzian version of $\Mmf^{\dag\, A}_{+1} $. 

Time reversal $\Tmc$ flips the $\Utop$ charge without changing the $ \SU(\Nf)_{\mmrm{f}} $ representation and if we assume $\Tmc^{2} = 1$ on monopoles, we can impose  
\begin{align}
\Tmc \cdot \Phi^{\dag\,A} \cdot\lrRb{\Tmc}^{-1} = \begin{cases}
+ E^{AB}\Phi_{\,B} &n =0 \mod 2,\\
+i E^{AB}\Phi_{\,B} &n =1 \mod 2.\\
\end{cases}
\end{align}
$\hat{\Tmc} = (-)^{M}\circ \Tmc$ is also a time reversal that squares to one on monopoles but the only difference is a unitary generated by a local current so it is enough to only consider $\Tmc$ for our purposes.

Similarly for $\CTmc$ we choose
\begin{align}
\CTmc \cdot \Phi^{\dag\,A} \cdot\lrRb{\CTmc}^{-1} = E_{AB}\Phi^{\dag\,B}.
\end{align}
Whatever phase we could have included can be absorbed in a redefinition of the monopole operator. 

Finally, $\CRmc$ leaves the gauge flux fixed but acts as conjugation in $\SU(\Nf)_{\mmrm{f}}$. Assuming that $(\CRmc)^{2}$ is a unitary generated by a local current, i.e. it is a $\Utop$ element when restricted to monopoles, we have
\begin{align}
\CRmc \cdot \Phi^{\dag\,A} \cdot\lrRb{\CRmc}^{-1} = \begin{cases}
+ E_{AB}\Phi^{\dag\,B} &n =0 \mod 2,\\
+iE_{AB}\Phi^{\dag\,B} &n =1 \mod 2.\\
\end{cases}
\end{align}
where again there was a phase ambiguity that corresponds to multiplying by some $\Utop$ element so that $(\CRmc)^{2}= 1$ on monopoles.

Then we find that $\Cmc$ and $ \Rmc $ act as 
\begin{align}
\Cmc \cdot \Phi^{\dag\,A} \cdot\lrRb{\Cmc}^{-1} &= \begin{cases}
+ \Phi_{\,A} &n =0 \mod 2,\\
+i \Phi_{\,A} &n =1 \mod 2.\\
\end{cases}\\
\Rmc\cdot \Phi^{\dag \, A}\cdot\lrRb{\Rmc}^{-1} &= (-1)^{n}E^{AB}\Phi_{B}.
\end{align}

The weird phases for odd $n$ were introduced to simplify actual calculations. The calculations are usually easier done by diagonalizing $E_{AB}$ and write the monopoles in this basis. Note that for $n$ even the representation is orthogonal and for $n$ odd is simplectic, which is why is better to treat the cases separately.

\subsection{ \texorpdfstring{$ \Nf = 2n = 0 \mod{4}  $}{ Nf =0 mod 4}}
Let's think of $E_{AB}$ as matrix acting on  $\bigwedge^{\frac{\Nf}{2}}\CC^{\Nf} \cong \CC^{\binom{\Nf}{\Nf/2}}$. It is symmetric, has trace zero and squares to the identity. Then there are some real orthogonal tensors $\hat{\x}^{A}_{\r}$ such that 
\begin{equation}
\begin{split}
\hat{\x}^{A}_{\r} E_{AB}\hat{\x}^{B}_{\s} &= \l_{\r}\d_{\r\s} \\
\hat{\x}^{A}_{\r} \hat{\x}^{A}_{\s} &= \d_{\r\s} \\
\hat{\x}^{A}_{\r} \hat{\x}^{B}_{\r} &= \hat{\d}_{B}^{A} \\
\end{split}
\end{equation}
for some numbers $\l_{\r} = \pm 1$, where half of them are positive and the other half are negative. We can then define ${\x}^{A}_{\r} = \sqrt{\l_{\r}}\hat{\x}^{A}_{\r}$ so that 
\begin{equation}
\begin{split}
{\x}^{A}_{\r} E_{AB}{\x}^{B}_{\s} = \d_{\r\s}\\
{\x}^{A}_{\r} \lrRb{{\x}^{A}_{\s} }^{*}= \d_{\r\s}\\
{\x}^{A}_{\r} \lrRb{{\x}^{B}_{\r} }^{*}= \hat{\d}^{A}_{B}
\end{split}
\end{equation}
This allow us to define 
\begin{align}
\Phi^{\dag}_{\r} &\equiv \x_{\r}^{A}E_{AB}\Phi^{\dag,B} \\
\Phi^{}_{\r} \equiv\lrRb{\Phi^{\dag}_{\r}}^{\dag} &= (\x_{\r}^{A})^{*}E_{AB}\Phi_{,B} = (\x_{\r}^{A})\Phi_{,A} 
\end{align}
Then 
\begin{equation}
	\begin{split}
			\Tmc: \Phi^{\dag}_{\r} \longrightarrow &  (O_{T})_{\r\s}\Phi_{\s} \\
		\Rmc:		\Phi^{\dag}_{\r} \longrightarrow  & \Phi_{\r}^{}\\
		\CTmc:		\Phi^{\dag}_{\r} \longrightarrow  &\Phi^{\dag}_{\r}\\
		\Cmc:		\Phi^{\dag}_{\r} \longrightarrow  	&  (O_{T})_{\r\s}\Phi_{\s} \\
		(O_{T})_{\r\s} = (\x_{\r}^{A})^{*} E_{AB} \x_{\s}^{B}& =(\x_{\r}^{A}\x_{\s}^{A})^{*} = \l_{\r} \d_{\r\s}
	\end{split}
\end{equation}
where $\l_{\r}$ are the eigenvalues $(\pm 1)$ of $E_{AB}$. If is useful to note that
\begin{subequations}
		\begin{align}
			O_{T}^{2} =\id \quad \text{and}\quad O_{T}^{*} = O_{T}.
		\end{align}
\end{subequations}

If an element  $U \in \SU(\Nf)_{\mmrm{f}}$ acts on Dirac fermions as $\psi_{\a} \longrightarrow U_{a}^{\,\,\b}\psi_{b}$, then $\Phi^{(\dag)}_{\r} \longrightarrow  U_{\r\s}\Phi^{(\dag)}_{\s}$ with
\begin{subequations}
	\begin{align}
	U_{\r\s} = \chi_{\r}^{[\a_{1}\dots \a_{n}]}			U_{\a_{1}}^{\,\,\b_{1}}\cdots U_{\a_{n}}^{\,\,\b_{n}} \lrRb{\chi_{\s}^{[\b_{1}\dots \b_{n}]}}^{*}.
	\end{align}
\end{subequations}
We can check that $U^{\top}\cdot U = \id$ and $ U  = U^{*}  $. This shows explicitly that the representation is orthogonal and real. We can also show that $O_{T}\cdot U \cdot O_{T} $ corresponds to the element $V \in \SU(\Nf)_{\mmrm{f}}$ that acts on the fermions as $V_{\a}^{\,\,\b} = (U_{\a}^{\,\,\b})^{*}$.

Sometimes it is useful to go to a basis of real (hermitian) monopoles as 
\begin{subequations}
	\begin{align}
	\Phi^{(\dag)}_{\r }&= \frac{1}{\sqrt{2}}\lrRb{			R_{\r} \pm  i I_{\r} }
	\end{align}
\end{subequations}
where the symmetry breaking is evident once the monopole is condensed
\begin{equation}
	\begin{split}
			\Tmc: (R_{\r},I_{\r}) &\longrightarrow \lrRb{O_{T}}_{\r\s}(R_{\s},I_{\s})\\
		\CTmc: (R_{\r},I_{\r}) &\longrightarrow (R_{\r},-I_{\r})\\
		\Rmc: (R_{\r},I_{\r}) &\longrightarrow (R_{\r},-I_{\r})\\
				\Cmc: (R_{\r},I_{\r}) &\longrightarrow \lrRb{O_{T}}_{\r\s}(R_{\s},-I_{\s})
	\end{split}
\end{equation}
It is customary to call $R$ and $I$ the real and imaginary part of the monopole.

\subsection{ \texorpdfstring{$ \Nf = 2 \mod{4}  $}{ Nf =2 mod 4}}

Let's think of $E_{AB}$ again as matrix acting on  $\CC^{\binom{\Nf}{\Nf/2}}$. It is antisymmetric and squares to minus the identity. Then there are some real orthogonal tensors ${\x}^{A}_{\r }$ such that 
\begin{equation}
\begin{split}
{\x}^{A}_{\r  } E_{AB}{\x}^{B}_{\s } &= J_{\r\s}\\
{\x}^{A}_{\r } {\x}^{A}_{\s } &= \d_{\r\s}\\
{\x}^{A}_{\r } {\x}^{B}_{\r } &= {\d}_{B}^{A}
\end{split}
\end{equation}
where $J_{\r \s} $ is the canonical antisymmetric tensor, that in matrix form takes the form of antidiagonal matrix and squares to $-\d_{\r\s}$. Then define 
\begin{align}
\Phi^{\dag}_{\r} &\equiv \x_{\r}^{A}E_{AB}\Phi^{\dag,B} \\
\Phi^{}_{\r} \equiv\lrRb{\Phi^{\dag}_{\r}}^{\dag} &= (\x_{\r}^{A})^{*}E_{AB}\Phi_{,B} = J_{\r\s}\x_{\s}^{A}\Phi_{,A} 
\end{align}
so that
\begin{equation}
\begin{split}
\Tmc: \Phi^{\dag}_{\r} \longrightarrow  &i J_{\r\s} \Phi_{\s}\\
\Rmc:		\Phi^{\dag}_{\r} \longrightarrow  &-J_{\r\s}\Phi^{}_{\s}\\
\CTmc:		\Phi^{\dag}_{\r} \longrightarrow  &J_{\r\s}\Phi^{\dag}_{\s} 
\end{split}
\end{equation}
The $\SU(\Nf)_{\mmrm{f}}$ group acts on $\Phi^{\dag}_{\r}$ by multiplication by 
\begin{align}
U_{\r\s} = \chi_{\r}^{A}U_{A}^{\,\,B}(\chi_{\s}^{B})^{*}.
\end{align}
We now have $U^{\top} \cdot J \cdot U = J$ 
\footnote{$  J_{\a\b}U_{\a\r}U_{\b\s} ={ \color{red}J_{\a\b}\chi_{\a}^{A}\chi_{\b}^{B}}\chi_{\r}^{C}\chi_{\s}^{D}U_{A}^{\,\,C}U_{B}^{\,\,D}= {\color{blue} E^{AB}U_{A}^{\,\,C}U_{B}^{\,\,D}}\chi_{\r}^{C}\chi_{\s}^{D} = {\color{green!50!black} E^{CD}\chi_{\r}^{C}\chi_{\s}^{D}} = J_{\r\s}.$}and $U^{\dag}\cdot U = \id$. This shows explicitly that the representation is symplectic.

In contrast to the previous section, we cannot define a real and imaginary parts that transform nicely under $\SU(\Nf)_{\mmrm{f}}$. Instead, consider
\begin{subequations}
		\begin{align}
			\aabs_{\r} = \begin{pmatrix}
			J_{\r\s}\Phi^{}_{\s}\\
			\Phi^{\dag}_{\r}
			\end{pmatrix},
		\end{align}
\end{subequations}
so that $(\aabs^{\dag}_{\r})^{\top}=\ve J_{\r\s}\aabs_{\s}$, where $\ve = i\t^{2}$ and $\t^{a}$ are Pauli matrices acting on the components of $\aabs_{\r}$. Then
\begin{equation}
	\begin{split}
			\Tmc: 
		\aabs_{\r}&\longrightarrow i \t^{1}\aabs_{\r}\\
		\Rmc:	\aabs_{\r}&\longrightarrow \t^{1}J_{\r\s}\aabs_{\s}\\
		\CTmc: 
		\aabs_{\r}&\longrightarrow J_{\r\s} \aabs_{\s}\\
		\Utop: 	\aabs_{\r}&\longrightarrow e^{-i\q\t^{3}}\aabs_{\r}\\
\SU(\Nf)_{\mmrm{f}}: \aabs_{\r}&\longrightarrow U_{\r\s}\aabs_{\s}
	\end{split}\quad.
\end{equation}

\section{Symmetry properties from zero modes { -- the \texorpdfstring{$N_f=8$}{Nf=8} case on honeycomb lattice}}\label{app:HCSU4_explicit}

Following the discussion in Sec \ref{sec:monopoleWF}, we find that the $\SU(8)$ representation of the monopoles $\mmbf{70}_8$ breaks into $\SU(4)_s \times \SU(2)_v$ representations as
\begin{equation}
\bf{70}_8 \longrightarrow (\bf{1}_4,\bf{5}_2) \oplus  (\bf{15}_4,\bf{3}_2) \oplus (\bf{20}_4,\mmbf{1}_2)
\end{equation}
The $\SU(4)_s$ representations can be understood by the identification $\SU(4)/\mathds{Z}_2 = \SO(6)$. $\mmbf{1}_4$, $ \mmbf{15}_4$ and $\mmbf{20}_4$ are the trace (scalar), antisymmetric part (adjoint) and traceless symmetric part of an $\SO(6)$ matrix obtained by taking the tensor product of two vectors. For the $\SU(2)$ case, we can similarly use the identification $\SU(2)/\mathds{Z}_2=\SO(3)$. 

Consider $\SU(4)$ tensors with two indices that transform as an $\SO(6)$ vector
\begin{equation}
(s_{\amf} ) = \left( \JJ\s^1, \JJ\s^2,\JJ\s^3,i \JJ\t^1,i \JJ\t^2,i \JJ\t^3 \right)
\end{equation}
where $\JJ=i^2 \s^2 \t^2 $ and $\bar{a}=1,\dots,6$ is an $\SO(6)$ index. The factor of $i$ in the last three components is important to make $s_{\amf}$ transform as an $\SO(6)$ vector under $s_{\amf} \rightarrow U^{\top} s_{\amf}  U =R_{\amf}^{\,\, \bmfk} s_{\bmfk}$, with $R\in \SO(6)$. Note that under the partial action of time reversal $\JJ^{\top}\cdot s_{\amf}^{*} \cdot \JJ = - \h_{\amf}^{\Tmc}s_{\amf} $ where $\h^{\Tmc} = (+1,+1,+1,-1,-1,-1)$.

Similarly for the $\SU(2)_v$ factor, we introduce 
\begin{equation}
(v_{\imf}) = \lrRb{\ve_\m \m^1,\ve_\m \m^2, \ve_\m \m^3 }
\end{equation}
with $\ve_\mu= i \m^2$ and $\bar{i}=1,2,3$ is an $\SO(3)$ index. Under partial time reversal $\ve_{\m}^{\top}\cdot v_{\imf}^{*}\cdot \ve_{\m} = -v_{\imf}$.

Next we consider the tensors $S_{\amf\bmfk} = s_{\amf}\otimes s_{\bmfk}$ and $V_{\imf\jmf} = v_{\imf}\otimes v_{\jmf}$. The monopole wave-functions are then obtained by antisymmetrizing the tensor product of the appropriate components (trace, symmetric or antisymmetric components) of $S_{\amf\bmfk}$ and $V_{\imf\jmf}$. We also need to introduce an extra normalization factors:
\begin{equation}
    \begin{split}
        \x_{\imf\jmf}^{(\Vmc)} &= \mathbb{c}_{\Vmc}\Amc\lrSb{S_{\amf\amf}\otimes V_{\langle\imf\jmf\rangle}}\\
        \x_{\amf\bmfk,\imf\jmf}^{(\Nmc)} &= i\mathbb{c}_{\Nmc}\Amc\lrSb{S_{[\amf\bmfk]}\otimes V_{[\imf\jmf]}}\\
        \x_{\amf\bmfk}^{(\Qmc)} &= \mathbb{c}_{\Qmc}\Amc\lrSb{S_{\langle\amf\bmfk\rangle}\otimes V_{\imf\imf}}
    \end{split}
\end{equation}
with $\mathbb{c}_{\Vmc/\Nmc/\Qmc}$ real numbers. In the above expression, $[\dots]$ denotes antisymmetrization of indices and $\langle\dots \rangle$ corresponds to the traceless symmetric combination of indices. The factor of $i$ in $\x^{(\Nmc)}$ is important to make the representation explicitly real and to make $(\x_{\r}^{A})^{*}E^{AB}=\x_{\r}^{B}$. From here we point out that the time-reversal matrix becomes $\lrRb{O_T}_{\r\s}=\pm \Tmc'[\chi_{\r}^{A}](\chi_{\s}^{A})^{*}$ (with sum over A), where $\Tmc'$ is the partial action of time-reversal we discussed previously in the section. We have $\Tmc'[\x_{\imf\jmf}^{(\Vmc)}] = \x_{\imf\jmf}^{(\Vmc)}$, $\Tmc'[\x_{\amf\bmfk,\imf\jmf}^{(\Nmc)}] = - \h^{\Tmc}_{\amf}\h^{\Tmc}_{\bmfk}\x_{\amf\bmfk,\imf\jmf}^{(\Nmc)}$ \footnote{The extra minus sign comes from the $i$ in the normalization.} and $\Tmc'[\x_{\amf\bmfk}^{(\Qmc)}] = \h^{\Tmc}_{\amf}\h^{\Tmc}_{\bmfk}\x_{\amf\bmfk}^{(\Qmc)}$. Lastly, we make a change of labels, $[\amf \bmfk]\rightarrow A $, for the $\Nmc$ monopoles to match the notation in the main text. Under this change of basis $-\h^{T}_{\amf}\h^{T}_{\bmfk}\rightarrow (O_{\Tmc})_{AA}$. To see this, note that the adjoint indices that get a minus sign are the ones that rotate only either $\vec{\s}$ or $\vec{\t}$. This exactly corresponds to the elements where $\h^{\Tmc}_\amf=\h^{\Tmc}_\bmfk$.

As mentioned in the main text, the $\imf$ index transforms as $F_2$. It is then easy to see that $V_{\imf\imf}$ transform as $A_{1}$ and $V_{[\imf\jmf]}$ transforms as $F_2$. The tensors $V_{\langle\imf\jmf\rangle}$ go into two different representations: the off-diagonal terms ($\imf\neq \jmf$) transform still as $F_2$ while the diagonal terms transform as $E_2$. More explicitly, if we define
\begin{equation}
\begin{alignedat}{4}
V_{0} &= \frac{V_{11}+V_{22}+V_{33}}{2\sqrt{3}},  &\quad 	V_{\imf} &= \frac{1}{2\sqrt{2}}\abs{\ve_{\imf\jmf\kmf}}V_{\jmf\kmf},\\
V_{\pm} & = \frac{V_{11}+\w^{\pm 1} V_{22}+\w^{\pm 2}V_{33}}{2\sqrt{3}},  & \quad V_{\imf A} &= \frac{1}{2\sqrt{2}}\ve_{\imf\jmf\kmf}V_{\jmf\kmf},
\end{alignedat}
\end{equation}
the transformation properties under $\SU(2)_v$ and $C_{6v}^{'''}$ are summarized in Tab. \ref{tab:repSU2vC6vp}. As mentioned in the main text, it is not necessary to build the explicit linear combinations that transform under a certain $C_{6v}^{'''}$ irreps. We can directly obtain the irreps by using the fact that restricting representations to subgroups commute with tensor product operations of irreps. Using Tab.~\ref{tab:tensorC6vvv}, we see that as $\mmbf{3}\rightarrow F_2$ and $\mmbf{1} \rightarrow 1$ then 
\begin{equation}
    \begin{split}
       \mmbf{5} = \left(\bigvee \mmbf{3}\right) \ominus \mmbf{1} 
       \longrightarrow& \left(\bigvee F_2 \right) \ominus A_1\\
       \longrightarrow& \left( A_1\oplus E_2 \oplus F_1 \right) \ominus A_1\\
       \longrightarrow& E_2 \oplus F_1
    \end{split}
\end{equation}
as we showed using the explicit construction.

\begin{table}[t]
	\centering
	\begin{tabular}{ | c || c | c|| c |c |c | c | }
		\hline
		\fbox{$\mathrm{R}_{SU(2)_{\mathrm{v}}}$} 	&	\fbox{$ \mathrm{R}_{C_{6v}^{'''}}$}	& \fbox{Basis} &	\multicolumn{1}{c|}{\fbox{$T_1$}}	&	\multicolumn{1}{c|}{\fbox{$T_2$}}	&	\multicolumn{1}{c|}{\fbox{$R_y$}}	&	\multicolumn{1}{c|}{\fbox{$C_{6h}$}} 
		\\\hlinewd{.8pt}\hline
		\fbox{$\mmbf{1}$} & \fbox{$A_1$} & \fbox{$V_0$}  & \fbox{$\Id$}& \fbox{$\Id$} & \fbox{$\Id$}& \fbox{$\Id$}  \\ \hline
		
		$\mmbf{3}$ & $F_{2}$	&$ (V_{{1}A},V_{{2}A},V_{{3}A})^{\top}$&	
		\fbox{$ \matTone $}	&	\fbox{$\matTtwo $	}&	\fbox{$-\matRy $}	& \fbox{	$ \matCsix$ }
		\\
		\hline
		&	$F_{1}$	&$ (V_{{1}S},V_{{2}S},V_{{3}S})^{\top}$&	
		\fbox{$\matTone$}	&	\fbox{$\matTtwo$}	&	\fbox{$ +\matRy $}	& \fbox{	$\matCsix$ }
		
		\\\cline{2-7}
		
		\multirow{-2}{*}{$\mmbf{5}$}	    & $E_{2}$
		&   $(V_+ ,V_-)^{\top}$ &   $\Id$  &  $\Id$ &  \fbox{  $ 
			\left( \begin{smallmatrix} 
			0 & 1 \\
			1 & 0 \\
			\end{smallmatrix} \right)$ }& $ 
		\left( \begin{smallmatrix} 
		\omega & 0 \\
		0 & \omega^2 \\
		\end{smallmatrix} \right)$
		
		\\\hline
		
	\end{tabular}
	
	\caption{\label{tab:repSU2vC6vp}Summary of the symmetry properties of the $V_{\imf\jmf}$ tensors under  $\SU(2)_{\mmrm{v}}$ and the extended point group $C_{6v}^{''''}$. We denote the valley irreps  ($\mathrm{R}_{\SU(2)_{\mmrm{v}}}$) by their dimension. The point group irreps ($\mathrm{R}_{C_{6v}^{'''}}$) are denote by the usual symbols explained in Sec.~\ref{app:C6vppp}. The symmetry operation $g \in \{T_{1},T_{2},R_{y},C_{6h}\}$} act on the vectors in the basis column by left multiplication.
\end{table}
\def\arraystretch{1.5}
\section{Representation theory of space groups}\label{app:RepresentationTheory}

In this section we spell out some details about the representation theory of extended point grounds used in the main text, focusing mainly in the extension of $C_{6v}$ to $C_{6v}^{'''}$. We will only talk about \emph{linear} representations as we only attempt to classify the representations of gauge invariant operators, which should transform linearly under the symmetries of the system. We restrict to unitary representations and leave time reversal on its own. 

\subsection{Finite groups}
Recall that the representation theory of a finite group boils down to find all the conjugacy classes and then construct the character table. A character of a representation $R$ is a function from the group ($G$) to the complex numbers given by $\chi_{R}(g) = \frac{1}{\dim{R}}\tr(\p_R(g))$ where $\p_R(g)$ is the matrix that represents the element $g$ in the representation $R$. From the definition it is clear that if $g' = h\cdot g \cdot h^{-1}$, then $\x_R(g)=\x_R(g')$, i.e. the characters are constants on conjugacy classes. We can define an inner product between characters given by $(\x_1,\x_2) =\frac{1}{\abs{G}} \sum_{g\in G}\bar{\x}_1(g)\x_2(g)$. It turns out that the characters of irreducible representations (irreps) are orthogonal under this product and form a basis for characters. There are as many irreps as conjugacy classes and a irrep is fully classified once we know the the value of its character on each conjugacy class. Which is why most of the information of the representation thoery is summarized on a table which gives a table with a list of the irreps with the value of its character next to it. 

Representations can also be combined to form new representations by taking direct sums or tensor products. It is useful to have a rule about how to combine irreps under tensor products as this will specify how to write tensor products of any representations. Furthermore, we can define symmetric square $\bigvee R$ and antisymmetric square $\bigwedge R$ of a representation $R$ by taking the symmetric and antisymmetric elements of the tensor product $R\otimes R$. 

The rules to take the product of representations can be figure out by noting that $\chi_{R_1\otimes R_2}(g)=\chi_{R_1}(g)\chi_{R_2}(g)$ and then taking inner products with respect to the irreps. For the symmetric and antisymmetric squares of a representations, we can use $\chi_{\bigvee R }(g) = \half\lrRb{\chi_{R}(g)^2 + \chi_R(g^2)}$ and $\chi_{\bigwedge R }(g) = \half\lrRb{\chi_{R}(g)^2 -\chi_R(g^2)}$. The last relations imply that 
\begin{equation}
    \begin{split}
    \bigvee(R_1\oplus R_2) &= \bigvee R_1 \oplus \bigvee R_2 \oplus (R_1\otimes R_2), \\
    \bigwedge(R_1\oplus R_2) &=\bigwedge R_1 \oplus \bigwedge R_2\oplus (R_1\otimes R_2),
    \end{split}
\end{equation}
and
\begin{equation}
    \begin{split}
    \bigvee(R_1\otimes R_2) &= (\bigvee R_1 \otimes \bigvee R_2)\oplus (\bigwedge R_1 \otimes \bigwedge R_2), \\
    \bigwedge(R_1\oplus R_2) &= (\bigvee R_1 \otimes \bigwedge R_2)\oplus (\bigvee R_1 \otimes \bigwedge R_2).
    \end{split}
\end{equation}

\subsection{Space groups}

Often we want to classify how objects transform under the symmetries of the underlying lattice which is encoded in the space group. The space group we encounter is often a semdirect prodcut of the point group $G_{\text{pt}}$ (symmetries of the unit cell) and translations ($\cong \ZZ^{2}$). This group is \emph{not} a finite group so we cannot directly apply the techniques of the previous section. Nevertheless, the action of the space group on the objects we care about is such that some combinations of translations act trivially. This allow us to study the representation theory of $G_{\text{pt}}^{(\L)}=(G_{\text{pt}}\ltimes \ZZ^{2})/( \L)$  where $\L$ is a set of translations that is closed under composition and conjugation by the point group. This condition ensures that the quotient is well defined. Notice that if $\L' \subset \L$, then $G_{\text{pt}}^{(\L')}$ is isomorphic to  $G_{\text{pt}}^{(\L)}\ltimes (\L/\L')$ and thus the irreps $G_{\text{pt}}^{(\L)}$ are also irreps of $G_{\text{pt}}^{(\L')}$.

We can understand $G_{\text{pt}}^{(\L)}$ as the point group of a super unit cell that depends on $\L$. This groups are thus often called the extended point groups. Another way to understand the group quotient is that we are imposing relations among the translations we are adding. 

\subsection{ The \texorpdfstring{$C_{6v} \ltimes \ZZ^2$}{C6v x ZZ2} case}

The triangular, Kagome and honeycomb lattice share the same space group, the difference is the location of the sites and the edges of the lattices. The point group is called $C_{6v}$. As our main interest on the main text is on the honeycomb lattice, we will have in mind said lattice in the examples below. We denote the extended point group $C_{6v}^{(\Lambda)}$ by $C_{6v}^{(\abs{\L})}$ where $\abs{\L}$ is the length of the smallest unit vector of $\L$\footnote{At least for the cases considered in this appendix there is no ambiguity in this notation.}. 

\subsubsection{ \texorpdfstring{$C_{6v}$}{C6}}
$C_{6v}$ is generated by a reflection ($\s$)\footnote{This is $R_y$ in the main text.} and a $\p/3$ rotation ($C_{6}$) around the hexagon center. They satisfy $\s^2=C_6^6=(\s C_6)^2=e$, where $e$ is the identity element. There are $6$ conjugacy classes represented by $\{e,C^{}_{6},C_3=C^{2}_{6},C_2=C^3_{6},\s,\s \cdot C_6\}$. The character table for the 6 irreps is shown in Tab.~\ref{tab:CharacterTableC6v}. The one dimensional representations are actually representations of the $\ZZ_2\times \ZZ_2$ subgroup generated by reflections $\s$ and inversion $C_2=C_6^3$. The irreps are generated by $A_2$ and $B_1$. The two dimensional representations correspond to objects with lattice angular momentum: $E_{l}$ get a $\exp(\pm \frac{2\p i }{6})l$ under $C_{6}$. The product rules are simple: all the one dimensional irreps square to the trivial irrep ($A_1$), while $A_2\otimes B_1=B_2$, $E_2 = B_1\otimes E_1$ and $\bigvee E_1 = A_1 \oplus E_2$ and $\bigwedge E_1 = A_2$.

\begin{table}[t]
    \centering
    \begin{equation*}
\begin{array}{|c||c|c|c|c|c|c|}\hline
\fbox{$\left[g\right]$}  &  \fbox{$ \lrSb{e} $}  & \fbox{$ \lrSb{C_{6}^{}} $} & \fbox{$\lrSb{C_{6}^{2}}$}  & \fbox{$ \lrSb{C_{6}^{3}} $}  & \fbox{$\lrSb{\sigma}$}  & \fbox{$\lrSb{\sigma\cdot{C_6}}$}  \\  \hline \hline
D_{[g]} & 1  & 2 & 2 & 1 & 3& 3 \\  \hline\hline
A_1 & 1 & 1 & 1 & 1 & 1 & 1  \\  \hline
A_2 & 1 & 1 & 1 & 1 & -1 & -1  \\  \hline
B_1 & 1 & -1 & 1 & -1 & 1 & -1  \\  \hline
B_2 & 1 & -1 & 1 & -1 & -1 & 1  \\  \hline
E_1 & 2 & 1 & -1 & -2 & 0 & 0  \\  \hline
E_2 & 2 & -1 & -1 & 2 & 0 & 0  \\  \hline 
\end{array}
\end{equation*}
    \caption{Character table for the point group $C_{6v}$. $[g]$ denote the conjugacy class and $D_{[g]}$ denote its dimension. }
    \label{tab:CharacterTableC6v}
\end{table}

\subsubsection{\texorpdfstring{$C_{6v}^{(2)}\left(C_{6v}^{'''}\right)$}{C6vppp}}\label{app:C6vppp}

The group $C_{6v}^{(2)}$ is what we called $C_{6v}^{'''}$ in the main text. The quotient corresponds to $\L=\lrRb{2\ZZ}^2$ which correspond to the constrains $T_{1}^{2}=T_{2}^{2}=e$. We have four new conjugacy classes represented by $T_1$, $T_1\cdot C_2$, $\s\cdot T_1$ and $\s\cdot C_{6}\cdot T_1$. The four new irreps are three dimensional and correspond to objects with momentum at the $\Mbs$ points (the middle points of the edges of the hexagonal BZ) with different behavior under $C_2$ and $\s$ - they can be obtained from the other three by tensoring with the one dimensional irreps of $C_{6v}$:
\begin{equation}
	\begin{split}
		F_2 & = A_2 \otimes F_1 \\
		F_3 & = B_1 \otimes F_1 \\
		F_4 & = B_2 \otimes F_1 
	\end{split}
\end{equation}
The character table for the new irreps of $C_{6v}^{(1,0)} $ is shown in Tab.~\ref{tab:charTableC6vppp}. The product rules for the representations can be obtained from Tab.~\ref{tab:tensorC6vvv} and the expressions of the other irreps in terms of "fundamental" irreps.
\begin{table}[t]
    \centering
    \begin{equation*}
	\begin{array}{|c||c|c|c|c|c|c|}\hline
	 R& R \otimes A_2 & R \otimes B_1 & R \otimes E_1 & R \otimes F_1 &  \bigvee R & \bigwedge R \\ \hhline{|=|=|=|=|=|=|=|}
	A_2 & A_1 & B_2 & E_1 & F_2 & A_{1} & \cdot \\\hline
	B_1 & B_2 & A_1 & E_2 & F_3 & A_{1} & \cdot \\\hline
	E_1 & E_1 & E_2 & A_1 \oplus A_2 \oplus E_2 & F_3\oplus F_4 & A_{1} \oplus E_{2} & A_{2} \\\hline
	F_1 & F_2 & F_3 & F_3\oplus F_4 & A_1 \oplus E_2 \oplus F_1 \oplus F_2 &A_{1} \oplus E_{2} \oplus F_{1} & F_{2} \\\hline
	\end{array}
\end{equation*}
    \caption{Rules for the tensor products of representations of $C_{6v}^{'''}$. The last two columns denote the symmetric and antisymetric squares of the representations.}
    \label{tab:tensorC6vvv}
\end{table}

\begin{table}[b]
    \centering
    \begin{equation*}
\begin{array}{|c||c|c|c|c|c|c|c|c|c|c|}\hline
\fbox{$[g]$}  &  \fbox{$ [e] $}  & \fbox{$ [C_{6}^{}] $} & \fbox{$ [C_{6}^{2}] $}  & \fbox{$ [C_{6}^{3}] $}  & \fbox{$[\s]$}  & \fbox{$[\sigma\cdot{C_6}]$} &
\fbox{$ [T_1] $} & \fbox{$ [C_6^3\cdot{T_1}] $} & \fbox{$ [\sigma \cdot T_2] $}  & \fbox{$ [\sigma \cdot C_6 \cdot T_1 ]$} \\  \hline \hline
\fbox{$D_{[g]}$}  & 1  & 8 & 8 & 1 & 6& 6& 3& 3& 6& 6 \\  \hline\hline
A_1 & 1 & 1 & 1 & 1 & 1 & 1 & 1 & 1 & 1 & 1 \\  \hline
A_2 & 1 & 1 & 1 & 1 & -1 & -1 & 1 & 1 & -1 & -1 \\  \hline
B_1 & 1 & -1 & 1 & -1 & 1 & -1 & 1 & -1 & 1 & -1 \\  \hline
B_2 & 1 & -1 & 1 & -1 & -1 & 1 & 1 & -1 & -1 & 1 \\  \hline
E_1 & 2 & 1 & -1 & -2 & 0 & 0 & 2 & -2 & 0 & 0 \\  \hline
E_2 & 2 & -1 & -1 & 2 & 0 & 0 & 2 & 2 & 0 & 0 \\  \hline\hline 
F_1 & 3 & 0 & 0 & 3 & 1 & 1 & -1 & -1 & -1 & -1 \\  \hline
F_2 & 3 & 0 & 0 & 3 & -1 & -1 & -1 & -1 & 1 & 1 \\  \hline
F_3 & 3 & 0 & 0 & -3 & 1 & -1 & -1 & 1 & -1 & 1 \\  \hline
F_4 & 3 & 0 & 0 & -3 & -1 & 1 & -1 & 1 & 1 & -1 \\  \hline
\end{array}
\end{equation*}
     \caption{Character table for the extended point group $C_{6v}^{'''}$. $[g]$ denote the conjugacy class and $D_{[g]}$ denote its dimension. }
    \label{tab:charTableC6vppp}
\end{table}

\section{Mean field, Projective symmetry group and low-energy basis for the \texorpdfstring{$\p$}{pi} flux state on the honeycomb lattice}\label{app:piFluxHC}

\subsection{Mean field details}
The mean field Hamiltonian in Eq.~\ref{eq:HpiFluxMF} in momentum space becomes (we are going to omit spin in this Appendix and set $t=1$)
\begin{equation}
\begin{split} 
H_{MF}&=\sum_{\kbs} f^{\dag}_{\kbs+\Qbs_{1} A} f^{}_{\kbs B} + (e^{i \kbs \cdot \abss_2}+e^{i \kbs \cdot \abss_3}) f^{\dag}_{\kbs A} f^{}_{\kbs B} +h.c. 
\end{split}
\end{equation}
where $ f_{\kbs u}=\frac{1}{\sqrt{N_{cell}}} \sum_{\rbs}e^{-i \kbs \cdot\rbs} f_{\rbs u} $ and $u=A,B$. As $ H_{MF} $ breaks translation by $\abss_{1}$,
it mixes partons with momenta $ \kbs $  and $ \kbs+\Qbs_1 $. Here $\Qbs_{i} = \Gbs_{i}/2$ where $\Gbs_{i}$ are the reciprocal lattice vectors. It is convinient to define a multicomponent fermion operator $F_{\kbs}$ by $F_{\kbs,un_{x}n_{y}}\equiv f_{\kbs + n_{x}\Qbs_{x} +n_{y}\Qbs_{y},u}$ where $n_{x},n_{y}\in \{0,1\}$ and restrict $\kbs$ to $ \half BZ $ which is the reduced BZ shown in Fig.~\ref{fig:HCPiFluxBandStructureContour}.

We introduce three sets of Pauli matrices. The first set $\n^{i}$ will act on the sublattice index $ u $ while the last two, $ \m^{i} $ and $\r^{i}$, act on the $n_{x}$ and $ n_{y} $ indices, respectively.  $H_{MF}$ becomes
\begin{equation}
\begin{split} 
H_{MF}&=\sum_{\kbf\in \half BZ} F_{\kbs}^{\dag} \hmf_{\kbs} F^{}_{\kbs} \\
\hmf_{\kbs}&= \n^{+}\lrRb{\m^{1}+\r^{3}e^{ik_{2}}+\m^{3}\r^{3}e^{ik_{3}}}+\hc
\end{split}
\end{equation}
where $k_{i}\equiv \kbs\cdot\abss_{i}$. The eigenvalues of $\hmf_{\kbs}$ can be easily found by first noting that $\r^{3}$ commutes with it so all the eigenvalues are at least double degenerated. Second, note that $ \hmf_{\kbs}^{2} = 3+2\lrRb{\cos(k_{1})\m^{3} +\cos(k_{2})\r^{3}\m^{1}-\sin(k_{3})\n^{3}\m^{2}\r^{3}} $.
The last three matrices anticommute with each other which implies that the eigenvalues are 
\begin{align}
E_{\kbs,s_{1}s_{2}} = s_{1}\sqrt{3+ 2 s_{2}\sqrt{\cos(k_{1})^{2}+\cos(k_{2})^{2}+\sin(k_{1}-k_{2})^{2}}} \quad, s_{1},s_{2}\in \{+1,-1\}.
\end{align}
Note that band touchings occur when the argument of the inner square root is either zero or $(3/2)^{2}$. The first option happens for $\kbs = \frac{\Qbs_{1}+\Qbs_{2}}{2}\equiv\bar{\GGbs}$ and the band touching occur at $ \frac{2\pm 1}{4}$ filling. While the second option happens for $ \kbs = \pm \frac{\Gbs_{1}-\Gbs_{2}}{12}\equiv \bar{\KKbs} $ at half filling.

\begin{figure}[t]
	\begin{subfigure}[t]{0.45\textwidth}
		\centering
		\includegraphics[height=6cm]{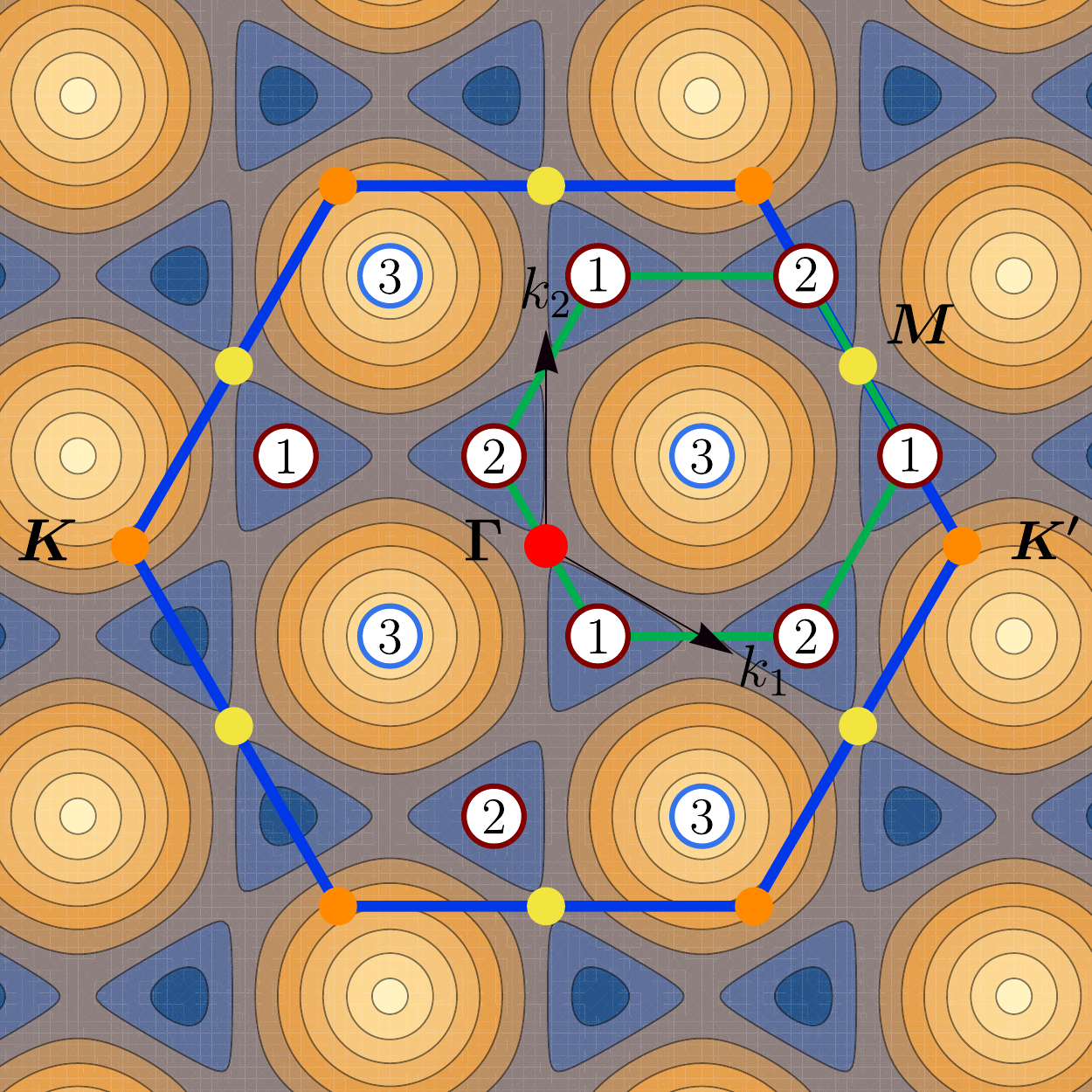}
		\caption{\label{fig:HCPiFluxBandStructureContour}}
	\end{subfigure}
	\hspace{0.5cm}
	\begin{subfigure}[t]{0.45\textwidth}
		\centering
		\includegraphics[height=6cm]{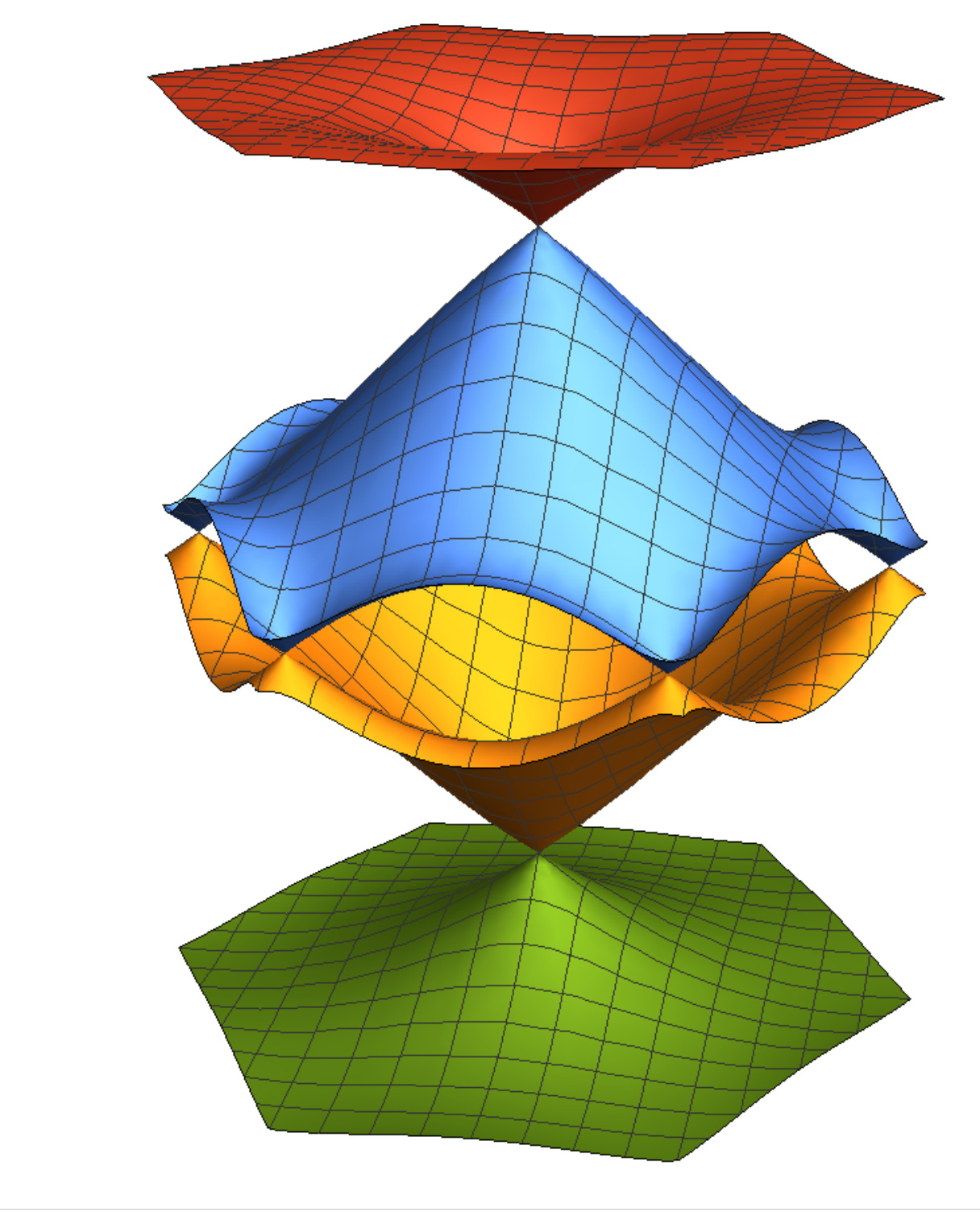}
		\caption{\label{fig:HCPiFluxBandStructure3D}}
	\end{subfigure}
	\caption{ \label{fig:HCPiFluxBandStructure}  
	(a) Contour plot for the first band above half-filling. The blue curve denote a boundary for the BZ while the green curve is the boundary of the reduced BZ. The Dirac points at half-filling correspond to the momenta labeled by $1$ and $2$ while the momentum $ 3 $ corresponds to the Dirac points at quarter filling. 
	(b) Band structure for the $\p$-flux mean field Hamiltonian centered around to the $\Gamma$ the center of the reduced BZ. }
\end{figure}

\subsubsection{ Quantum Hall mass}

In this section we explore the Hall mass for the $\p$-flux Hamiltonian, i.e. a hopping pattern that breaks time reversal and reflections but preserve their composition together with translations and rotations.

The idea here is to introduce imaginary second-nearest hoppings such that there is a $\p/2$ half flux around triangles with two nearest neighbor bonds and one second-nearest neighbor bond. This can be achieved by a term 
\begin{equation}
	 H' = \frac{i\x}{2} \sum_{\langle\langle i ,j\rangle\rangle	}(-1)^{s_{ij}} f_{i}^{\dag}f_{j}^{}
\end{equation}
with 
\begin{equation}
\begin{split}
s_{\rbs,\rbs+\abss_1}^{(u)}  & = u+1, \\
s_{\rbs,\rbs+\abss_2}^{(u)}  & = r_1+u, \\
s_{\rbs,\rbs+\abss_3}^{(u)}  & = r_1,
\end{split}
\end{equation}
$ s_{ij}+s_{ji} = 1 $	 and $\x$ real. In momentum space it reads
\begin{equation}
\begin{split}
H'  &= \x\sum_{\kbs\in \half BZ} F_{\kbs}^{\dag}\pmf_{\kbs} F_{\kbs}^{}\\ 
\pmf_{\kbs}  =& \lrSb{ \n^{3}\m^{3}\sin(k_1) - \n^{3}\m^{1}\r^{3}\sin(k_2) + \m^{2}\r^{3}\cos(k_3)}.
\end{split}
\end{equation}
After some algebra we find that the eigenvalues of $t \hmf_{\kbs} + \x \pmf_{\kbs}$ are 
\begin{equation}
\begin{split}
E_{\kbs,s_{1}s_{2}}^{\text{Hall}} &= \sqrt{3}s_{1}\sqrt{t^{2}+\x^{2}\lrRb{1-3g_{\kbs}^{2}}+ 2t s_{2}\sqrt{t^{2}g_{\kbs}^{2}  + \x^{2}\lrRb{1-4g_{\kbs}^{2}}  }} \quad, s_{1},s_{2}\in \{+1,-1\}, \\
g_{\kbs} &= \frac{1}{3}\sqrt{\cos(k_{1})^{2}+\cos(k_{2})^{2}+\sin(k_{3})^{2}}.
\end{split}
\end{equation}
Note that the gap for small $\x$ at the $\bar{\GGbs}$ point is $2 \sqrt{3}|\x|$ while at the $\bar{\Kbs}$ point is $\sqrt{3}|\x|$. 

\subsection{Calculation of the Projective Symmetry Group}\label{sec:PSGcalculation}
From Eq.~\ref{eq:HpiFluxMF}, we read the hopping amplitudes as 
\begin{alignat}{2}
	t_{\rbs,\rbs}^{AB}&= t_{\rbs,\rbs}^{BA} &&= e^{i\p r_{1}}  \notag\\
	t_{\rbs,\rbs+\abf_2}^{AB}&= t_{\rbs,\rbs-\abf_2}^{BA}&&= 1  \\
	t_{\rbs,\rbs+\abf_3}^{AB}&= t_{\rbs,\rbs-\abf_3}^{BA}&&= 1  \notag
\end{alignat}
with all other hopping terms vanishing. Next we compute the $\U(1)$ gauge rotations $e^{i\phi(\rbs)}$ associated with the physical symmetries.

\subsubsection{\texorpdfstring{$ T_{1} $}{Translation 1}}
Under translation $ T_1 $ we have 
\begin{equation}
\begin{split}
t_{\rbs,\rbs}^{AB}= e^{i\p r_{1}}& \,\,\,\, \longrightarrow \,\,\,\, t_{\rbs-\abf_1,\rbs-\abf_1}^{AB}=-e^{i\p r_{1}}  \\
t_{\rbs,\rbs+\abf_2}^{AB}= 1& \,\,\,\, \longrightarrow \,\,\,\, t_{\rbs-\abf_1,\rbs-\abf_1+\abf_2}^{AB}=1  \\
t_{\rbs,\rbs+\abf_3}^{AB}= 1& \,\,\,\, \longrightarrow \,\,\,\, t_{\rbs-\abf_1,\rbs-\abf_1+\abf_3}^{AB}=1  
\end{split}
\end{equation}
This gives
\begin{equation}
\begin{split}
\phi(\rbs,B)-\phi(\rbs,A) &= \pi  \mod 2\p\\
\phi(\rbs+\abf_2,B)-\phi(\rbs,A) &= 0 \mod 2\p\\
\phi(\rbs+\abf_3,B)-\phi(\rbs,A) &=  0 \mod 2\p
\end{split}
\end{equation}
%
%
whose solution is
\begin{equation}
\begin{split}
\phi(\rbs,A) &= \pi r_2 \\
\phi(\rbs,B) &= \pi (r_2+1) 
\end{split}.
\end{equation}
\subsubsection{\texorpdfstring{$ T_{2} $}{Translation 2}}
$ T_2 $ does not require any gauge transformation so 
\begin{align}
\phi(\rbs,u) = 0.
\end{align}

\subsubsection{\texorpdfstring{$\s$}{Reflection} }

We consider the reflection $ \sigma$ ($ \mathcal{R}_y  $ in the main text) that acts as $ (r_1,r_2) \rightarrow (r_1+r_2,-r_2) $ and $ A \leftrightarrow B $.

\begin{equation}
\begin{split}
t_{\rbs,\rbs}^{AB}= e^{i\p r_{1}}& \,\,\,\, \longrightarrow \,\,\,\, t_{\tilde{\rbs},\tilde{\rbs}}^{BA}=e^{i\pi (r_1+r_2)}  \\
t_{\rbs,\rbs+\abf_2}^{AB}= 1& \,\,\,\, \longrightarrow \,\,\,\, t_{\tilde{\rbs},\tilde{\rbs}-\abf_3}^{BA}=1  \\
t_{\rbs,\rbs+\abf_3}^{AB}= 1& \,\,\,\, \longrightarrow \,\,\,\, t_{\tilde{\rbs},\tilde{\rbs}-\abf_2}^{BA}=1  
\end{split}
\end{equation}
where $ \tilde{\rbs}= \sigma(\rbs) $.

\begin{equation}
\begin{split}
\phi(\rbs,B)-\phi(\rbs,A) &= \pi r_2  \mod 2\p\\
\phi(\rbs+\abf_2,B)-\phi(\rbs,A) &= 0 \mod 2\p\\
\phi(\rbs+\abf_3,B)-\phi(\rbs,A) &=  0 \mod 2\p
\end{split}
\end{equation}
%
whose solution is
\begin{equation}
\begin{split}
\phi(\rbs,A) &= \pi\frac{ r_2( r_2+1)}{2} \\
\phi(\rbs,B) &= \pi\frac{ r_2( r_2-1)}{2}
\end{split}
\end{equation}
%

\subsubsection{\texorpdfstring{$ C_{6} $}{Rotation} }
The $ C_6 $ rotation acts as $ (i_1,i_2,A) \rightarrow (-i_2,i_1+i_2,B) $ and $ (i_1,i_2,B) \rightarrow (1-i_2,i_1+i_2-1,A) $ and the inverse $ C_6^{-1} $ acts as $ (i_1,i_2,A) \rightarrow (i_1+i_2,1-i_1,B) $ and $ (i_1,i_2,B) \rightarrow (i_1+i_2,-i_1,A) $.

\begin{equation}
\begin{split}
t_{\rbs,\rbs}^{AB}= e^{i\p r_{1}}& \,\,\,\, \longrightarrow \,\,\,\, t_{\tilde{\rbs}+\abf_2,\tilde{\rbs}}^{BA}=1\\
t_{\rbs,\rbs+\abf_2}^{AB}= 1& \,\,\,\, \longrightarrow \,\,\,\, t_{\tilde{\rbs}+\abf_{2},\tilde{\rbs}+\abf_1}^{BA}=  t_{\tilde{\rbs}+\abf_{2},\tilde{\rbs}+\abf_2-\abf_{3}}^{BA}=1\\
t_{\rbs,\rbs+\abf_3}^{AB}= 1& \,\,\,\, \longrightarrow \,\,\,\, t_{\tilde{\rbs}+\abf_2,\tilde{\rbs}+\abf_2}^{BA}=e^{i \pi (r_1+r_2)}
\end{split}
\end{equation}
where $ \tilde{\rbs}= C_{6}^{-1}(\rbs) $.
\begin{equation}
\begin{split}
\phi(\rbs,B)-\phi(\rbs,A) &= \pi i_1 \quad  \mod 2\p\\ 
\phi(\rbs+\abf_2,B)-\phi(\rbs,A) &= 0 \quad\quad\mod 2\p\\
\phi(\rbs+\abf_3,B)-\phi(\rbs,A) &=  \pi ( i_1+ i_2) \mod 2\p
\end{split}
\end{equation}
%
%
Whose solution is
\begin{equation}
\begin{split}
\phi(\rbs,A) &= \pi\frac{ r_1( r_1+1)}{2} + \pi r_1 r_2 ,\\
\phi(\rbs,B) &= \pi\frac{ r_1( r_1-1)}{2} + \pi r_1 r_2 , \\
\phi(\rbs,u) &= \pi\frac{ r_1( r_1+1)}{2} +\p r_{1} u+ \pi r_1 r_2 .
\end{split}
\end{equation}
%
%

We can rewrite this in in the enlarged unit cell by writing $r_{i} = 2s_{i}+ e_{i}$ where $s_{i}\in \ZZ$ and $e_{i}=0,1$ so 
\begin{subequations}
		\begin{align}
			\phi(\sbs,u,e_{1},e_{2})  &= \p \lrRb{s_{1} + e_{1}+ e_{1} u + e_{1}e_{2}}
		\end{align}
\end{subequations}

\subsubsection{\texorpdfstring{$ \Tmc_{0} $}{TRS} }
As the hopping amplitudes are real numbers, there is no gauge transformation needed so we can take
\begin{align}
\phi(\rbs,u) = 0.
\end{align}

\subsubsection{PSG action in momentum space}
The previous transformation properties in momentum space read
\begin{equation}\label{eq:PSGMomentum}
\begin{split}
T_1 :F^{}_{\kbf} &\longrightarrow e^{i k_1} \mu^3 \r^1 \n^3 F^{}_{\kbf}  \\
T_2 :F^{}_{\kbf} &\longrightarrow e^{i k_2} \r^3 F^{}_{\kbf} \\
\sigma :F^{}_{\kbf} &\longrightarrow  \n^1 \exp(-i \frac{\pi}{2}  \frac{1-\rho^1}{2} \mu^3 \n^3) F^{}_{\sigma(\kbf)+\frac{\Qbf_2}{2}}  \\
C_6: F^{}_{\kbf}  &\longrightarrow \lrRb{ \n^+ + e^{-ik_{2}}\r^3 \n^- }\cdot \frac{1+\vec{\r}\cdot\vec{\m}}{2} \cdot \exp(i\pi 
\frac{1-\mu^1}{2}\frac{1-\r^{2} \n^3}{2}) F^{}_{C_6[\kbf]+\frac{\Qbf_1}{2}} \\
\mathcal{T}_0: F_{\kbs} &\longrightarrow F_{-\kbs}.
\end{split}
\end{equation}

\subsection{Low energy Hamiltonian}

We find the low-energy Hamiltonian at quarter-filling by defining  $\hat{\psi}_{\qbs} = F_{\GGbs'+\qbs}$ with $\GGbs' = \frac{\Gbs_1+\Gbs_2}{4}$ and expand for small $ \qbs $
\begin{equation}
	\begin{split}
P^{\dag}_{}\cdot\hmf_{\GGbs'+\qbs}\cdot P^{}_{} = -\sqrt{3} \id_{4\times 4} +\frac{1}{\sqrt{2}}\qbs\cdot\aabs + \Omc(\qbs^{2})
	\end{split}
\end{equation}
where $P$ is an $8$ by $ 4 $ matrix whose columns are the eigenvectors of $\hmf_{\GGbs'}$ with eigenvalue $ -\sqrt{3} $. 

In particular, we can factor $P= \Umc\cdot P_{0}$ for $\Umc$ a unitary and $P_{0}$ another projector. $ \Umc $ is given by $ \Umc = \Umc_{1}\cdot\Umc_{2}\cdot \Umc_{3}\cdot \Umc_{4}$ where 
\begin{equation}
	\begin{split}
					\Umc_{1} &= \exp(i\p\frac{\id+\m^{3}\n^{3}}{2}\frac{\id+\r^{3}}{2} )\\
		\Umc_{2} &= \exp(-i\frac{\p}{8}\m^{2})\\
		\Umc_{3} &= \exp(-i\frac{\arctan(\sqrt{2})}{2}\m^{3}\n^{3})\\
		\Umc_{4} &=i\r^{3} e^{i\frac{\p}{4}\n^{1}} e^{i\frac{\p}{3}\m^{2}}\frac{1+\vec{\r}\cdot\vec{\m}}{2} 
	\end{split}
\end{equation}
so that $ \Umc^{\dag}\cdot \hmf_{\GGbs'}\cdot\Umc^{}=-\sqrt{3}\n^{3}$. We can then simply take $P_{0}=\frac{1+\n^{3}}{2}$ which ultimately leads to $\aabs^{1}=+\r^{1}$ and $ \aabs^{2}=-\r^{3} $.

The Dirac fermions are then given by $\psi_{\qbs}= P^{\dag}\hat{\psi}_{\qbs}$. The Hamiltonian in Eq.~\ref{eq:HpiDirac} can be reproduced by identifying $(\g^{0},\g^{1},\g^{2})=(-i\r^{2},\r^{3},\r^{1})$ and introducing back the spin-orbital indices. The action on the Dirac fermions are then easily found from the explicit expression of $\psi_{\qbs}$ and Eq.~\ref{eq:PSGMomentum} evaluated at $\kbs=\GGbs'$. For example, $ \Tmc[\psi_{\qbs}] = \Tmc[P^{\dag} F_{\GGbs'+\qbs}]=P^{\top} F_{-\GGbs'-\qbs}=P^{\top} \m^{1}\r^{1}F_{\GGbs'-\qbs}$. One can check that $P^{\top} \m^{1}\r^{1}=-\m^{2}\r^{2}P_{}^{\dag}$ so that $ \Tmc[\psi_{\qbs}]=-i\r^{2}\g^{0}\psi_{-\qbs} $ which matches the result in the main text if we introduce back the spin-orbital action $\s^{2}\t^{2}$. 

\section{Wannier centers from PSG}\label{app:Wannier}
\subsection{Introduction}

In this appendix, we use the techniques developed in Refs.~\cite{Cano2018, Po17, Po18} to decompose the band structure of the $\p$-flux state on the Honeycomb lattice. We follow the approach used in Ref.~\cite{song2018spinon}. 

The idea is that in order to find the decomposition of the quantum spin Hall bands we need to calculate the eigenvalue spectrum of the rotations at high-symmetry points and then compare it with the spectra of the Wannier bands we define below. In particular, we consider three rotations: hexagon centered 6-fold rotation $C_{6h}$, hexagon centered 3-fold rotation $C_{3h}=C_{6h}^{2}$ and a bond (between sites 1 ad 6) centered 2-fold rotation $C_{2e}$ \footnote{We could have also included site centered 3-fold rotations but they do not give any new information in the present case.}.

We will compare the spinon band representation to those of Wannier insulator centered on site ($\G^v$), hexagon centers ($\G^{h}$) and bond centers $\G^{b}$, respectively. The representation of $\G^v$ is straightforward as these are our original basis. For the other two insulators, we take them to be equal-amplitude superpositions of spinons at the sites on the boundary of the plaquette or bond. As we have different choices for the phases between sites, these insulators will have internal quantum numbers. In particular, the hexagon-centered Wannier insulators come in six different flavours $\G_{\h}^{h}$, with $\h =\pm 1,\pm 2,\pm 3$, corresponding the different eigenvalues under $C_{6h}$. Similarly, the bond-centered insulators come in two flavours $\G_{\l}^{b}$ corresponding $\l =\pm 1$. Note that in order to make a meaningful comparison between the eigenvalue spectra we need to pick a gauge and use it for all the bands. 

As we will see below, the Dirac points of the spinon bands occur at the high-symmetry points. Therefore, it is \emph{fundamental} that we include the quantum spin Hall mass to split the spinon band into four Krammers sub-bands with well-defined eigenvalue spectra. The sub-bands will be denoted by $\G_{\text{PSG}}^{a}$ with $a=+2,+1,-1,-2$, where $a$ denotes the position of the band with respect to zero energy. Note that each Krammers band corresponds to two Krammers-pairs of spinons per enlarged unit cell.

\begin{figure}[b]
    \centering
\hspace{-0.25cm}    
    \begin{subfigure}[t]{0.25\textwidth}
        \centering
        \includegraphics[height=6cm]{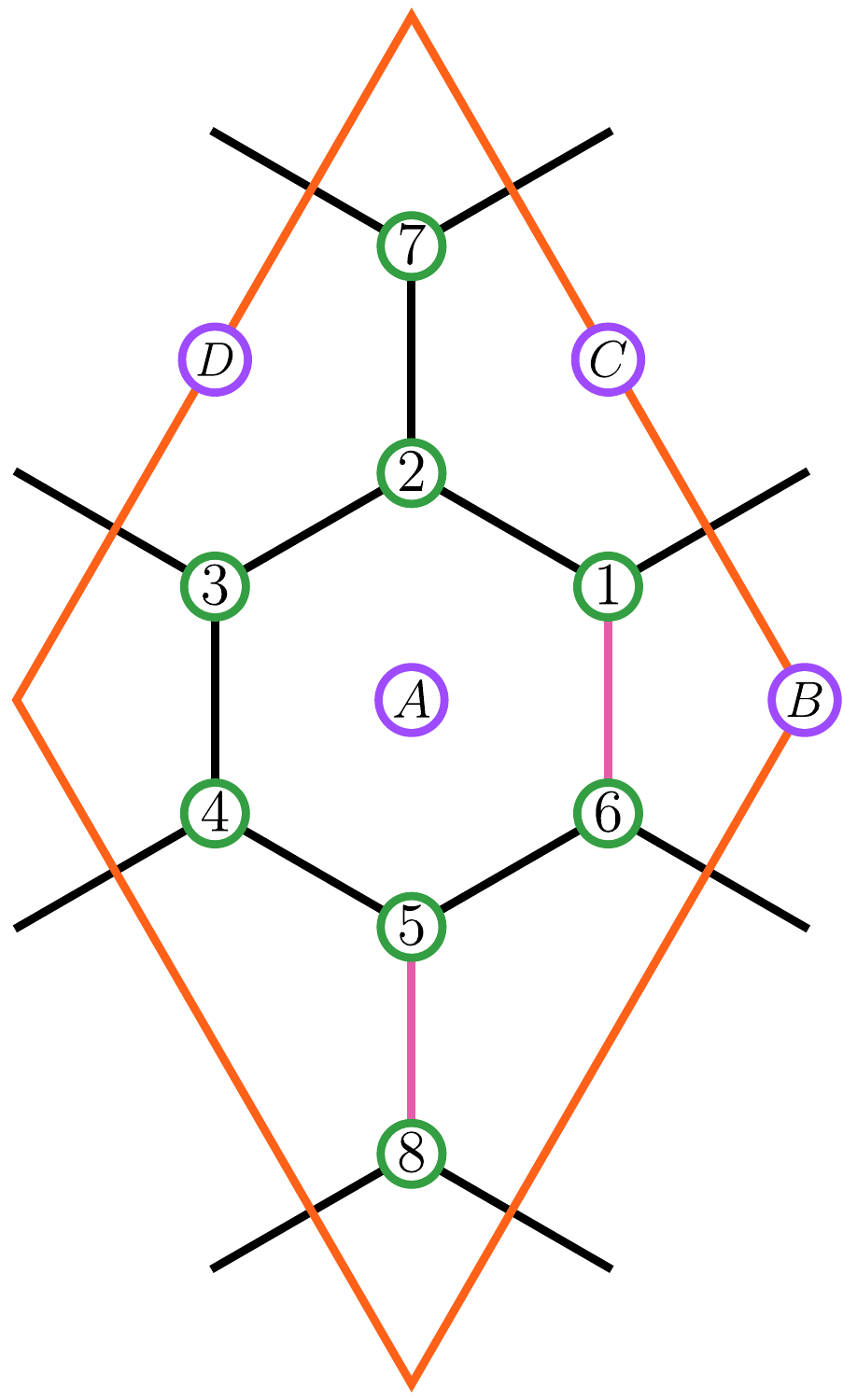}
	\caption{\label{fig:HCUnitCell}}
    \end{subfigure}
	\begin{subfigure}[t]{0.35\textwidth}
		\centering
		\includegraphics[height=6cm]{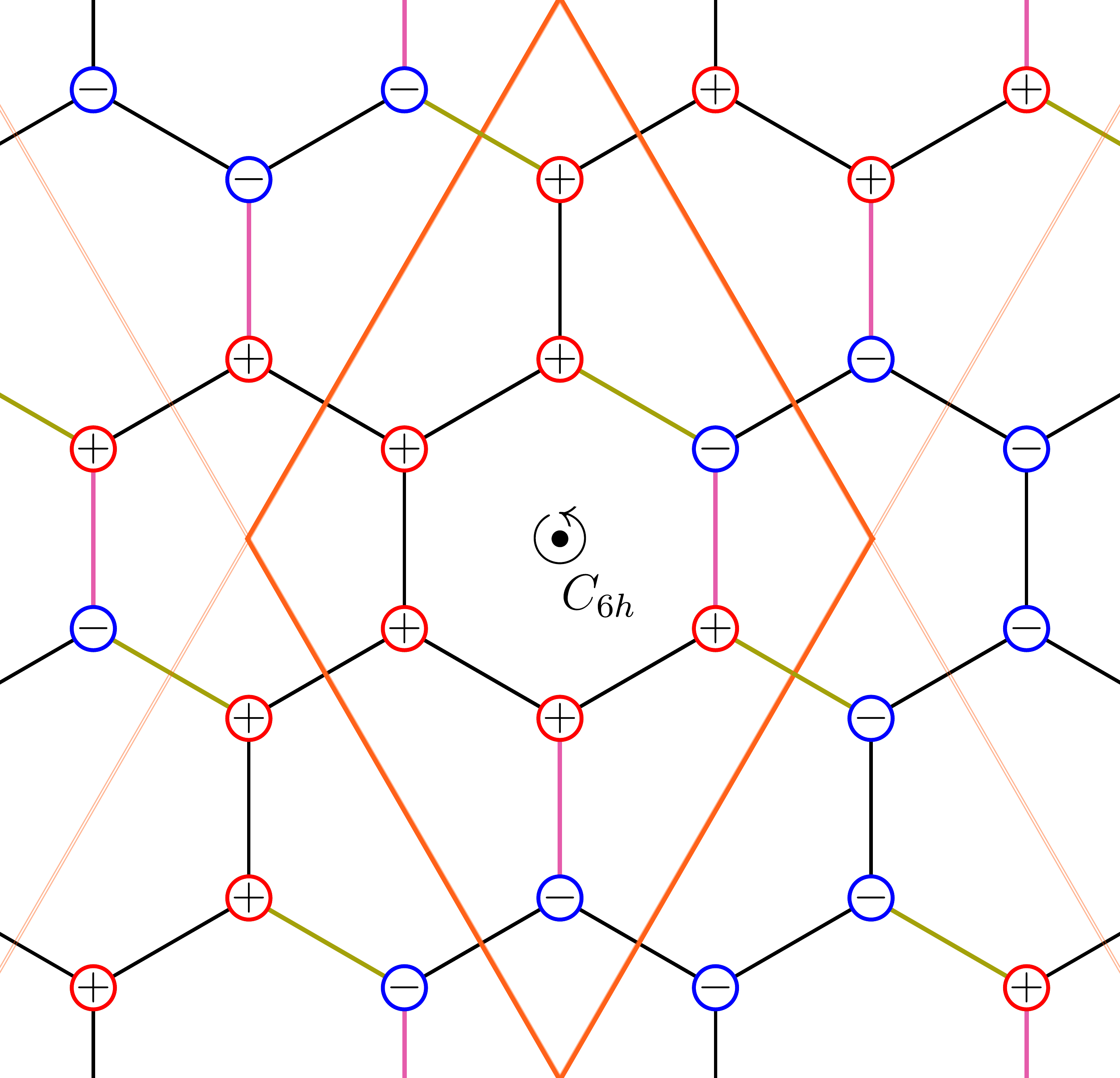}
		
		\caption{\label{fig:C6WannierLabeled}}	
	\end{subfigure}
\hspace{0.5cm}
	\begin{subfigure}[t]{0.35\textwidth}
		\centering
		\includegraphics[height=6cm]{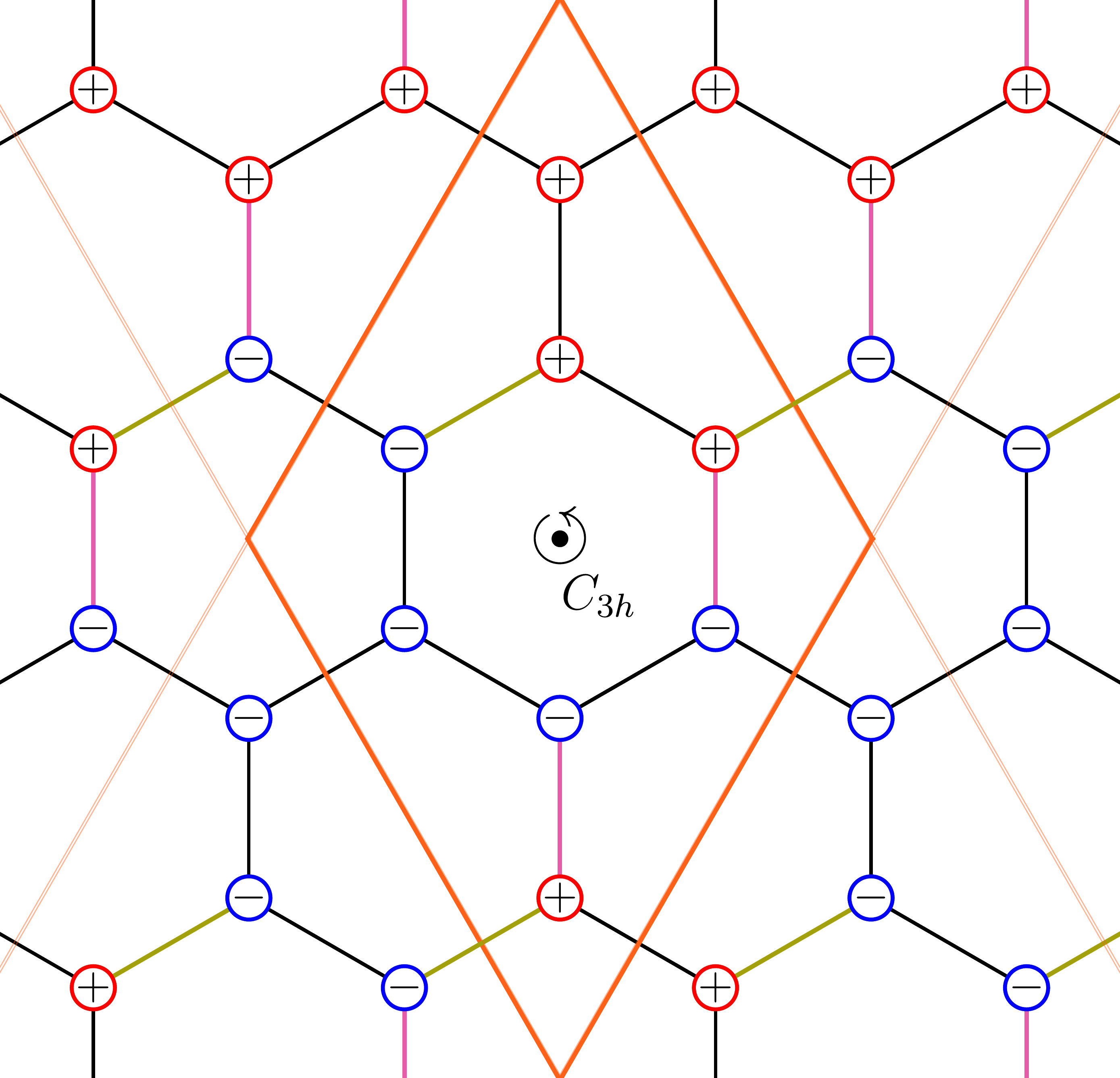}
		\caption{\label{fig:C3WannierLabeled}}	
	\end{subfigure}
	\caption{ \blue{(a)} Enlarged unit cell used in the band topology calculation. Pink bonds correspond to negative hoppings on those bonds and numbers correspond to the labeling of sites in the enlarged unit cell. \blue{(b) \& (c)} Choice of gauge transformation for $C_{6h}$ and $C_{3h}$: the plus or minus signs correspond to a gauge transformation $f_{i} \rightarrow \pm f_{i}$ for $C_{3h}$ and $f_{i} \rightarrow \pm i f_{i}$ for $C_{6h}$. }
\end{figure}

\subsection{Band topology calculation}

We shall denote the Wannier insulator on the unit cell labeled by $\rbs$ on the sublattice $a=1,\dots,8$ by $\ket{\rbs,a}$. Using the gauge choice for $C_{6h}$ depicted in Fig.~\ref{fig:C6WannierLabeled} we find that 
$\ket{\rbs,6} = (-1)^{r_1+1}i\ket{R_{\p/3}\rbs,1}$, $\ket{\rbs,7} = (-1)^{r_1}i\ket{R_{\p/3}\rbs+\abss_3,8}$, $\ket{\rbs,8} = (-1)^{r_1+1}i\ket{R_{\p/3}\rbs-\abss_3,7}$ and $C_{6h}\ket{\rbs,a} = (-1)^{r_1}i\ket{R_{\p/3}\rbs,a+1}$, where $R_{\p/3}$ is the $\p/3$ clockwise rotation matrix.
Going to momentum space, we find that $C_{6h}$ sends $\kbs \rightarrow \kbs' =R_{\p/3}\kbs +\Qbs_1$ and 
\begin{equation}
    \begin{split}
        C_{6h}\ket{\kbs,a} & = +i\ket{\kbs',a+1} \quad , a=1,\dots,5 \\
        C_{6h}\ket{\kbs,6} & = -i\ket{\kbs',1} \\
        C_{6h}\ket{\kbs,7} & = +ip_3^{*}\ket{\kbs',8} \\
        C_{6h}\ket{\kbs,8} & = -ip_3^{}\ket{\kbs',7} 
    \end{split}
\end{equation}
where $p_3 = \exp(i\kbs' \cdot \abss_3)$. The invariant momentum is simply $\kbs = \Qbs_1+\Qbs_2$ and thus $p_3 =1$. The eigenvalues are $[1,\W,\W^2,,\W^3,\W^4,\W^5] \oplus [+1,-1]$, where  $\W= e^{\frac{\p i}{3}}$.

For $C_{3h}$ we find 
\begin{equation}
    \begin{split}
        C_{3h}\ket{\kbs,a} & = -\ket{\kbs',a+2} \quad , a=1,\dots,4 \\
        C_{3h}\ket{\kbs,a} & = +\ket{\kbs',a-4}  \quad , a=5,6 \\
        C_{3h}\ket{\kbs,7} & = -p_2^{}\ket{\kbs',7} \\
        C_{3h}\ket{\kbs,8} & = -p_2^{*}\ket{\kbs',8} 
    \end{split}
\end{equation}
where $\kbs' =R_{2\p/3}\kbs+\Qbs_2 $ and $p_2=\exp(i\kbs'\cdot \abss_2)$. The invariant momenta are $\kbs = \Qbs_1 +\Qbs_3 :=\Mbs$ and $\kbs = \frac{\Gbs_1-\Gbs_2}{6} =\Kbs =- \Kbs'$. The eigenvalues in this case are $[1,\w,\w^2]\oplus[1,\w,\w^2]\oplus p_3 \oplus p_3^{*}$ \footnote{Note that for the invariant momenta $-p_2 =p_3^{*} $ with $p_3$ as previously defined.}.
As for $C_{2e}$ there is no gauge transformation needed, it is easy to see that the eigenvalues are four pairs of $[+1,-1]$.

We next construct the hexagon centered Wannier insulators. We start by taking the eigenvectors of $C_{6h}$ with eigenvalue $\W^{2\h}$ with weight on hexagon A, denoted by $\ket{\text{hex},A}$ (omitting the position dependence). The other three states are taken by translating the previous state by half lattice vectors. We shall denote the representation by $\G^{h}_{\h}$. From the construction we have $C_{6h}\ket{\text{hex},A} = \W^{2\n-1}\ket{\text{hex},A}$. Next, one can check that  $C_{6h}^3=(-1)^{\h}$ when acting on  $\ket{\text{hex},B/C/D}$ and furthermore with the appropiate phase conventions, the matrix becomes 
\def\arraystretch{1}
\begin{equation}
    (-1)^{\h}
\left(\begin{array}{ccc} 0&1&0\\ 0&0&1\\1&0&0\end{array}\right )
\end{equation}
\def\arraystretch{1.5}
whose eigenvalues are $(-1)^{\h}[1,\w,\w^2]=[\W^{\h},\W^{\h+2},\W^{\h+4}]$. Similarly, we find $C_{3h}\ket{\text{hex}_\h,A} = \w^\h\ket{\text{hex}_\h,A}$ and $C_{3h}^{3}=1$ on the other hexagons. Thus the eigenvalues are $\w^{\h}\oplus[1,\w,\w^2]$. As none of the hexagons are invariant under $C_{2e}$, the eigenvalues are $[+1,-1]\oplus [+1,-1]$.

Similarly the bond Wannier insulators come in two flavours $\G^b_{\l}$ with $\l=+1,-1$ which correspond to the equal-amplitude symmetric or antisymmetric superposition of the onsite insulators at the boundary of the bonds. By a similar calculation as for the previous bands we can fill Tab.~\ref{tab:WannierTable}.

\begin{table}[t]
	\centering
	\begin{tabular}{ |c||c|c|c||c|} 
		\hline
		$ $ & $ \Gamma^{v}_{} $ & $ \Gamma^{e}_{\lambda} $ & $ \Gamma^{h}_{\eta} $ & \text{HSM}\\[5pt] \hline\hline
		$ C_{6h} $ & $ [\text{Doublet}]\oplus[\text{Sextet}] $ & $ 2[\text{Sextet}] $ & $ \Omega^\eta \oplus [\text{Triplet}_{\eta}]$ & $\Mbs$	\\[5pt] \hline
		$ C_{3h} $ & $ p_3^{}\oplus p_3^{*} \oplus 2[\text{Triplet}_{0}] $ & $ 4[\text{Triplet}_{0}] $ & $ \omega^\eta \oplus [\text{Triplet}_{0}]$ & $\Mbs,\Kbs,\Kbs'$	\\[5pt] \hline
		$ C_{2e} $ & $ 4[\text{Doublet}]  $ & $ \lambda(1\oplus p^{}_{1} \oplus p^{}_{2} \oplus p^{}_{3}) \oplus 4[\text{Doublet}]$ & $ 2[\text{Doublet}]$& $\Mbs,\GGbs$	\\[5pt] \hline
		\hline
	\end{tabular}
	\caption{\label{tab:WannierTable} Space group representation of rotations of the Wannier insulators of the $\p$-flux state on the honeycomb lattice at high symmetry points. The multiplets are denoted by $ [\text{Sextet}]=[1,\Omega^1,\Omega^2,\Omega^3,\Omega^4,\Omega^5] $,  $ [\text{Triplet}_\eta]=(-)^\eta[1,\omega^1,\omega^2] $ and $ [\text{Doublet}]=[+1,-1] $. $p_i=e^{i\kbs\cdot \abss_i}$ is the phase factor obtained under $T_{i=1,2,3}$ translation. The high-symmetry momenta (HSM) for each group operations are recorded in the last column. They are $\Mbs=\frac{\Gbs_1+\Gbs_2}{2}$, $\Kbs=\frac{\Gbs_1-\Gbs_2}{6}$ and $\Kbs'=-\Kbs$, where $\Gbs_{1,2}$ are the reciprocal lattice vectors. }
\end{table}

We find that the spectrum of the PSG bands by straightforward diagonalization of the mean-field Hamiltonian at high-symmetry points and then finding the eigenvalues of the rotations acting on them. The result is shown in Tab.~\ref{tab:WannierTablePSG}. As previously mentioned, we need to introduce the quantum spin Hall mass to open a gap and be able to decompose the resulting bands in terms of the Wannier insulators. We find that the gap splits the eigenvalues equally between the bands below and above the previous Dirac points. This can be understood by remaining that the $C_{6h}$ rotation can be written (in the appropriate basis) as $\exp(-\frac{\p}{6}\g^0)$ times some matrix $R_{C_{6h}}$ acting on valley indices. As the QSH mass gives opposite $\g_0$ eigenvalues to spin up and spin down, the eigenvalues come in pairs.
\begin{table}[b]
	\centering
	\begin{tabular}{ |c||c|c|c| } 
		\hline
		$ $ & $ \G_{PSG(\Mbs)}  $ & $ \G_{PSG(\GGbs)} $ & $ \G_{PSG(\Kbs)}  $	\\[5pt] \hline\hline
		$ C_{6h} $ & $ \{ 1,\W^{1},\W^{2},\W^3 \}  \oplus \{1, \W^{3},\Omega^{4},\Omega^{5} \}  $ & $ -$ & $ -$\\[5pt] \hline
		$ C_{3h} $ & $ \{ 1 \oplus [\text{Triplet}_0]\} \oplus \{1 \oplus [\text{Triplet}_0]\} $ & $ - $ & $ \{[\w,\w^{2}]\} \oplus \{1 \oplus[\text{Triplet}_0]\} \oplus \{[\w,\w^{2}]\} $\\[5pt] \hline
		$ C_{2e} $ & $ \{2[\text{Doublet}]\}\oplus \{2[\text{Doublet}]\} $ & $ \bigoplus_{n=1}^{4}\{[\text{Doublet}]\} $ & $ -$\\[5pt] \hline
		\hline
	\end{tabular}
	\caption{\label{tab:WannierTablePSG} Space group representation of rotations of the spinon bands of the $\p$-flux state on the honeycomb lattice at high symmetry points. The eigenvalues are ordered from higher to lower bands. The multiplets are denoted by $ [\text{Sextet}]=[1,\Omega^1,\Omega^2,\Omega^3,\Omega^4,\Omega^5] $,  $ [\text{Triplet}_\eta]=(-)^\eta[1,\omega^1,\omega^2] $ and $ [\text{Doublet}]=[+1,-1] $. $p_i=e^{i\kbs\cdot \abss_i}$ is the phase factor obtained under $T_{i=1,2,3}$ translation. The high-symmetry momenta (HSM) for each group operations are recorded in the last column. They are $\Mbs=\frac{\Gbs_1+\Gbs_2}{2}$, $\Kbs=\frac{\Gbs_1-\Gbs_2}{6}$ and $\Kbs'=-\Kbs$, where $\Gbs_{1,2}$ are the reciprocal lattice vectors. }
\end{table}

A decomposition for the $PSG$ bands is 
\begin{equation}\label{eq:WannierPiHC}
\begin{split}
2 \Gamma^{+2}_{PSG} &= \Gamma^v+\left( \Gamma^v- \Gamma^h_{\eta=0}-\Gamma^h_{\eta=3} \right)+ \left(\Gamma^{h}_{\eta=1}+ \Gamma^{h}_{\eta=2} - \Gamma^{h}_{\eta=4}- \Gamma^{h}_{\eta=5}\right ) \\
2 \Gamma^{+1}_{PSG} &= \Gamma^v-\left( \Gamma^v- \Gamma^h_{\eta=0}-\Gamma^h_{\eta=3} \right)+ \left(\Gamma^{h}_{\eta=1}+ \Gamma^{h}_{\eta=2} - \Gamma^{h}_{\eta=4}- \Gamma^{h}_{\eta=5}\right ) \\
2 \Gamma^{-1}_{PSG} &= \Gamma^v-\left( \Gamma^v- \Gamma^h_{\eta=0}-\Gamma^h_{\eta=3} \right)- \left(\Gamma^{h}_{\eta=1}+ \Gamma^{h}_{\eta=2} - \Gamma^{h}_{\eta=4}- \Gamma^{h}_{\eta=5}\right ) \\
2 \Gamma^{-2}_{PSG}&= \Gamma^v+\left( \Gamma^v- \Gamma^h_{\eta=0}-\Gamma^h_{\eta=3} \right)- \left(\Gamma^{h}_{\eta=1}+ \Gamma^{h}_{\eta=2} - \Gamma^{h}_{\eta=4}- \Gamma^{h}_{\eta=5}\right ) \\
\end{split}
\end{equation}
This decomposition is unique modulo addition of multiples of $ \G^h_{\h=0}+\G^h_{\h=1}+\G^h_{\h=2}+\G^h_{\h=3}+\G^h_{\h=4}+\G^h_{\h=5}-\G^{b}_{\l=+1}-\G^{b}_{\l=-1}$ which is irrelevant for the monopole quantum numbers. Notice that under time reversal $ \eta \rightarrow3-\eta $ because as $ C_{6h} $ involves a factor of $ i $, time reversal involves a minus sign ($ -1=\Omega^3 $) on top of complex conjugation of eigenvalues.

\section{Numerical determination of the Berry phase}\label{app:berryNumerics}

\subsection{Monopole configuration on zero flux background}\label{app:berryNumericsHC}
We first find the gauge to introduce the $ 2\pi $ flux on the (0-flux) HC lattice and the gauge transformation needed to implement the lattice symmetries (in the $L_{1}, L_{2}\rightarrow \infty$). The Hamiltonian as usual is on a $ L_1  $ by $ L_2 $ lattice:
\begin{equation}
H=\sum_{\rbf \mu} e^{i A_{\mu}(\rbf) }f_{\rbf A}^{\dag}f_{\rbf+b_\mu B}^{} + h.c.
\end{equation}
where $ b_{1}= \boldsymbol{0} $, $ b_{2}= \abss_2 $ and $ b_{3}= \abss_{3} $. We introduce the $ 2\pi $ flux by choosing 
\begin{equation}\label{eq:defAmonopole}
\begin{split}
A_1(\rbf) &= \frac{2\pi}{L_1 L_2} (r_1 + s) \\
A_2(\rbf) &= 0 \\
A_3(\rbf) &= \frac{2\pi}{ L_2}( C-r_2  )\delta_{r_1, 0} \\
\end{split}
\end{equation}
we have to restrict $ 0 \leq r_{i=1,2} <L $. $s$ and $C$ are constants that will be constrained in the following.

Notice that under $ T_2 $ there is no gauge transformation needed. For $ T_1 $, $ \delta A_{\mu}(\rbf)=A_{\mu}(X^{-1}[\rbf])-A_{\mu}(\rbf)$ with $ X=T_1 $ so that
\begin{equation}
\begin{split}
\delta A_1(\rbf)&= A_1(\rbf-\abf_1)-A_1(\rbf)=-\frac{2\pi}{L_1 L_2}\left(1- \delta_{r_1,0} L_1 \right), \\
\delta A_2(\rbf)&= 0 ,\\
\delta A_3(\rbf)&= \frac{2\pi( r_2-C)}{L_2}(\delta_{r_1,0}-\delta_{r_1,1}).
\end{split}
\end{equation}
The translated Hamiltonian is related to the initial one by threading Wilson Loops (WLs) along the $ \abss_2 $ direction with total flux $ e^{\frac{2\pi i}{L_1}} $ in addition to some boundary terms in the $\abss_{1}$. In order to remove this WL, a gauge transformation$ f^{}_{u \rbf} \rightarrow  e^{i\phi_u(\rbf)} f^{}_{u \rbf}$ with 
\begin{equation}
\begin{split}
\phi_A(\rbf) &= \frac{2\pi (r_2-C)}{L_2}\delta_{r_1,0} \\
\phi_B(\rbf) &= \frac{2\pi (r_2-1-C)}{L_2}\delta_{r_1,0}
\end{split}
\end{equation}
is used. The net change in the phases is given by
\begin{equation}
\begin{split}
\delta A_1(\rbf)+ \phi_B(\rbf)-\phi_A(\rbf) &= -\frac{2 \pi}{L_1 L_2} \\
\delta A_2(\rbf)+ \phi_B(\rbf+\abf_2)-\phi_A(\rbf) &= 0 \\
\delta A_3(\rbf)+ \phi_B(\rbf+\abf_3)-\phi_A(\rbf) &= 0
\end{split}
\end{equation}
Note that the remaining WLs in the $2$ direction have a flux $\exp(\pm\frac{2\p i}{L_{1}})$ which goes to zero in the $L_1\rightarrow \infty$ limit. This means that in later limit the $T_1$ (projective) symmetry is recovered. 

Next, we take $ L_1=L_2=L $ in order to be able to do rotations. Consider
\begin{equation}
\begin{split}
C_6[H]&= \sum_{\rbs,\m} e^{i A_{\mu}(\rbf)}f_{B C_6[\rbf]}^{\dag}  f_{A C_6[\rbf+b_\mu-\abf_2]}^{} +h.c. 
\end{split}
\end{equation}
where $ C_6[\rbf] $ is clockwise rotation by $ \pi/3 $ around the plaquette to the right of $\rbs=(0,0)$. Let $ \tilde{b}_\mu =C_{6}[\abf_2-b_\mu]  $ ( $ \tilde{b}_1=b_3 $, $ \tilde{b}_2=b_1 $ and $ \tilde{b}_3=b_2 $). Therefore
\begin{equation}
\begin{split}
C_6[H]&= \sum_{\rbs,\m} e^{i A_{\mu}(C_6^{-1}[\rbf])}f_{B \rbf}^{\dag}  f_{A \rbf-\tilde{b}_\mu}^{} +h.c.  \\
&= \sum_{\rbs,\m} e^{-i A_{\mu}(C_6^{-1}[\rbf+\tilde{b}_\mu])}f_{A \rbf}^{\dag}  f_{B \rbf+\tilde{b}_\mu}^{} +h.c.  
\end{split}
\end{equation}

The change in the gauge field is 
\begin{equation}
\begin{split}
\delta A_1(\rbf) &= -\frac{2\pi}{L^2} (r_1 + s) \\
\delta A_2(\rbf) &= -\frac{2\pi}{L} (r_1+C) \delta_{r_1+r_2 ,L-1}\\
\delta A_3(\rbf) &= -\frac{2\pi}{ L^2}(r_1+ \tilde{r}_2+s  ) +\frac{2\pi}{ L}( r_2 - C  )\delta_{r_1, 0}\\
\end{split}
\end{equation}
where $\tilde{r}_{2}=\textit{mod}(r_{1}+r_{2},L)-r_{1}$.

The field $ \delta A_{\mu }$ varies slowly in the bulk but has jumps along the lines $ r_1=0 $ and $ r_1+r_2=L-1 $. In order to eliminate them, we perform a gauge transformation that imposes $ \phi_B(r+b_\mu)-\phi_A(r)+\delta A_{\mu} (\rbf)=0 $ away from the jumps. A solution is given by 
\begin{equation}
\begin{split}
\phi_A(\rbf) &= -\frac{\pi }{L^2}({r_1(2\tilde{r}_2+r_1+1)}+s(2(r_1+\tilde{r}_2)+1)  ), \\
\phi_B(\rbf) &= -\frac{\pi }{L^2}({r_1(2\tilde{r}_2+r_1-1)}+s(2(r_1+\tilde{r}_2)-1)  ). 
\end{split}
\end{equation}
The boundary terms become
\begin{equation}\begin{split}
\phi_B(L-1,x+1-L)-\phi_A(0,x)+\delta A_{3}(x,0) = \frac{\pi  (L-1-2C)}{L} \\
\phi_B(L-1,x+1-L)-\phi_A(L+1-x,x)+\delta A_{2}(L+1-x,x) = \frac{2 \pi  (s-C)}{L}
\end{split}
\end{equation}
Setting $ C=\frac{L-1}{2} $ and $ s=-\frac{L+1}{2} $ kills both terms. The $C_6$ being discussed is hence preserved. The other rotations can be obtained by combination with translations which we already know how to implement.

A problem we face now is how to decide the overall phase of the Gauge transformation in order to fix the Berry phase. To see that this is an issue, consider a uniform gauge transformation, say $ e^{i \q/L^2} $, with $ N=n L^2 $ occupied fermion modes. The expectation value of the symmetry operation gets an extra $ e^{i n \q} $ phase. 

In order to fix the phase, we impose that the total phase $\exp(i \sum_{\rbf u}\phi_u(\rbf))$ is $ 1 $. This condition comes from the requirement that our state is gauge invariant. 
%
Consider the sum
\begin{equation}
\Theta=\sum_{r_1,r_2=0 }^{L-1} \sum_{u=A,B} \phi_u(\rbf) =2\pi \frac{(2L+5)(L-1)}{6}.
\end{equation}
and then redefine $ \phi \rightarrow \phi-\frac{\Theta}{2L^2} $. The gauge transformation now reads
\begin{equation}
\begin{split}
\phi_A(\rbf) &= -\frac{\pi }{L^2}({r_1(2\tilde{r}_2+r_1+1)}-\frac{L+1}{2}(2(r_1+\tilde{r}_2)+1)  ) -2\pi \frac{(2L+5)(L-1)}{12 L^2},\\
\phi_B(\rbf) &= -\frac{\pi }{L^2}({r_1(2\tilde{r}_2+r_1-1)}-\frac{L+1}{2}(2(r_1+\tilde{r}_2)-1)  ) -2\pi \frac{(2L+5)(L-1)}{12 L^2}.
\end{split}
\end{equation}
We numerically checked that this gauge transformation gives us the same $\p$ Berry phase as found in  Ref.~\cite{song2018spinon}.

\subsection{Monopole configuration on \texorpdfstring{$\p$}{pi} flux background}

The monopole configuration of the $\pi$ flux state is obtained by taking
\begin{equation}
H=\sum_{\rbf \mu} t_{\rbf\sbf}^{AB}e^{i A_{\mu}(\rbf) }f_{\rbf A}^{\dag}f_{\rbf+b_\mu B}^{} + h.c.
\end{equation}
where $A$ is the same as before (Eq.~\ref{eq:defAmonopole}) and $ t_{\rbf\sbf}^{AB}$ are the same hopping parameters (Eq.~\ref{eq:HpiFluxMF}) used in the main text. It is easy to see that the symmetries can be implemented by simply combining the gauge transformation of the previous section with the gauge transformation from Sec.~\ref{sec:PSGcalculation}. We numerically checked that these gauge transformations give consistent phases with the  Berry phases obtained in the main text using the band calculation.

\section{Mathematical backgrounds for anomaly calculation}
\label{app:math}
\subsection{General constructions}

In this section, we review math constructions used with cohomology groups. 

First, recall that the cup product gives a graded bilinear product between $(\psi_{i},\phi_{j})\in H^{i}(G) \times H^{j}(G)$ to $H^{i+j}(G)$ denoted as $\psi_{i} \cup \phi_{j}$. Here $G$ is a Abelian group that in our cases is given simply $\ZZ$ or $\ZZ_{n}$ (integers modulo $n$). Note that given a group homomorphism $\rho: G \rightarrow G'$ which abusing notation we denote also by $\r$, at the cohomology level. For example, if we take $G=\ZZ$ and $G'=\ZZ_{n}$ then the induced map is modulo $n$ reduction. Let us denote the group and cohomology maps by  $ \r_{n} $. Similarly, we can take $G=G'=\ZZ\,\text{or}\,\ZZ_{n}$ and consider the map corresponding to multiplication by some element in $G$. At the level of cohomology this operation is also multplication by said element.

Note that we can combine the two previous operations at the group level to obtain the short exact sequence (SEC)
\[0 \longrightarrow	\ZZ \overset{n\cdot}{\longrightarrow}\ZZ \overset{\rho_{n}}{\longrightarrow}  \ZZ_{n} \longrightarrow 0 . \]
From this SEC, we can obtain a long exact sequence at the cohomology level that allow as to define the Bockstein homomorphism $ \b_{n} $
\[  \dots {\longrightarrow} H^{i}(\ZZ) \overset{n\cdot}{\longrightarrow} H^{i}(\ZZ)  \overset{\rho_{n}}{\longrightarrow}  H^{i}(\ZZ_{n}) \overset{\b_{n}}{\longrightarrow} H^{i+1}(\ZZ)  \longrightarrow \dots \quad. \]
By looking at the long exact sequence, we see that $\b_{n}$ can be though of as the obstruction to lifting a $\ZZ_{n}$ cochain to a $\ZZ$ one, which is why sometimes it is denoted by $\b_{n}(x) = \frac{1}{n}\dd{x}$.

Following Ref.~\cite{kapustin2013topological}, we define the Pontraygin square operation $\Pmc: H^{i}(G) \rightarrow H^{2i}(\G(G))$\footnote{Note that there is a different $\Pmc$ for each $G$ but for ease of reading we omit an explicit dependence on $G$.}, where $\G(G)$ is the universal quadratic group $G$. $\Pmc$ is universal in the sense is the only map $ H^{i}(G) \rightarrow H^{2i}(\G(G)) $ such that $\Pmc(x+y)-\Pmc(x)-\Pmc(y)  = 2 x \cup y$. Note that $\G(\ZZ_{2n})=\ZZ_{4n}$ and  $\G(\ZZ_{2n-1})=\ZZ_{2n-1}$. In particular, for $G=\ZZ_{2n}$ we define $\Pmc(x) = x\cup x + x \cup_{1} \dd{x}$ and for $G=\ZZ_{2n-1}$ we can simply take $\Pmc(x)=x\cup x$. Here $\cup_{1}$ is a generalized bilinear product of degree -1 defined in the appendix of Ref.~\cite{kapustin2013topological}.

From now on, we will focus on the case where $G=\ZZ_{2n}$ as our anomaly calculations only require these cases. 

Note that the property $\Pmc(x+y)-\Pmc(x)-\Pmc(y)  = 2 x \cup y$ implies that $\Pmc(\sum_{i=1}^{N}x_{i})= \sum_{i=1}^{N}\Pmc(x_{i}) + \sum_{1\le i<j \le N} 2 x_{i} \cup x_{j}$ which we will frequently use. By the explicit formula for $\Pmc$ for $G=\ZZ_{2n}$, if $ x = \r_{2n} \hat{x}$  , for some integer class $\hat{x}$, then $\Pmc(x) = \r_{4n} \lrRb{\hat{x} \cup \hat{x}}$.

We actually need a bit more of relationships between the previously defined operations which can be found in Ref.~\cite{Whitehead1949OnSC4dPolyhedra}. We accept that $\ZZ_0 = \ZZ$. First of all, let us define $\m_{p,q}:H^{*}(\ZZ_{q}) \rightarrow H^{*}(\ZZ_{p})$ by sending $x$ to the cohomology class of $\frac{p}{(p,q)} x'$  where $x'$ is a cochain in cohomology class of $x$. $(p,q)$ is largest common divisor of $p$ and $q$. From the definition, we have $\m_{p,q}(a\cdot x) = (a\!\!\mod p)\cdot\m_{p,q}(x)$ for $a\in \ZZ_{q}$. Note that $\m_{p,0} = \r_{p}$ and , if $r \neq 0$,  $ \m_{p,r p}$ is reduction modulo  $r$ and $ \m_{r q, q}$ is multiplication by $r$. For ease of reading, we denote the former operations by $\r_{p}$ as in the $r=0$ case. The following relations can be found in Sections 2 and 4 of Ref.~\cite{Whitehead1949OnSC4dPolyhedra} \footnote{Note also that the $\D_{q} $ defined in Ref.~\cite{Whitehead1949OnSC4dPolyhedra} is our $ \b_{q} $ .}  $ (r,s \in \ZZ-\{0\}) $:
\begin{subequations}
		\begin{align}
			\b_{p}\circ \m_{p,q} & =   \frac{q}{(p,q)}\b_{q} \\
			\Pmc \circ \m_{2r,m}& =  \begin{cases}
				\frac{2r}{(2r,m)} \m_{4r,2m} \,\circ \,\Pmc \quad &, m\,\,\text{even} \vspace{0.25cm}\\	
				\frac{r}{(r,m)} \,\m_{4r,\,\,m} \,\circ \, \Pmc \quad &, m\,\,\text{odd}\\
			\end{cases} \\
			\Pmc(s \cdot x)&= \begin{cases}
			s^{\,\,}\cdot \Pmc(x) \quad &, x \in H^{2i-1}(\ZZ_{2r}) \vspace{0.25cm}\\	
			s^{2}\cdot \Pmc(x) \quad &, x \in H^{\,\,2i\,\,}(\ZZ_{2r}) \\	
			\end{cases}
		\end{align}
\end{subequations}
In particular, we can set $m = 2n $ and $r=2mn$ in the second relation to obtain $ \Pmc\circ\m_{2mn,2n} = m \m_{4mn,4n}\circ \Pmc $ which can be sloppily written as $\Pmc(m\cdot x) = m^{2}\cdot \Pmc(x) \mod 4mn$ for $x\in H^{*}(\ZZ_{2n})$. {For our manipulations, there formulas simply say that $\Pmc$ behaves as a square and we are allowed to move integers outside of the argument.}

Eq.~3.3 of Ref.~\cite{Whitehead1949OnSC4dPolyhedra} tells us that for $x \in H^{*}(\ZZ_{p})$, $y \in H^{*}(\ZZ_{q})$ and if $d = (p,q)$ and $r = \mmrm{lcd}(p,q)=\frac{pq}{d}$, then $\m_{r,p}(x) \cup \m_{r,q}(y) = \m_{r,d}\lrRb{x \cup y}$ where $x\cup y =\m_{d,p}(x)\cup \m_{d,q}(y)$. We can combine this expression with the linearity of $\m$ to obtain $   \lrRb{a \cdot x} \cup \lrRb{ b\cdot y}  =  ab \cdot \lrRb{ x \cup y}  $ where $a p, b q |r $.

Now we introduce the Steenrod squares $ \Sq^{i}: H^{*}(\ZZ_{2}) \rightarrow H^{*+i}(\ZZ_{2}) $ that can be axiomatically defined, see  \cite{milnor1974characteristic}. These operations satisfy $\Sq^{i}(x) = x \cup x$ if $x\in H^{i}(\ZZ_{2})$ and are $\ZZ_{2}$ linear. $\Sq^{0}$ is the identity map and $\Sq^{1}(x\cup y) = x \cup \Sq^{1}(y) + \Sq^{1}(x)\cup y$. There is an interesting connection between $\Sq^{1}$ and $\b_{2}$ which can be written as $\Sq^{1} = \r_{2}\circ \b_{2} $.

Additionally, on a compact $d$ dimensional manifold $X_{d}$, there exist the Wu classes $\n_{i} \in H^{i}(X_{d},\ZZ_{2})$ that for every $x \in H^{d-j}(X_{d},\ZZ_{2})$ satisfy $\int_{X_{d} } \n_{j} \cup x  = \int_{X_{d}} \Sq^{j}(x) \mod 2$. In particular, for 4-dimensional closed manifolds $\n_{1} = w_{1}^{TM}$, $ \n_{2}=w_{2}^{TM} + w_{1}^{TM} \cup w_{1}^{TM} $, where $w_{i}^{TM}$ are the i-th Stiefel-Whitney classes of  the Tangent bundle of the manifold. For orientable manifolds $w_{1}^{TM} = 0$. If the manifold is additionally spin, $w_{2}^{TM} = 0$. We point that if $x \in H^{1}(X_{2n},\ZZ_{2})$ then $x^{2m} = \Sq^{1}(x^{2m-1})$.  Therefore, for $X_{2n}$ orientable  $\int_{X_{2n}} x^{2n} = \int_{X_{2n}} x^{2n-1} \cup w_{1}^{TM} = 0$.
\subsection{\texorpdfstring{$ \Om(N) $}{O(N)} bundles}\label{app:OmBundles}

We start by introducing the characteristic classes of the $\Om(N)$ bundles we use: the Stiefel-Whitney (SW) characteristic classes denoted by $w_{i}\in H^{i}(\ZZ_{2})$ and the Pontraygin classes $p_{i} \in H^{4i}(\ZZ)$. As we working in $4$ dimensions, we are only interested in the first four SW classes and the first Pontraygin class. Note that for $\Om(N)$ bundles, we have $w_{i}=0$ if $i>N$. Furthermore, if our bundle can be restricted to a $\SO(N)$ bundle, then  $w_{1} = 0 $ and the Pontraygin class $p_{1}$ is related to the gauge curvature $ F $ by
\begin{align}
p_{1}[\SO(N)] = \frac{1}{8\p^{2}}\int \Tr_{\SO(N)}[F\wedge F],
\end{align}
where the trace is evaluated in the vector representation of $\SO(N)$.

There is a relation between $ \Sq^{i} $ and the SW classes $ w_{j} $ of a general $ \Om(N)$ bundle (see \cite{milnor1974characteristic}, §8):
\[ \Sq^{j} w_{i} = \sum_{k = 0 }^{j} \binom{j-i}{k}w_{j-k} \cup w_{i+k}\]
where $w_{0 } = 1$ and $\binom{x}{i} = x(x-1)\dots(x-i+1)/i!$. In particular, $\Sq^{1}(w_{2n}) =w_{2n }\cup w_{1} + w_{2n+1}$ and $ \Sq^{1}(w_{2n+1}) =w_{1}\cup w_{2n+1}$.

The following identity is valid (see Theorem C of Ref.~\cite{CohomologyRealGrassmann}):
\be\label{eq:PontrayginSWRelation}
	p_{1}  =\Pmc(w_{2}) +  2 \cdot \lrRb{ w_{1} \cup \Sq^{1}\lrRb{w_{2}} + w_{4}}  \mod 4 
\ee
which can be equivalently written as $ p_{1}  =\Pmc(w_{2}) +  2 \cdot \lrRb{w^{2}_{1} \cup{w_{2}} +  \Sq^{1}\lrRb{w_{3}} + w_{4}}  \mod 4. $

Given an $\Om(A)$-bundle $ \a $ and an $\Om(B)$-bundle $ \b $, we can construct the Whitney sum $\a \oplus \b$ which is a $\Om(A+B)$. Naturally, the classes of $\a \oplus \b$ can be written in terms of the classes of $\a$ and $\b$ (see e.g. Ref.~\cite{brown1982cohomology}):
\begin{subequations}\label{WhitneySumPlaneBundles}
		\begin{align}
			w_{i }[\a \oplus \b]   &= \sum_{j=0}^{i} w_{j}[\a] \cup w_{i-j}[\b]\\
			p_{1}[\a \oplus \b] &= p_{1}[\a] + p_{1}[\b]  + \b_{2}(w_{1}[\a]) \cup \b_{2}(w_{1}[\b])
		\end{align}
\end{subequations}

\subsection{\texorpdfstring{$ \PSU(N) $}{PSU(N)} bundles}\label{app:PSUnbBundles}

In Ref. \cite{woodward1982classification} (and  reviewed in Ref. \cite{duncan2013components}), the author classified $\PSU(N) (\equiv \PU(N))$ bundles over spaces of dimension up to 4 by assigning two cohomology classes $ t_2 \in H^{2}(X,\ZZ_N )   $ and  $q_4 \in H^{4}(X,\ZZ)$ to each principal $ \mathrm{PU}(N) $-bundle $P\rightarrow  X $. In our notation, $ t_2 $ corresponds to $ u_2 $ , the second Stifel-Whitney class of the $ \PSU(N) $-bundle and $ q_4 $ is defined as the second Chern number of the complexified adjoint bundle that ends up being minus the instanton number of the $ \PSU(N) $-bundle\footnote{the relative minus sign with respect to Ref. \cite{woodward1982classification} is due to the convention of the Lie algebra. Here we take it to be Hermitian while in the original work they take a anti-Hermitian Lie algebra.	 }
\begin{equation}
\begin{split}
q_4(P):=&-ch_2^{\PSU(N)}(P\otimes \CC)\\
=&-\frac{1}{8\pi^2}\int_{X} Tr_{\mmrm{Adj}}[{F}_{}\wedge {F}_{}\,] =: -p_{1}[\PSU(N)]\\
\end{split}
\end{equation}
where $ F $ is the curvature of the $\PSU(N)$ bundle and the trace is taken over the adjoint representation. 

We restrict to the case where $N$ is even from now on.  The classes $q_{4}$ and $t_{2}$ are related by $ q_4(P)= (N+1)\Pmc\lrRb{t_{2}} \mod 2N  $ or in our notation
\begin{equation}\label{eq:P1PSUN}
 p_{1}[\PSU(N)] = (N-1)\Pmc(u_{2}) \mod 2N
\end{equation}
Notice that $\r_{2}\Pmc(u_{2}) =  \r_{2}\r_{N}\Pmc(u_{2}) = \r_{2}(u_{2} \cup u_{2}) = \r_{2}(u_{2}) \cup \r_{2}(u_{2})$, then $ \frac{1}{2}\int_{X_{4}} \Pmc(u_{2}) = \int_{X_{4}}\r_{2}(u_{2}) \cup w_{2}^{TM} = 0 $ when $X_{4}$ is a closed spin oriented manifold (This matches Witten's result \cite{witten2000InstantonNumbersInTorus}). 

For $N=2$ the formulas reduce to $\PSU(2) = \SO(3)$ where $u_{2} = w_{2}$ is the SW class. Eq.~\ref{eq:P1PSUN} tells us then $p_{1} = \Pmc(w_{2}) \mod 4$ which matches with Eq.~\ref{eq:PontrayginSWRelation} for $N=3$ and $w_{1}=0$.

\subsection{PSU(4) and SO(6) bundles}\label{app:PSU4vsSO6}
It is known that $\SU(4)/\ZZ_{2} = \SO(6)$, where the $\ZZ_{2}$ is generated by $-\id \in \SU(4)$. If a $\PSU(4)$-bundle can be lifted to a $\SO(6)$-bundle there must be a relation between the characteristic classes. Specifically, we can lift the $\PSU(4)$ structure when $2 \cdot u_{2} = 0$ or in other words there exists a class $w_{2} \in H^{2}(\ZZ_{2})$ such that $u_{2} = \m_{4,2}w_{2} = 2\cdot w_{2}$. This condition can be understood by thinking of $u_{2}$ as the obstruction of objects charged under the $\SU(4)$ center, more precisely, if a object of charge $q$ goes around a 2-cycle $\g$ the object picks a phase $\exp(i\frac{2\p q}{4} \int_{\g}u_{2})$. Now note that if $u_{2}=2w_{2}$, every object of even charge does not feel $w_{2}$. The charge two objects precisely corresponds to $\SO(6)$ representations. Therefore, the gauge structure can be lifted to a $\SO(6)$ bundle.

In the previous situation $  p_{1}[\PSU(4)] = \Pmc(2\cdot w_{2 }) = 4\cdot\Pmc(w_{2}) = 4 \cdot w_{2} \cup w_{2}  \mod 8 $ which means $p_{1}$ is divisible by $4$. The Pontraygin classes are related as $ p_{1}[\SO(6)] =\frac{1}{4}p_{1}[\PSU(4)] $\footnote{See the definition in terms of $F$ and use the following observation: $Tr_{\bigwedge^{2}f_{N} }[U]=\frac{\Tr[U]^{2}-\Tr[U^{2}]}{2}$, then by setting $U = e^{i\a_{a}T^{a} }e^{i\b_{b}T^{b}}$ and expanding to order $\Omc(\a_{a}\b_{b})$, we get that $Tr_{\bigwedge^{2}f_{N} }[T^{a}T^{b}]=(N-2)\Tr_{f_{N}}[T^{a}T^{b}]$. Now notice that the $\SO(6)$ vector representation corresponds to the antisymmetric square of the fundamental of $\SU(4)$($\bigwedge^{2}f_{4}$).}. This relation is consistent with the previous condition by recalling that $p_{1}[\SO(6)] = w_{2} \cup w_{2} \mod 2$. 

\section{DSL in rectangular representations}\label{sec:DSLrect}

\subsection{Emergent compact \texorpdfstring{$\U(p)_g \, \QCD_3$}{U(p)g QCE3} }

We now extend the parton construction to spin systems where the on-site $\sus$ representations with rectangular Young tableaux with $n$ rows and $p$ columns. In this cases, we can represent the spin system generators of $\sus$ as
\begin{equation}
S^{\alpha}_{\,\, \beta} = \sum_{s=1,\dots,p} f^{\dag \alpha s}_{} f^{}_{\beta s}
\end{equation}
where $s=1,\dots, p$ and $\alpha=1,\dots, N$. We then put the fermions at $n/N$ filling by gauging the $U(1)$ symmetry. At this point, the physical states are in the antisymmetric representation $(\varpi_{np})$ of $\SU(Np)$. Next, we gauge the $\SU(p)$ symmetry corresponding to rotations in the $p$ index as this extracts the states that are trivial under $\SU(p)$. In other words, it picks the irreps in the decomposition of $(\varpi_{np})$ under the branching $\SU(Np)\longrightarrow \SU(N)\times \SU(p)$ that have a trivial irrep in the second factor. From Eq.~\ref{eq:Decomp}, we see that the only factor we keep has a rectangular YT with $n$ columns and $p$ rows in the second factors and thus have the YT we want in the first factor.

We can consider now $p$ copies of the mean-field Hamiltonians with $IGG=U(1)_g$ and $N_f$ low energy Dirac fermions. We  then obtain $\U(p)\, \QCD_3$ with $\Nf$ fundamental fermions. According to Ref.~\cite{DyerMonopoleTaxonomy} and references therein, the monopoles are classified by the topological flux along the Cartan directions of the gauge field, i.e. along the diagonal $\U(1)^{p} $ subgroup of $\U(p)$, modulo the action of the Weyl group that in this case is just permutation of the $\U(1)$'s. The monopole with the lowest scaling dimension corresponds to inserting a $2\p$ flux along a single direction of the $\U(p)$ gauge field. The symmetry properties of this monopoles are similar to those of $\U(1)\,\, \QED_{3}$, the difference being the Berry phase factors comming from the filled bands. 

\subsection{Emergent \texorpdfstring{$\Sp(p)_g \, \QCD_3$}{Sp(p)g QCD3} }\label{sec:emergentSpQCD}

Consider now spin $S=\frac{p}{2}$ representations of $\SU(2)_s$. In this section, we will discuss a parton decomposition that has a larger gauge redundancy than $\U(p)$, namely $\Sp(p)$ ($\Sp(1)=\SU(2) $). The enhancement of the gauge redundancy is a generalization of the $S=\frac{1}{2}$ \cite{PhysRevX.8.011012, PhysRevX.7.031051} and $S=1$ \cite{PhysRevLett.108.087204, PhysRevB.91.195131} cases. 

Consider $2p$ fermionic partons $f_{\alpha s}$ where $\alpha=\ua,\da$ and $s=1,\dots,p$. Introduce the parton operators
\begin{equation}
{F} =\left(
\begin{array}{cccccc}
f_{\ua 1}^{} & \dots & f_{\ua p}^{}  & -f_{ }^{\dag \da 1} & \dots & -f_{}^{\dag \da p}   \\
f_{\da 1}^{}  & \dots  & f_{\da p}^{}  & f_{}^{\dag \ua 1} & \dots & f_{}^{\dag \ua p}   \\
\end{array}
\right),
\end{equation}
then the spin operators are written as 
\begin{equation}\label{eq:majoranaParton}
\mmbf{S}= \frac{1}{2} f^{\dag \alpha s}_{} \vec{\s}_{\alpha}^{\,\,\,\beta}f_{\beta s}^{} = \frac{1}{4}\Tr[{F}^{\dag}_{}\vec{\s}F^{}_{}  ].
\end{equation}
We see that there is an apparent larger $\U(2p)$ redundancy $F\rightarrow F U^{\dag}$ with $U\in \U(2p)$. Nevertheless, the $F$ operators satisfy a reality condition $F^{\dag}=\s^2\,F^{\top}\rho^2$\,\footnote{$\rho^a$ are Pauli matrices acting on particle-hole channel.}, then $U$ must satisfy $U^{\top}\rho^2 U=\rho^2$ and thus the gauge symmetry is $\Sp(p)$. If we consider $p$ copies of a mean-field Hamiltonian with $IGG=\Sp(1)=\SU(2)$, the resulting mean-field Hamiltonian has $IGG=\Sp(p)$. This happen for the $\p$-flux state on the square lattice and zero flux state of the honeycomb lattice.

\subsection{\texorpdfstring{$ \Sp(N)_s $}{Sp(N)s} spin systems}

Let $(\vartheta^{}_{k})$ be the heighest wight representations of $\Sp(N)_s$ that corresponds to the $k$-fold antisymmetric product of the defining representations modulo contraction with the invariant antisymmetric tensor\footnote{Equivalently $(\vartheta_{k})$ is the largest representation when we restrict the $(\varpi_k)$ representation of $\SU(2N)$ to its $\Sp(N)$ subgroup.}. For a parton construction for the case the irreps with heighest weight $p\vartheta_{N}$, we introduce the parton operator 
\begin{equation}
F = \left(
\begin{array}{cccccc}
f_{1, 1}^{} & \dots & f_{1, p}^{}  & -f_{ }^{\dag N+1, 1} & \dots & -f_{}^{\dag N+1, p}   \\
\vdots & \ddots& \vdots & \vdots & \ddots& \vdots\\
f_{N, 1}^{} & \dots & f_{N, p}^{}  & -f_{ }^{\dag 2N 1} & \dots & -f_{}^{\dag 2N, p}   \\
f_{N+1, 1}^{} & \dots & f_{N+1, p}^{}  & f_{ }^{\dag 1, 1} & \dots & -f_{}^{\dag 1, p}   \\
\vdots & \ddots& \vdots & \vdots & \ddots& \vdots \\
f_{2N, 1}^{}  & \dots  & f_{N+1, p}^{}  & f_{}^{\dag N, 1} & \dots & f_{}^{\dag N, p}   \\
\end{array}
\right),
\end{equation}
where $f_{\a,s}$, with $\a=1,\dots,2N$ and $s =1,\dots,p$, are fermionic annhilation operators. We then gauge the $\Sp(p)$ symmetry that acts as $F\longrightarrow F\cdot U^{\top}$. In order to prove that gauging this symmetry gives the right Hilbert space, recall that the Hilbert space whose algebra is generated by $f_{\a,s}$ is the sum of the spinor representations of $\Spin(4pN)$. This is so because we can construct $4pN$ Majorana operators out of the $2pN$ complex fermions. Then, we need to consider the branching under $\Spin(4pN)\longrightarrow \Sp(N)\times\Sp(p)$ and find the factors that are trivial in the second factor. We have not been able to find a proof but using SAGE \cite{sagemath}, we verified that gauging  gives the right answer for all pairs of $(N,p)$ such that $ pN\le16 $. 

We can consider $pN$ copies of a mean-field Hamiltonian for $S=\half$ systems with $IGG=\SU(2)_g$ and $2M$ low energy Dirac fermions. We expect an emergent $\Sp(p)_g\, \QCD_3 $ with $\Nf=NM$. If we instead consider copies of mean-field Hamiltonians with $IGG=\U(1)_{g}$, we expect to find emergent $\U(p)_g\, \QCD_3 $ with $\Nf=2NM$. 

This motivate the possibility of a stable $\U(1)$ DSL on the vector $\SO(5)(\cong \Sp(2)/\ZZ_2)$ Heisenberg model on the triangular lattice as the bare monopoles still carry $\Kbs$ momenta\cite{song2018spinon,song2018spinonNumeric} that cannot be cancelled by the zero-modes contributions. The monopole quantum numbers can be easily found by starting from \ref{eq:Rep708} and restricting the $\SO(6)$ reps to $\SO(5)$ reps: 
\begin{equation}
    \begin{split}
        \mmbf{20} & \longrightarrow \mmbf{14} \oplus \mmbf{5} \oplus \mmbf{1} \\
        \mmbf{15} & \longrightarrow \mmbf{10} \oplus \mmbf{5}\\
        \mmbf{1} & \longrightarrow  \mmbf{1}
    \end{split}
\end{equation}
where $\mmbf{1}$ is the scalar, $\mmbf{5}$ is the vector, $\mmbf{10}$ is the adjoint and $\mmbf{14}$ is the symmetric traceless representation.

\end{widetext}

\newpage

\end{document}